\documentclass[10pt,a4paper]{article}
\usepackage{setspace}
\usepackage{float}

\usepackage{natbib}
\usepackage{amsmath}
\usepackage{amsfonts}
\usepackage{amssymb}
\usepackage{amsthm}
\usepackage{graphicx}
\usepackage{hyperref}
\usepackage{bbm}
\usepackage{cancel}
\usepackage{caption}
\usepackage{xcolor}
\usepackage{geometry}
\usepackage{algorithm}
\usepackage{algpseudocode}
\usepackage[bb=boondox]{mathalfa}
\usepackage{tikz}
\usepackage{pgfplots}
\usepgfplotslibrary{groupplots,fillbetween}
\definecolor{ORcolor}{RGB}{255,127,14}
\definecolor{IPWcolor}{RGB}{31,119,180}
\definecolor{DRcolor}{RGB}{44,160,44}
\definecolor{PIGcolor}{RGB}{214,39,40}
\usepackage{multirow}

\usepackage{apptools}  % appendix theorems
\usepackage{chngcntr}  % appendix theorems
\AtAppendix{\counterwithin{lemma}{section}}
\AtAppendix{\counterwithin{assumption}{section}}

\usepackage{booktabs}       % professional-quality tables
\usepackage{algorithm}
\usepackage{algpseudocode}

\theoremstyle{plain}
\newtheorem{lemma}{Lemma}
\newtheorem{theorem}{Theorem}
\newtheorem{assumption}{Assumption}
\newtheorem{prop}{Proposition}
\newtheorem{cor}{Corollary}
\newtheorem*{remark}{Remark}
\newtheorem{example}{Example}

\newcommand{\tbl}[1]{\caption{#1}}
\newcommand{\E}{\mathbb{E}}
\newcommand{\pr}{P}%{\mathbb{P}}
\newcommand{\ind}{\perp\!\!\!\!\perp}

\newcommand{\full}{\mathrm{full}}

\DeclareMathOperator{\argmax}{argmax}

\usepackage{contour}
\contourlength{0.03em}
\newcommand{\boldwidth}{\contour{black}}

\title{Sharp sensitivity bounds and maximin policy learning \\under violations of principal ignorability}
\author{Sihyung Park \quad Wenbin Lu \quad Shu Yang\\[0.5em]{\normalsize Department of Statistics, North Carolina State University}}
\date{}

\begin{document}

	\maketitle
	
\begin{abstract}
Identification of assignment effects among compliers often relies on the untestable assumption of principal ignorability. Sensitivity analyses based only on outcome means and ranges can admit complier distributions incompatible with the observed data. We develop a distributional sensitivity model preserving the full observed outcome distributions and identified complier proportions. The challenge is to recover latent complier distributions within observed mixtures while ensuring compatibility across assignment arms. We characterize exactly which pairs arise from a single compatible full-data law, obtaining sharp bounds on effects and policy values. This characterization shows how complier proportions constrain departures from principal ignorability and why informative bounds remain without a finite sensitivity restriction. Accounting for principal scores in both population weights and sensitivity limits, we establish conditions for doubly robust consistency, semiparametric efficiency, and conservative bounds under quantile misspecification. Double robustness requires consistent principal score estimation. We also develop maximin decision-tree policies with value consistency and inference guarantees. Numerical studies demonstrate gains from retaining distributional information, and an Oregon Health Insurance Experiment application examines implications for complier effects and assignment policies.
\end{abstract}

\section{Introduction}

When treatment assignment affects uptake, an important question is its effect among individuals who would receive treatment only if offered. Principal stratification defines these individuals as compliers through their potential treatment receipt under each assignment \citep{psog2,ps1,ps2}. Even in a randomized trial, however, each individual is observed under only one assignment. For example, those who receive treatment in the offer arm include both compliers and individuals who would have received treatment without the offer. Their observed outcomes therefore do not identify the outcomes of compliers alone, and randomization does not, by itself, identify the mean outcome under either assignment among compliers.

The Oregon Health Insurance Experiment illustrates the practical importance of this distinction. A lottery gave selected households the opportunity to apply for Medicaid \citep{finkelstein2012oregon}. Among individuals whose enrollment depended on lottery selection, what effect did this opportunity have on mental health? A related policy question is to whom such an opportunity should be offered when its effect is uncertain. Both questions concern the effect of assignment rather than treatment receipt, allowing the offer itself to influence outcomes through channels other than enrollment.

Principal ignorability (PI) permits identification by relating complier outcomes to those in the observed assignment--receipt groups, conditional on baseline covariates. It underlies principal-score and semiparametric methods for estimating principal effects without an exclusion restriction \citep{ding2017principal,ps3,lu2026principal,chen2026principal}. Its credibility, however, depends heavily on the information captured by the covariates. Individuals with similar recorded health histories may face different barriers to care that affect both enrollment and mental health. Although randomization balances assignment groups, it does not eliminate outcome differences between latent compliance strata. Moreover, because stratum membership is only partially observed, PI cannot be verified from the data. Sensitivity analysis therefore provides a way to assess how departures from PI affect conclusions about complier effects and preferred assignment policies.

Earlier work examines sensitivity of principal score estimators to PI violations \citep{stuart2015assessing}, while joint modeling approaches allow departures through multiple imputation \citep{wang2023sensitivity}. For one-sided noncompliance, \citet{nguyen2024sensitivity} relate stratum-specific outcome means through odds ratios, generalized odds ratios, mean ratios or standardized mean differences, and develop regression, weighting and influence-function-based estimators. \citet{wu2024partial} derive partial identification regions under joint models that specify how outcomes depend on the potential intermediate variables. Our objective is to characterize sharp bounds on complier effects and policy values without specifying parametric models for outcomes or intermediate variables.

Alongside specifying departures from PI, sensitivity analysis must determine which candidate complier distributions remain compatible with the observed mixtures. \citet{chen2026beyond} incorporate observed cell means and principal-stratum mixing proportions into their bounds, but assess sensitivity to PI violations indirectly through restrictions involving an unmeasured confounder. We seek a sensitivity model that directly restricts departures of the complier outcome distributions from the observed mixtures while preserving the identified complier proportions and the full observed outcome distributions.

We propose a distributional sensitivity model that represents each observed outcome distribution as a mixture with an identified complier fraction. The model bounds the conditional density of complier outcomes relative to the observed mixture, with a sensitivity parameter that controls departures from PI. This restriction admits an interpretation through the odds of complier membership, connecting our approach to marginal sensitivity analysis \citep{tan2006distributional,zhao2019sensitivity,dorn2023sharp,dorn2025doubly}. Our main result characterizes exactly which pairs of complier outcome distributions are jointly compatible with a single full-data law consistent with the observed data and the underlying causal model. This characterization yields sharp bounds on complier effects and policy values and reveals how the complier fraction constrains departures from PI in each arm. We further show that, under the same sensitivity restriction, retaining only outcome means and ranges can produce substantially wider bounds.

We develop estimators of these bounds and use them to learn treatment assignment policies. Our endpoint estimators account for the principal scores in both the complier weights and the density ratio limits. We establish conditions for doubly robust consistency and efficient inference, and show when misspecified quantile models nevertheless yield conservative endpoints. For policy learning, we build on empirical welfare maximization \citep{kitagawa2018welfare,athey2021policy,sverdrup2020policytree} and methods for principal strata \citep{chu23c,park2026evaluating} to select the policy that maximizes the worst-case expected outcome among compliers. Related work optimizes worst-case regret under unmeasured confounding \citep{kallus2018confounding}; here, uncertainty concerns latent complier outcomes, and the objective is their worst-case policy value. For fixed-depth decision trees, we establish value consistency and inference for the optimal value under separation and nuisance-rate conditions.

The remainder of the paper is organized as follows. Section~\ref{sec:setup} introduces the setting. Sections~\ref{sec:msm} and~\ref{sec:id} develop the sensitivity model, characterize compatible distributions, and derive sharp bounds and their estimators. Section~\ref{sec:itr} addresses policy learning, while Section~\ref{sec:sim} evaluates endpoint estimation and worst-case policy performance through simulations. Section~\ref{sec:oregon} applies the proposed methods to the Oregon Health Insurance Experiment, where modest departures from PI allow for a null complier effect, whereas larger departures favor universal access within the considered policy class in the absence of access costs or capacity constraints. The Appendix provides proofs and implementation details.

\section{Problem setup}  \label{sec:setup}
\subsection{Noncompliance problem} \label{sec:noncompliance}
Following standard notation in causal inference and the individualized treatment regime literature, let $X$ denote baseline covariates, $Z\in\{0,1\}$ the randomized treatment assignment, and $S\in\{0,1\}$ the treatment actually received. An individual is considered compliant with the assigned treatment when $Z=S$ and noncompliant otherwise. Let $Y$ denote the observed outcome, oriented so that larger values indicate more favorable outcomes. The observed data are denoted by $O=(X,Z,S,Y)$.

We define the assignment propensity scores as $e_z(X)=\pr(Z=z\mid X)$ for $z\in\{0,1\}$, so that $e_0(X)=1-e_1(X)$, and the conditional treatment receipt probabilities as $p_{z,s}(x)=\pr(S=s\mid X=x,Z=z)$. For notational convenience, we write $p_z(x)=p_{z,z}(x)$. We further define the outcome regression function $\mu_z(x)=\E[Y\mid X=x,Z=S=z]$ for individuals whose received treatment agrees with assignment $z$.

We adopt the potential outcomes framework. For $T\in\{S,Y\}$, let $T(z)$ denote the potential value of $T$ under assignment $Z=z$. In particular, $S(z)$ represents the treatment that would be received under assignment $z$, and $Y(z)$ represents the corresponding potential outcome. We assume consistency and no interference, such that $S=S(Z)$ and $Y=Y(Z)$.

\begin{assumption}[Stable unit treatment value assumption (SUTVA)] \label{A:consistency}
    For $T\in\{S,Y\}$, the potential outcomes $T(0)$ and $T(1)$ are well-defined, with no relevant hidden versions of treatment assignment. Furthermore, one individual's treatment assignment does not affect another individual's potential outcomes. The observed variables satisfy
    $
        T=ZT(1)+(1-Z)T(0).
    $
\end{assumption}
\noindent

\begin{assumption}[Conditional exchangeability] \label{A:nuc}
    $(S(0),S(1),Y(0),Y(1)) \ind Z \mid X$.
\end{assumption}
\noindent
This assumption requires that treatment assignment be independent of the potential treatment receipt and outcomes conditional on baseline covariates. It holds by design in randomized trials when assignment is randomized conditional on $X$. In observational studies, it is plausible when $X$ includes all common causes of treatment assignment and the potential outcomes, including potential treatment receipt.

\subsection{Principal stratification}  \label{sec:ps}
To account for noncompliance, we adopt the principal stratification framework, which classifies individuals according to their potential treatment receipt under each assignment. Let $U=(S(0),S(1))$ denote the principal stratum membership, defining four latent strata as summarized in Table \ref{table:strata}. {We note that although receipt $S$ is observed after assignment, $U$ is an assignment-invariant latent variable, not an observed post-assignment subgroup.} In particular, $U=01$ identifies compliers, who would receive treatment if and only if assigned to it.

Rather than focusing on the population average treatment effect (ATE), we target the potential outcome means within the complier stratum,
\begin{gather*}
\psi_0=\E[Y(0)\mid U=01], \qquad
\psi_1=\E[Y(1)\mid U=01],
\end{gather*}
and the corresponding complier average treatment assignment effect,
\begin{equation*}
\psi_C = \E[Y(1)-Y(0)\mid U=01]=\psi_1-\psi_0.
\end{equation*}

\begin{table}[H]
\tbl{Description of Principal Strata}
\centering
{\footnotesize	\begin{tabular}{crrrcrrr}
	$U$  & Stratum type    & Description \\ 
	\hline
	11 & Always-taker & Patients who would be treated regardless of assignment. \\
	01 & Complier     & Patients who comply with assigned treatment. \\
	10 & Defier          & Patients who would defy assigned treatment. \\
	00 & Never-taker  & Patients who would be untreated regardless of assignment.
	\end{tabular} }
	\label{table:strata}
\end{table}

Because principal strata are defined by joint potential treatment receipt under both assignments, stratum membership is generally not fully observed. Identifying complier-specific potential outcome means therefore requires additional assumptions. We first introduce the monotonicity assumption, commonly imposed in the identification of local causal effects \citep{ps1,ps2}.

\begin{assumption}[Monotonicity] \label{A:monoton} \ %\label{A2} \
	$S(1) \ge S(0)$ almost surely.
\end{assumption}
\noindent
Monotonicity rules out defiers ($U=10$) and is therefore also referred to as the no-defiers assumption. Under monotonicity, the observed group with $Z=S=1$ consists of compliers ($U=01$) and always-takers ($U=11$), whereas the group with $Z=S=0$ consists of compliers and never-takers ($U=00$). Together with Assumptions \ref{A:consistency} and \ref{A:nuc}, monotonicity identifies the conditional probability of complier membership as
\begin{align*}
	\pi_{C}(X) := \pr(U=01 | X) = p_1(X) - p_{0,1}(X) = p_0(X) - p_{1,0}(X).
\end{align*}
However, identifying the proportion of compliers does not suffice to identify their potential outcome means, because the observed assignment--receipt groups contain mixtures of principal strata. To distinguish the complier-specific means from those of the other strata, we consider the following assumption: 
for each $x$ in the relevant covariate support,
\begin{enumerate}
	\item[(i)] $\E[Y(1) \mid X=x, U=01] = \E[Y(1) \mid X=x, U=11]$.
	\item[(ii)] $\E[Y(0) \mid X=x, U=01] = \E[Y(0) \mid X=x, U=00]$.
\end{enumerate}
This assumption, mean principal ignorability, requires that, conditional on baseline covariates, the potential outcome mean under treatment assignment does not differ between compliers and always-takers, and the potential outcome mean under control assignment does not differ between compliers and never-takers. Under Assumptions \ref{A:consistency} and \ref{A:nuc}, this implies
\begin{gather*}
	\E[Y|X=x, Z=1, U=01] = \E[Y|X=x, Z=S=1] = \mu_1(x), \\
	\E[Y|X=x, Z=0, U=01] = \E[Y|X=x, Z=S=0] = \mu_0(x).
\end{gather*}
Consequently, the complier-specific potential outcome means are identified by standardizing the observed outcome regressions over the covariate distribution of compliers.

Principal ignorability requires that baseline covariates adequately account for differences in potential outcome means across principal strata. In practice, however, the observed covariates may fail to capture important factors associated with both compliance behavior and outcomes, making this assumption difficult to justify. Moreover, because principal ignorability cannot generally be tested using observed data, violations may remain undetected, raising concerns about the credibility of causal conclusions that rely on it.

	\section{Distributional sensitivity to principal ignorability}\label{sec:msm}
\subsection{Observed mixtures and admissible complier distributions}

Under Assumptions~\ref{A:consistency}, \ref{A:nuc}, and \ref{A:monoton}, the group $Z=S=1$ consists of compliers and always-takers, whereas $Z=S=0$ consists of compliers and never-takers. The conditional proportion of compliers in group $z$ is identified as $\gamma_z(x)=\pi_C(x)/p_z(x)$. For $0<\gamma_z(x)<1$, the observed conditional outcome distribution can be expressed as
\[
P_{Y\mid X=x,Z=S=z}
=\gamma_z(x)P_{Y(z)\mid X=x,U=01}
+\{1-\gamma_z(x)\}P_{Y(z)\mid X=x,U=u_z},
\qquad u_1=11,\quad u_0=00.
\]
Although the mixing proportion and observed mixture distribution are identified, their component distributions are not. Any sensitivity analysis that varies the complier component must therefore remain compatible with the observed mixture. We represent a candidate complier distribution by its Radon--Nikodym derivative
\[
W_z(x,y)=\frac{d\widetilde P_{Y(z)\mid X=x,U=01}}{dP_{Y\mid X=x,Z=S=z}}(y).
\]
Compatibility with the mixture identity requires $W_z\ge0$, $\E[W_z(X,Y)\mid X,Z=S=z]=1$, and $\gamma_z(X)W_z(X,Y)\le1$ almost surely. The last inequality ensures that the implied distribution of the other stratum is nonnegative.
%: its density relative to the observed distribution is $\{1-\gamma_zW_z\}/(1-\gamma_z)$. 
%Together, these restrictions preserve the entire observed outcome distribution, including its support and the probability mass available to each principal stratum.

For bounds $L_z$ and $U_z$, we restrict departures of the complier distribution from the observed mixture by requiring its density ratio to belong to the following class:
\begin{align}
\mathcal E_z(L_z,U_z) = \{W_z:\; &L_z(X)\le W_z(X,Y)\le U_z(X), \label{def:MSM2} \\
&\E[W_z(X,Y)\mid X,Z=S=z]=1 \}.  \label{eq:weights}
\end{align}
%\begin{equation}
%\E[W_z(X,Y)\mid X,Z=S=z]=1.  \label{eq:weights}
%\end{equation}
To respect the mixture constraint $\gamma_z(x)W_z(x,y)\le1$, we specify density ratio bounds that depend on the identified complier fraction. For a fixed finite sensitivity parameter $\Lambda_z\ge1$, we choose
\[
L_z(X)=\frac{1}{\gamma_z(X)+\Lambda_z\{1-\gamma_z(X)\}},\qquad
U_z(X)=\frac{1}{\gamma_z(X)+\Lambda_z^{-1}\{1-\gamma_z(X)\}}.
\]
We denote the class with these bounds by $\mathcal E_z(\Lambda_z)$. The parameter $\Lambda_z$ controls how far the complier density may depart from the observed mixture: $W_z(x,y)>1$ assigns relatively more density to outcome $y$ among compliers, and $W_z(x,y)<1$ assigns less. For example, $\Lambda_z=2$ permits density ratios between $1/\{2-\gamma_z(x)\}$ and $2/\{1+\gamma_z(x)\}$. As the complier fraction approaches one, both bounds approach one, reflecting the diminishing scope for departures from the observed mixture. Moreover, since $\gamma_zU_z\le1$, every normalized candidate also yields a nonnegative companion distribution. The identity $W_z=1$ is admissible at every sensitivity level. When $\Lambda_z=1$, it is the unique admissible candidate, implying equality between the complier distribution and the observed mixture. This distributional equality implies the mean principal ignorability condition in Section~\ref{sec:ps}. If $\gamma_z(x)=1$, the observed group contains only compliers, so its outcome distribution is identified without sensitivity analysis. This includes treated recipients under one-sided noncompliance.

When the candidate distribution coincides with the true complier distribution, a change of measure gives
\[
\E[Y(z)\mid X,U=01]=\E[W_z(X,Y)Y\mid X,Z=S=z].
\]
Optimizing this expectation over the normalized candidate class yields bounds on the complier mean. For these bounds to be sharp for causal effects, we must establish that the candidates in both arms can be simultaneously realized under a single full-data distribution that preserves the observed law. We establish this joint compatibility in Section~\ref{sec:datacompatible}.

\subsection{Observable constraints and the model's behavior}\label{sec:distribution_information}
For $z\in\{0,1\}$, define the observed conditional outcome distribution and its quantile function by $F_z(y\mid x)=P(Y\le y\mid X=x,Z=S=z)$ and $Q_{z,x}(u)=\inf\{y:F_z(y\mid x)\ge u\}$, $0<u<1$.
In the following proposition, optimization over $\mathcal E_z(\Lambda_z)$ is conditional on $X=x$: the bounds and normalization in \eqref{def:MSM2}--\eqref{eq:weights} apply to $W_z(x,\cdot)$ under $F_z(\cdot\mid x)$.

\begin{prop}[Compliance information and limiting identification regions]\label{prop:mixture_limits}
For $z=0,1$ and $P_X$-almost every $x$, suppose $\E[|Y|\mid X=x,Z=S=z]<\infty$ and $0<\gamma_z(x)<1$.
\begin{enumerate}
\item With $F_z(\cdot\mid x)$ and finite $\Lambda_z\ge1$ held unchanged, increasing $\gamma_z(x)$ narrows the admissible weight interval at $x$. Consequently,
\[
\inf_{W_z\in\mathcal E_z(\Lambda_z)}\E[W_z(X,Y)Y\mid X=x,Z=S=z]
\]
is nondecreasing in $\gamma_z(x)$, and the corresponding supremum is nonincreasing. Both converge to $\mu_z(x)$ as $\gamma_z(x)\uparrow1$.
\item With $F_z(\cdot\mid x)$ and $\gamma_z(x)$ held unchanged, the set of admissible weights at $x$ expands as $\Lambda_z$ increases. At $\Lambda_z=1$, both conditional endpoints equal $\mu_z(x)$; both converge to $\mu_z(x)$ as $\Lambda_z\downarrow1$.
\item As $\Lambda_z\to\infty$,
\begin{align*}
&\inf_{W_z\in\mathcal E_z(\Lambda_z)}\E[W_z(X,Y)Y\mid X=x,Z=S=z] \to \frac{1}{\gamma_z(x)}\int_0^{\gamma_z(x)}Q_{z,x}(u)\,du,\\
&\sup_{W_z\in\mathcal E_z(\Lambda_z)}\E[W_z(X,Y)Y\mid X=x,Z=S=z] \to \frac{1}{\gamma_z(x)}\int_{1-\gamma_z(x)}^1Q_{z,x}(u)\,du.
\end{align*}
These limits are the smallest and largest possible values of $\E[Y(z)\mid X=x,U=01]$ over pairs of conditional outcome laws satisfying
\[
P_{Y\mid X=x,Z=S=z}
=\gamma_z(x)P_{Y(z)\mid X=x,U=01}
+\{1-\gamma_z(x)\}P_{Y(z)\mid X=x,U=zz}.
\]
The quantile integrals allow fractional allocation at atoms.
\end{enumerate}
\end{prop}
The limiting endpoints have the tail-trimming form used in \citet{lee2009training}. Here, the identified complier fraction determines the trimming mass in each assignment arm.

\begin{figure}[!htbp]
\centering
\includegraphics[width=\textwidth]{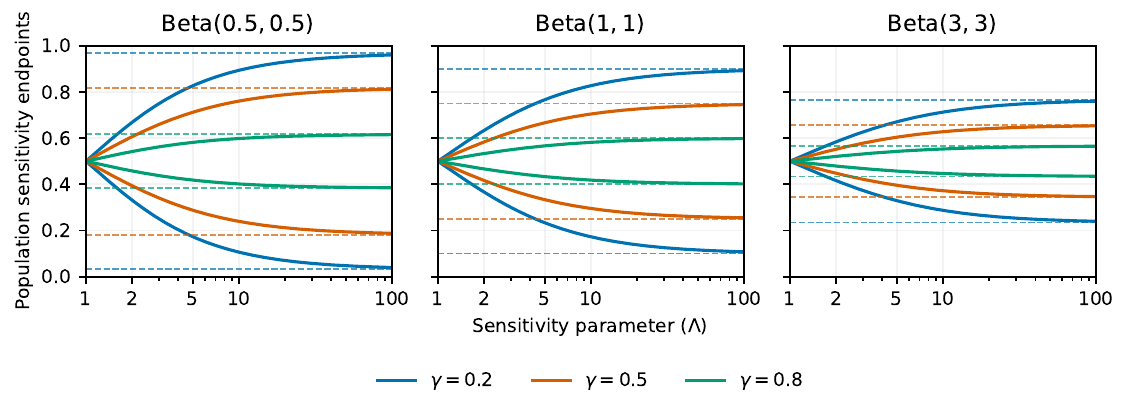}
\caption{Population sensitivity endpoints for three symmetric beta outcome distributions. All have mean $1/2$ and support $[0,1]$; Beta$(1/2,1/2)$ places more mass near the boundaries, Beta$(1,1)$ is uniform, and Beta$(3,3)$ concentrates mass near the mean. Colors identify complier fractions $\gamma_z(x)\in\{0.2,0.5,0.8\}$. Solid curves show the lower and upper endpoints; horizontal dashed lines of the same color give their $\Lambda_z\to\infty$ limits. The horizontal axis is logarithmic.}
\label{fig:population_mixture}
\end{figure}

As $\Lambda_z\to\infty$, the sensitivity bounds approach the full range of complier means compatible with the observed outcome distribution and its identified complier fraction. Importantly, these limits can remain informative even without restrictions on departures from PI. To maximize the complier mean, the highest $\gamma_z(x)$ fraction of the observed outcome distribution is assigned to compliers; assigning the lowest fraction minimizes it. The corresponding upper- and lower-tail averages give the endpoints in Proposition~\ref{prop:mixture_limits}. Both endpoints remain finite for integrable outcomes, even with unbounded support. Thus, the observed mixture alone can provide informative bounds on complier means despite arbitrarily large departures from PI.
Figure~\ref{fig:population_mixture} shows how the interval narrows as the complier fraction increases and how its width depends on where the observed distribution places its mass. 
%The beta distributions share a mean and support but differ in both shape and dispersion.

\begin{cor}\label{cor:information_relaxation}
For $z=0,1$ and $P_X$-almost every $x$, suppose $0<\gamma_z(x)<1$ and the outcome range is $[0,1]$. Allow the conditional outcome law to vary over distributions on $[0,1]$ with the same mean $\mu_z(x)$, retaining the same $\gamma_z(x)$, $\Lambda_z$, weight limits, and conditional normalization. The sharp lower and upper conditional means over this relaxed class are
\begin{align*}
&\max\{L_z(x)\mu_z(x),\,1-U_z(x)\{1-\mu_z(x)\}\}, \quad  \min\{U_z(x)\mu_z(x),\,1-L_z(x)\{1-\mu_z(x)\}\},
\end{align*}
and the interval preserving $F_z(\cdot\mid x)$ is contained in this relaxed interval.
\end{cor}

Corollary~\ref{cor:information_relaxation} highlights that for a fixed sensitivity level and complier fraction, replacing this distribution with only its mean and range can yield wider bounds, as the relaxed analysis permits allocations to compliers that are incompatible with the observed outcome probabilities. 
Figure~\ref{fig:targeted_comparisons}(a) compares the full-law endpoints with the mean-and-range relaxation in Corollary~\ref{cor:information_relaxation}. At $\gamma=0.5$ and $\Lambda=5$, all three beta distributions have mean $0.5$ and natural range $[0,1]$, yielding the same relaxed interval $[0.167,0.833]$. Retaining the full observed distribution instead yields intervals of $[0.288,0.712]$, $[0.333,0.667]$, and $[0.396,0.604]$ for Beta$(1/2,1/2)$, Beta$(1,1)$, and Beta$(3,3)$, respectively. These intervals are approximately 36\%, 50\%, and 69\% narrower than the relaxed interval.

\subsection{Odds ratio interpretation and connection to marginal sensitivity analysis}
The sensitivity model also admits an interpretation in terms of complier membership odds. Define the candidate outcome-dependent complier probability within group $z$ as
$\widetilde\gamma_{z}(x,y)=\widetilde P(U=01\mid X=x,Y=y,Z=S=z)$.
By Bayes' rule, $\widetilde\gamma_{z}(x,y)=\gamma_z(x)W_z(x,y)$. For interior mixtures, the density ratio restriction is equivalent to
\begin{align}
\Lambda_z^{-1}\le
\frac{\widetilde\gamma_{z}(X,Y)/\{1-\widetilde\gamma_{z}(X,Y)\}}
{\gamma_z(X)/\{1-\gamma_z(X)\}}
=\frac{W_z(X,Y)\{1-\gamma_z(X)\}}{1-\gamma_z(X)W_z(X,Y)}
\le\Lambda_z.  \label{def:MSM1}
\end{align}
The normalization condition is equivalent to $\E[\widetilde\gamma_{z}(X,Y)\mid X,Z=S=z]=\gamma_z(X)$. Thus, $\Lambda_z=2$ allows the outcome-dependent odds of complier membership to range from one-half to twice the odds based on covariates alone, while preserving the identified complier fraction. Allowing separate $\Lambda_0$ and $\Lambda_1$ accommodates different departures from PI in the two observed mixtures. 

The odds restriction connects our model to marginal sensitivity analysis \citep{tan2006distributional,zhao2019sensitivity}. \citet{dorn2023sharp} establish data compatibility and joint attainability for population treatment effects, while \citet{dorn2025doubly} derive sharp bounds through distributionally robust optimization and develop robust, efficient estimators. Their counterfactual density ratio representation compares outcome distributions across assignment groups, with bounds $[\Lambda^{-1},\Lambda]$. In contrast, our target is a latent component of the observed mixture, where assigning mass to compliers constrains the companion distribution through its density $(1-\gamma_z(x)W_z(x,y))/(1-\gamma_z(x))$. Although the optimization principle remains applicable, characterizing the admissible component distributions requires additional arguments beyond optimizing over fixed density ratio bounds.

%%
%\subsection{Interpreting and selecting sensitivity levels}\label{sec:calibration}
%
%The sensitivity parameter bounds the odds of complier membership conditional on the outcome, rather than specifying an outcome effect or an absolute change in membership probability. 
%To calibrate sensitivity levels, one can examine representative covariate profiles, compute their identified complier fractions, and assess the plausibility of outcome-dependent membership probabilities near these limits. Conversely, accommodating a proposed probability $r\in(0,1)$ at a given profile requires at least $\max\{\operatorname{odds}(r)/\operatorname{odds}(\gamma_z(x)),\operatorname{odds}(\gamma_z(x))/\operatorname{odds}(r)\}$. These comparisons help interpret plausible departures from PI.
%
%Because compliance mechanisms and mixture proportions may differ across arms, we recommend assessing sensitivity levels separately before adopting a common value. The level at which an estimated effect interval first includes zero provides a summary of sensitivity along the chosen parameter path. 

%
\subsection{Characterization of compatible complier distributions}  \label{sec:datacompatible}

Fix the observed law $P$. For each arm, let $\mathcal G_z(P,\Lambda_z)$ consist of the candidate conditional laws that assign to every Borel set $B$ the probability $\int_B W_z(x,y)\,dP_{Y\mid X=x,Z=S=z}(y)$ for $W_z\in\mathcal E_z(\Lambda_z)$.
Each weight defines one conditional law, since \eqref{eq:weights} gives total mass one, and the probability of each Borel set varies measurably with $x$. We characterize conditional laws that agree $P_X$-almost surely.

\begin{theorem}[Characterization of compatible complier distributions]\label{thm:compat_plus}
Let $P$ have real-valued integrable outcomes, a standard Borel covariate space, $0<e_1(X)<1$, and $0<\pi_C(X)<\min\{p_1(X),p_0(X)\}<1$. Let $\mathfrak P(P;\Lambda_0,\Lambda_1)$ be the class of all full-data laws preserving $P$ and satisfying Assumptions~\ref{A:consistency}, \ref{A:nuc}, and \ref{A:monoton}, together with \eqref{def:MSM1}. The pairs of conditional complier laws induced by these full-data laws satisfy
\begin{equation}
\left\{\bigl((P_{\full})_{Y(0)\mid X,U=01},(P_{\full})_{Y(1)\mid X,U=01}\bigr):
 P_{\full}\in\mathfrak P(P;\Lambda_0,\Lambda_1)\right\}
=\mathcal G_0(P,\Lambda_0)\times\mathcal G_1(P,\Lambda_1).
\label{eq:exact_complier_characterisation}
\end{equation}
In particular, every admissible pair is realized by one full-data law, with
\[
\E_{P_{\full}}[Y(z)\mid X,U=01]
=\E_P[W_zY\mid X,Z=S=z],\qquad z=0,1.
\]
\end{theorem}
The forward inclusion establishes necessity, while the reverse inclusion establishes sufficiency. Thus, armwise optimization captures all compatible pairs without admitting any that cannot be jointly attained. Appendix~\ref{app:supporting_representations} shows how the remaining principal-stratum means are identified.
%The Appendix explicitly constructs the receipt response types and both potential outcomes.

\section{Identification and estimation of sensitivity bounds}   \label{sec:id}
We first derive conditional endpoints and then average them over the covariate distribution of compliers.

\subsection{Partial identification of $\psi_{1,X}$ and $\psi_{0,X}$}

\begin{assumption}[Overlap and moments]\label{A:positivity_boundedmean}
For constants $\epsilon>0$ and $M<\infty$, assume almost surely $ E[Y^2\mid X,Z=z,S=s]\le M$ and $\epsilon \le e_z(X) \le 1-\epsilon$, $\epsilon \le p_{z,s}(X) \le 1-\epsilon$, and $\pi_C(X) \ge \epsilon$.
\end{assumption}
Write $\pi_C:=\E[\pi_C(X)]$, and take the sensitivity parameters $\Lambda_0$ and $\Lambda_1$ to be fixed and finite throughout. These conditions are imposed for estimation and inference, whereas identification requires only the weaker positivity and integrability conditions in Theorem~\ref{thm:compat_plus}.

We begin by partially identifying the conditional complier-specific potential outcome mean $\psi_{1,X} := \E[Y(1) | X,Z=1,U=01]$.
Let $\psi_{1+,X}$ and $\psi_{1-,X}$ denote its sharp upper and lower bounds under the sensitivity model \eqref{def:MSM2}, respectively. By Theorem \ref{thm:compat_plus}, these bounds can be characterized as
\begin{align*}
	\psi_{1+,X} = \sup_{W_1 \in \mathcal{E}_1(\Lambda_1)} \E[W_1Y|X,Z=S=1], \quad 
	\psi_{1-,X} = \inf_{W_1 \in \mathcal{E}_1(\Lambda_1)} \E[W_1Y|X,Z=S=1].
\end{align*}

If $L_z=U_z=1$, the endpoint is $\mu_z(X)$. Otherwise define
\[
\tau_{z-}(X)=\frac{1-L_z(X)}{U_z(X)-L_z(X)}
=\frac{1+(\Lambda_z-1)\gamma_z(X)}{\Lambda_z+1},\qquad
\tau_{z+}(X)=1-\tau_{z-}(X).
\]
Applying the quantile optimization characterization of \citet{rockafellar2000optimization} gives the following representations.

Write $Q_{z\pm}(x):=Q_{z,x}(\tau_{z\pm}(x))$ for the true quantile functions corresponding to the upper and lower bounds, and denote their fitted or candidate counterparts by $\hat Q_{z\pm}(x)$. For any candidate quantile function $\hat Q$, not necessarily equal to the true quantile function, define
\[
 \rho_{z\pm}(x;\hat Q) = \E[\{\pm(Y - \hat Q(x))\}_+ \mid X=x,Z=S=z].
\]
For an outcome $y$ at $x$, define the transformed outcome and its conditional expectation
\begin{align*}
    \nu_{z\pm}(x,y;\hat Q)
    &= L_z(x)y+(1-L_z(x))\hat Q(x)
    \pm (U_z(x)-L_z(x))\{\pm(y-\hat Q(x))\}_+, \\
	\omega_{z\pm}(x;\hat Q_{z\pm})
    &= \E[\nu_{z\pm}(X,Y;\hat Q_{z\pm})\mid X=x,Z=S=z]\\
    &= L_z(x)\mu_z(x)+(1-L_z(x))\hat Q_{z\pm}(x)
    \pm (U_z(x)-L_z(x))\rho_{z\pm}(x;\hat Q_{z\pm}).
\end{align*}

\begin{theorem}[Inverse probability weighting]\label{thm:ipw1x}
For $z \in \{0,1\}$ and any candidate quantile $\hat Q_{z\pm}$,
\[
%\E[\nu_{z-}(X,Y;\hat Q_{z-})\mid X,Z=S=z]
\E\left[\frac{\mathbb{1}\{Z=S=z\}}{e_z(X)p_z(X)}\nu_{z-}(X,Y;\hat Q_{z-}) ~\bigg|~ X\right]
\le \psi_{z-,X} \le \psi_{z+,X}
\le \E\left[\frac{\mathbb{1}\{Z=S=z\}}{e_z(X)p_z(X)}\nu_{z+}(X,Y;\hat Q_{z+}) ~\bigg|~ X\right].
%\E\left[\frac{\mathbb{1}\{Z=S=z\}}{e_z(X)p_z(X)}\nu_{z+}(X,Y;\hat Q_{z+}) \mid X\right]
\]
The leftmost equality holds when $\hat Q_{z-}=Q_{z-}$, and the rightmost equality holds when $\hat Q_{z+}=Q_{z+}$.
\end{theorem}

Theorem~\ref{thm:ipw1x} gives the conditional bounds. This gives a simple interpretation of the weighted transformed outcome and recovers ordinary complier-weighted estimation when $\Lambda_z=1$. Marginal bounds average it over the covariate distribution of compliers. Appendix~\ref{app:supporting_representations} gives the outcome-regression formula (OR) and shows how averaging over the covariate distribution of compliers yields the marginal bounds. 
At the true quantiles, IPW and OR bounds are
\begin{align*}
\psi_{z\pm}
= \frac{1}{\pi_C}\E\left[\frac{\mathbb{1}\{Z=S=z\}\gamma_z(X)}{e_z(X)}
\nu_{z\pm}(X,Y; Q_{z\pm})\right]
= \frac{\E[\pi_C(X)\omega_{z\pm}(X; Q_{z\pm})]}{\pi_C}.
\end{align*}

For either representation, the lower and upper expressions bound $\psi_z$ from below and above, respectively, attaining the sharp endpoints when the candidate quantiles are correctly specified. The corresponding estimators replace population expectations and nuisance functions with their empirical counterparts and estimates. OR relies on conditional outcome regressions and principal scores, whereas IPW incorporates the assignment propensity through inverse-probability weighting. 
By Theorem~\ref{thm:compat_plus}, the sharp effect endpoints are $\psi_{C\pm}=\psi_{1\pm}-\psi_{0\mp}$.

\subsection{Robust estimation with estimated mixture proportions} \label{sec:TR}

The endpoint functional depends on the principal scores through both the complier weights $\pi_C(X)/\pi_C$ and the density-ratio limits $L_z,U_z$. Accounting only for the former would overlook how variations in the principal scores affect the admissible distributions. We derive estimating functions that account for both dependencies and establish robustness and inference guarantees under the following conditions.

\begin{assumption}[Sample splitting]  \label{A:samplesplitting}
	The nuisance estimator $\hat\theta = (\hat e_1, \hat{p}_{z,s}, \hat\mu_z, \hat\rho_{z\pm}, \hat Q_{z\pm})$ is fitted independently of observations $\{O_i=(X_i,Z_i,S_i,Y_i)\}_{i=1,\dots,n}$.
\end{assumption}

Throughout, set $\hat e_0(X)=1-\hat e_1(X)$ and $e_0^*(X)=1-e_1^*(X)$, where $e_1^*$ is the probability limit of $\hat e_1$. Thus, conditions stated for $\hat e_z$ apply to the fitted propensity model for either arm. Let $P_X$ denote the marginal law of $X$, and write $\|f\|_2:=\|f\|_{L^2(P_X)}=\{\int f(x)^2\,dP_X(x)\}^{1/2}$.
For the observations $O_1,\ldots,O_n$ independent of the nuisance estimator, define the finite-sample mean operator by $\E_n[f]:=n^{-1}\sum_{i=1}^n f(O_i)$.
We next impose regularity conditions on the fitted nuisance functions to ensure that the resulting weights and regression predictions remain well behaved.

\begin{assumption}[Fitted overlap and bounded regressions]\label{A:hat_positivity_boundedmean}
With probability tending to one, $\hat e_z$ and $\hat{p}_{z,s}$ satisfy the same overlap bounds as in Assumption \ref{A:positivity_boundedmean}, possibly with a smaller $\epsilon>0$, and $\hat \pi_C := \hat p_1-\hat p_{0,1}\ge\epsilon$ almost surely. $\hat\mu_z$, $\hat{Q}_{z\pm}$, and $\hat\rho_{z\pm}$ are uniformly bounded by a fixed constant.
\end{assumption}

We also characterize the probability limits of the nuisance estimators, allowing for potential model misspecification.

\begin{assumption}[Probability limits]\label{A:plim}
$\hat{e}_z$, $\hat{p}_{z,s}$, $\hat\mu_z$, $\hat\rho_{z\pm}$, $\hat Q_{z\pm}$ converge in $L^2(P_X)$ in probability to ${e}^*_z$, ${p}_{z,s}^*$, $\mu_z^*$, $\rho_{z\pm}^*$, $Q^*_{z\pm}$, respectively. These limits are deterministic and bounded.
\end{assumption}

Let $e_1,p_{z,s},\mu_z,\rho_{z\pm},Q_{z\pm}$ denote the true nuisance functions. Here, `correct' refers to equality of a fitted function and its target, whereas `correctly specified' refers to equality of its probability limit and its target. When the candidate quantile function converges to a possibly incorrect limit $Q^*_{z\pm}$, correct outcome regression specification means $\mu_z^*=\mu_z$, $\rho_{z\pm}^* = \rho_{z\pm}(\cdot\,;Q^*_{z\pm})$.

For the asymptotic results, we further require convergence rates that control the remainder terms, together with regularity of the conditional outcome distribution near the relevant quantiles.

\begin{assumption}[Error rates]\label{A:rate}
Define the errors:
\begin{align*}
 r_p&=\sum_{z,s}\|\hat p_{z,s}-p_{z,s}\|_2,& r_e&=\|\hat e_1-e_1\|_2,\\
 r_m&=\sum_{z,\pm}\big\{\|\hat\mu_z-\mu_z\|_2+
 \|\hat\rho_{z\pm}-\rho_{z\pm}(\cdot;\hat Q_{z\pm})\|_2\big\},&
 r_Q&=\sum_{z,\pm}\|\hat Q_{z\pm}-Q_{z\pm}\|_2.
\end{align*}
Each error is $o_P(1)$ and
\(
 R_n:=r_p^2+(r_e+r_p)r_m+r_e r_p+r_Q^2=o_P(n^{-1/2}).
\)
For every arm with $\Lambda_z>1$, the conditional law $P_{Y|X,Z,S}(y|X,Z=S=z)$ has a uniformly bounded density on the interval containing the true and fitted quantile values. Each $Q_{z\pm}(x)$ is unique, and the density is bounded away from zero in its neighborhood.
\end{assumption}
%The propensity errors coincide across arms: $\|\hat e_0-e_0\|_2=\|\hat e_1-e_1\|_2=r_e$. The remainder rate $R_n$ is common to both arms because the regression and principal-score errors above sum over the arms.

For a candidate quantile $\hat Q_{1\pm}$, define the uncentered estimating functions
\begin{align}
		D 
		&= {\pi}_C(X) + \frac{Z}{{e_1}(X)}(S - {p}_1(X)) - \frac{1-Z}{1-{e_1}(X)}(S - {p}_{0,1}(X)),  \label{eq:D} \\
		N_{1\pm} 
		&= \pi_C(X)\omega_{1\pm}(X;\hat Q_{1\pm})  \label{eq:N1} \\
		&\quad+ \Big( {\omega}_{1\pm}(X;\hat Q_{1\pm}) + {\gamma}_1(X)(1-{\gamma}_1(X)){\omega'}_{1\pm}(X;\hat{Q}_{1\pm}) \Big) \frac{Z}{{e_1}(X)}(S - {p}_1(X))  \label{eq:N2} \\
		&\quad- \Big( {\omega}_{1\pm}(X;\hat Q_{1\pm}) + {\gamma}_1(X){\omega'}_{1\pm}(X;\hat{Q}_{1\pm}) \Big) \frac{1-Z}{1-{e_1}(X)}(S - {p}_{0,1}(X))  \label{eq:N3} \\
		&\quad+ \frac{Z S \gamma_1(X)}{e_1(X)} \big\{\nu_{1\pm}(X,Y;\hat Q_{1\pm})-\omega_{1\pm}(X;\hat Q_{1\pm})\big\},  \label{eq:N4} \\
		&\omega'_{1\pm}(X;Q) := \frac{\partial \omega_{1\pm}(X;Q)}{\partial \gamma_1(X)} = L_1'(X)\big(\mu_1(X) - Q(X)\big) \pm \big(U_1'(X)-L_1'(X)\big)\rho_{1\pm}(X;Q).
	\end{align}
Let $\hat N_{1\pm},\hat D$ denote these functions evaluated at fitted nuisance functions, and define the doubly robust estimator (DR)
\[
\hat\psi_{1\pm}^{\mathrm{DR}}=\frac{\E_n[\hat N_{1\pm}]}{\E_n[\hat D]}.
\]

\begin{theorem}[Robust estimation and efficient inference]\label{thm:TR}
Under \eqref{def:MSM2} and Assumptions~\ref{A:consistency}--\ref{A:positivity_boundedmean},
\begin{align}
\psi_{1-}^{\mathrm{DR}}:=\frac{\E[N_{1-}]}{\E[D]}
\le\psi_{1-}\le\E[Y(1)\mid U=01]\le\psi_{1+}
\le\frac{\E[N_{1+}]}{\E[D]}=:\psi_{1+}^{\mathrm{DR}}.\label{eq:TR1}
\end{align}
The outer inequalities are equalities at the true quantiles. Suppose the fitted nuisances satisfy Assumptions~\ref{A:samplesplitting} and \ref{A:hat_positivity_boundedmean}.

\begin{enumerate}
\item[(i)] Under Assumption \ref{A:plim}, if $\hat{p}_{z,s}$'s are correctly specified and either $\hat e_1$ or $(\hat\mu_1, \hat{\rho}_{1\pm})$ are correctly specified, the estimator converges to the valid but possibly conservative endpoint associated with $Q^*_{1\pm}$. If $Q^*_{1\pm} = Q_{1\pm}$, this is the sharp endpoint.
\item[(ii)] With $\hat e_1$ and $(\hat\mu_1, \hat{\rho}_{1\pm})$ correct at a fixed candidate quantile function $Q$, perturbations on $\hat{p}_{z,s}$'s produce a locally quadratic bias relative to that endpoint. At $Q = Q_{1\pm}$, this is a local $O(r_p^2)$ bias relative to the sharp endpoint.
\item[(iii)] Under Assumptions \ref{A:positivity_boundedmean}, \ref{A:samplesplitting}, \ref{A:hat_positivity_boundedmean}, and \ref{A:rate},
\[
 \sqrt n(\hat\psi_{1\pm}^{\mathrm{DR}}-\psi_{1\pm})
  = n^{-1/2}\sum_{i=1}^n\varphi_{1\pm}(O_i)+o_P(1) \xrightarrow{d} \mathcal N(0,E[\varphi_{1\pm}^2]),
\]
where $\varphi_{1\pm}=\{N_{1\pm}-\psi_{1\pm}D\}/\pi_C$ uses the true nuisance models and true quantiles.
\item[(iv)] Under the assumptions of part (iii), at an observed-data distribution satisfying the stated strict overlap and quantile regularity conditions, $\varphi_{1\pm}$ is the efficient influence function, and thus the estimator attains the semiparametric efficiency bound.
\end{enumerate}
\end{theorem}

\begin{figure}[!htbp]
\centering
\includegraphics[width=\textwidth]{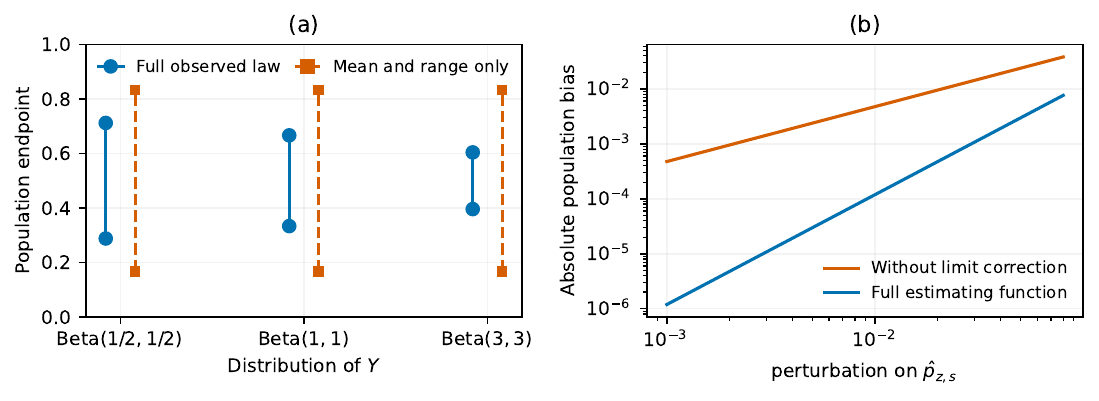}
\caption{Panel (a) retains the same $\gamma=0.5$, $\Lambda=5$, mean and range of $Y$ while comparing the fixed observed distribution with a relaxation allowing all distributions with those summaries. Panel (b) perturbs only $p_{0,1}$ by $\delta>0$ and compares absolute population bias with and without the correction for estimated density-ratio limits; both axes are logarithmic. We compute both panels directly from population formulas.}
\label{fig:targeted_comparisons}
\end{figure}

The $\omega'_{1\pm}$ terms in \eqref{eq:N2}--\eqref{eq:N3} account for the dependence of the density ratio limits on the principal scores. Their combined contribution is
\[
\gamma_1(X)\omega'_{1\pm}(X;\hat Q_{1\pm})
\left\{(1-\gamma_1(X))\frac{Z}{e_1(X)}(S-p_1(X))
-\frac{1-Z}{e_0(X)}(S-p_{0,1}(X))\right\}.
\]
Treating $L_1,U_1$ as fixed would omit this contribution. Although its expectation is zero under the true nuisance functions, its inner product with the principal scores captures the variation in the density ratio limits. Omitting it therefore fails to account for the full pathwise derivative. 
Figure~\ref{fig:targeted_comparisons}(b) examines the effect of accounting for the dependence of the density ratio limits on estimated principal scores. Here, we set $p_1=0.7$, $p_{0,1}=0.3$, a uniform treated-recipient outcome distribution, and $\Lambda_1=5$, and perturb the fitted $p_{0,1}$ to $p_{0,1}+\delta$. All other fitted components are exact at the resulting candidate quantile. We compare the expected full estimating function with its counterpart omitting only the $\omega'_{1-}$ terms. Omitting these terms produces bias proportional to $\delta$, whereas including them reduces the bias to order $\delta^2$. At $\delta=0.04$, the corresponding absolute biases are approximately $0.0190$ and $0.0019$, respectively.
The correction vanishes at $\Lambda_1=1$ and yields the local guarantee in part (ii). The corresponding control-arm functions follow by relabeling assignment and receipt.

The corresponding doubly robust estimators of the complier average treatment effect bounds are
$
\hat\psi_{C\pm}^{\mathrm{DR}} := \hat\psi_{1\pm}^{\mathrm{DR}} - \hat\psi_{0\mp}^{\mathrm{DR}}.
$
Under Theorem~\ref{thm:TR}(iii)--(iv), their efficient influence functions are $\varphi_{C\pm}=\varphi_{1\pm}-\varphi_{0\mp}$.

\section{Maximin assignment policies}\label{sec:itr}
We now use the same uncertainty set to choose an assignment policy. Joint compatibility makes the average of its selected arm-specific lower endpoints an attainable worst-case complier value.

Let $\pi:\mathcal X\to\{0,1\}$ be a deterministic treatment rule, where $\mathcal X$ is the space of covariates. Let $Y(\pi) = Y(1)\pi(X) + Y(0)(1-\pi(X))$ be the counterfactual outcome had the subject followed the decision rule. Define its complier value as
\[
V(\pi)
:= \E[Y(\pi)\mid U=01]
 = \frac{1}{\pi_C} \E\left[ \pi_C(X) \left\{ \pi(X)\psi_{1,X} + (1-\pi(X))\psi_{0,X} \right\} \right].
\]
By Theorem \ref{thm:compat_plus}, the sharp lower bound on $V(\pi)$ is
\begin{align*}
	V_-(\pi) := \E\left[ \frac{\pi_C(X)}{\pi_C} \{ \pi(X) \psi_{1-,X} + (1-\pi(X)) \psi_{0-,X} \} \right].
\end{align*}

For a fixed treatment rule $\pi$, define the uncentered estimating function $N_\pi(O):=\pi(X)N_{1-}(O)+\{1-\pi(X)\}N_{0-}(O)$, where the nuisance functions and quantiles are evaluated at their true values. Under the conditions of Theorem \ref{thm:TR}(iii)--(iv), the efficient influence function of $V_-(\pi)$ and its variance are
\[
\varphi_\pi(O)
=
\frac{N_\pi(O)-V_-(\pi)D(O)}{\pi_C},
\qquad
\Upsilon(\pi):=\E[\varphi_\pi(O)^2].
\]

Over the class of all measurable treatment rules, the optimal rule $\pi^*$ maximizing $V_-(\pi)$ coincides with the pointwise maximin rule $\pi^*(x) = \mathbb{1}\{\psi_{1-,X=x}\ge\psi_{0-,X=x}\}$.
We restrict optimization to an interpretable policy class, including constant policies.

For fixed-depth trees, maximizing the criterion below has the policy-tree form of \citet{sverdrup2020policytree}, with rewards replaced by estimated contributions to the sharp lower-bound value.

Define the doubly robust estimator of the lower bound on the complier value as
\begin{align}
\hat{V}_{-}(\pi)
:= \frac{\E_n[\hat N_{1-}(O)\pi(X) + \hat N_{0-}(O)(1-\pi(X))]}{\E_n[\hat D(O)]},  \label{eq:V_DR}
\end{align}
where $\hat N_{z-}$ and $\hat{D}$ denote the estimating functions $N_{z-}$ and $D$ evaluated at the fitted nuisance functions.
We then learn the maximin treatment rule over a prespecified policy class $\Pi$ by solving
$$
\hat\pi\in\argmax_{\pi\in\Pi}\hat V_{-}(\pi).
$$

We consider the class $\Pi = \Pi_H^{\mathrm{tree}}$ of binary decision tree policies with maximum depth $H$, where $X\in\mathbb R^p$ and both $p$ and $H$ are fixed independently of the sample size. Each internal node $v$ splits on $X_{j_v}\le t_v$, with $j_v\in\{1,\ldots,p\}$ and $t_v\in\mathbb R$, while each terminal node assigns treatment 0 or 1. Thus, each tree has at most $2^H$ terminal nodes.

When $\E_n[\hat D]>0$, maximization is equivalent to maximizing $\E_n[\pi(X)\{\hat N_{1-}-\hat N_{0-}\}]$. Since the policy class induces only finitely many distinct assignment vectors on the sample, an empirical maximum is attained. We select split cutoffs from ordered observed coordinate values, including a cutoff below the minimum, and resolve ties deterministically to obtain a measurable $\hat\pi$. Our theoretical results assume a global empirical maximizer. The fixed-depth policy class has finite Vapnik--Chervonenkis (VC) dimension \citep{leboeuf2020decision}, and policies agreeing $P_X$-almost surely are considered equivalent.

Assume a population maximizer $\pi^*\in\argmax_{\pi\in\Pi}V_-(\pi)$ exists. The sharp criterion uses the true quantiles. For possibly incorrect quantile limits $Q^*=(Q^*_{0-},Q^*_{1-})$, write $V_{Q^*}$ for the corresponding conservative criterion.

\begin{theorem}[Value consistency]\label{thm:opl_param}
Under Assumptions \ref{A:consistency}-\ref{A:plim}, 
suppose the principal scores are correctly specified and, for each arm, either the assignment propensity or its outcome regressions are correctly specified. Then, $\sup_{\pi\in\Pi}V_{Q^*}(\pi)-V_{Q^*}(\hat\pi)=o_P(1)$. If $Q^*=Q$, the learned policy is consistent in value for the optimal sharp lower-bound criterion.
\end{theorem}
Inference on the optimal sharp value additionally requires convergence of the assignment decisions.
Define the policy distance $\delta(\pi,\pi') = \{P_X(\pi(X) \ne \pi'(X))\}^{1/2}$. The regret of a policy $\pi\in\Pi$ is defined as $R(\pi) := V_-(\pi^*)-V_-(\pi)$, i.e., its loss in sharp lower-bound value relative to an optimal policy within the tree class $\pi^* \in \Pi$.

\begin{assumption}[Separation] \label{A:opl}
There exists a deterministic maximizer $\pi^*\in\Pi$ such that, for every $\eta>0$, $\inf_{\substack{\pi\in\Pi:\delta(\pi,\pi^*)\ge\eta}}
 R(\pi) > 0$, with the convention that the infimum of the empty set is $+\infty$.
\end{assumption}
Assumption \ref{A:opl} requires a strictly positive value gap between $\pi^*$ and any policy that differs from it on a set of nonnegligible probability. It concerns the population value rather than the tree representation, allowing distinct trees to represent the same policy.
A sufficient condition for a tree class is that $X$ is supported on $[-K,K]^p$, each coordinate of $X$ has a continuous marginal distribution, and every maximizing policy in $\Pi_H^{\mathrm{tree}}$ agrees with $\pi^*$ almost surely. For each tree structure and choice of split coordinates and leaf assignments, the cutoffs can be restricted to the compact set $[-K,K]$. Continuous marginal distributions make the map from cutoffs to policies continuous under $\delta$. Since there are finitely many such discrete choices at fixed depth, the policy class is compact under $\delta$. Continuity of the value and uniqueness of the maximizing policy then imply separation by the argument in \citet{newey1994large}.

\begin{lemma}[Policy convergence and negligible regret]\label{lem:policy_convergence}
Let $\Pi = \Pi_H^{\mathrm{tree}}$, with fixed depth $H$ and covariate dimension $p$. Suppose Assumption \ref{A:opl} holds, $\hat\pi$ is the empirical maximizer defined above, and
\[
 \sup_{\pi\in\Pi}|\hat V_-(\pi)-V_-(\pi)|=o_P(1).
\]
Then $\delta(\hat\pi,\pi^*)=o_P(1)$. If additionally Assumptions \ref{A:positivity_boundedmean}-\ref{A:hat_positivity_boundedmean} and \ref{A:rate} hold, then
$R(\hat\pi) = o_P(n^{-1/2})$.
\end{lemma}

\begin{theorem}[Inference on the optimal lower-bound value] \label{thm:opl}
Let $\Pi = \Pi_H^\textrm{tree}$, with fixed depth $H$ and covariate dimension $p$. Under Assumptions \ref{A:consistency}, \ref{A:nuc}, \ref{A:monoton}, \ref{A:positivity_boundedmean}, \ref{A:samplesplitting}, \ref{A:hat_positivity_boundedmean}, \ref{A:rate}, and \ref{A:opl}, the empirical maximizer $\hat\pi \in \Pi$ satisfies
\[
 \sqrt n\{\hat V_{-}(\hat\pi) - V_{-}(\pi^*)\}
 \ \overset d\longrightarrow\ \mathcal N(0,\Upsilon(\pi^*)).
\]
\end{theorem}
Theorem~\ref{thm:opl} permits Wald-type inference with the same asymptotic variance as if $\pi^*$ were known, provided $\Upsilon(\pi^*)$ is consistently estimated.

{\remark
As an alternative to decision trees, we consider linear policies in a fixed feature space, of the form $\pi_\beta(x) = \mathbb{1}\{\beta^\top\Phi(x)>0\}$ with $\|\beta\|_2=1$, where $\Phi$ is a fixed reproducing kernel Hilbert space (RKHS) feature map. Appendix \ref{AA:alternative_policy} studies empirical maximization of the estimated sharp lower-bound value over this class, with ordinary linear policies as a special case. Under the corresponding separation and nuisance rate conditions, Theorem \ref{thm:rkhs_opl} establishes asymptotic normality of the estimated optimal value within this policy class.
}

	\section{Numerical studies}\label{sec:sim}

We conduct Monte Carlo experiments to evaluate endpoint estimation accuracy, robustness to nuisance-model misspecification, and the performance of learned policies.

\subsection{Simulation design}
We generated 500 independent data sets for each sample size $n\in\{2000,5000\}$. The baseline covariate and treatment assignment were generated as
$X\sim\mathcal N(1.5,0.5^2)$ and $Z\mid X\sim\operatorname{Bernoulli}\{\operatorname{expit}(3X-4.5)\}$.
To generate potential treatment receipt, we drew $R\sim\operatorname{Uniform}(0,1)$ independently and set
\begin{gather*}
	S(0)=\mathbb{1}\{R<p_{0,1}(X)\},\quad S(1)=\mathbb{1}\{R<p_1(X)\},\\
p_{0,1}(X)=\operatorname{expit}(0.5X-0.5),\quad p_1(X)=\operatorname{expit}(0.5X+0.5).
\end{gather*}
This construction ensures monotonicity, with complier probability $\pi_C(X)=p_1(X)-p_{0,1}(X)$.

Within the observed groups $(Z,S)=(z,z)$, outcomes followed normal distributions with standard deviation $1.5$ and conditional means $\mu_0(X)=1.5+1.5X$ and $\mu_1(X)=1+2X$.
We generated stratum-specific potential outcomes by reweighting these distributions to preserve the observed outcome mixtures while violating PI. The reweighting was constructed at $\Lambda_0=\Lambda_1=5$, with the complier means attaining the upper bound under assignment $0$ and the lower bound under assignment $1$. For $\Lambda<5$, the sharp bounds remain well-defined functionals of the observed distribution, but the sensitivity model need not contain the data-generating law. Further implementation details are provided in Appendix~\ref{app:sim_impl}.

We evaluated the OR, IPW, and DR estimators across sensitivity levels $\Lambda_0=\Lambda_1=\Lambda\in\{1.001,2,3,5,8\}$. Four nuisance model specification settings were considered, each under correctly specified and misspecified quantile regression models. Assignment propensity and principal scores were estimated using logistic regression, while outcome means were estimated using linear regression within each assignment--receipt group. Misspecification was introduced by using intercept-only models for the assignment propensity and outcome means and omitting $X$ from the principal score models. Quantile regression models included either an intercept and $X$ or an intercept alone. The hinge expectations were estimated using linear working models, which were not necessarily correctly specified even when the other nuisance models were correctly specified.

For policy learning, we additionally considered a principal-ignorability-guided (PIG) outcome regression criterion \citep{ps3,chu23c}. Each method optimized its respective criterion over the class of linear policies $\pi(X)=\mathbb{1}\{\beta_0+\beta_1X>0\}$. Nuisance estimation and policy optimization were performed using the same sample, without sample splitting, whereas our theoretical results assume independent nuisance training and evaluation. Thus, the simulations assess finite-sample performance without satisfying all theoretical assumptions. 
We computed reference bounds using an independent sample of one million observations. The true worst-case policy values and oracle maxima were computed by numerical integration of the known outcome laws and principal scores.

\subsection{Estimation of sensitivity bounds}
\begin{figure}[!htbp]
\centering
\captionsetup{skip=4pt}
\includegraphics[width=\textwidth]{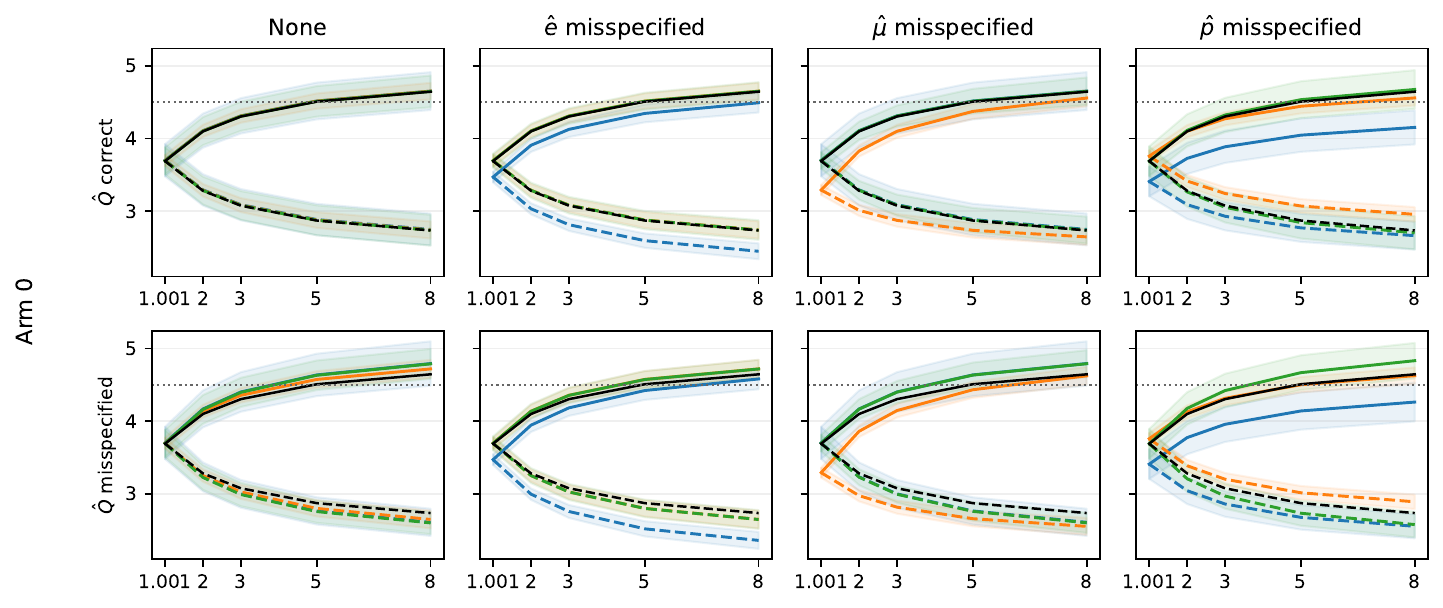}
\includegraphics[width=\textwidth]{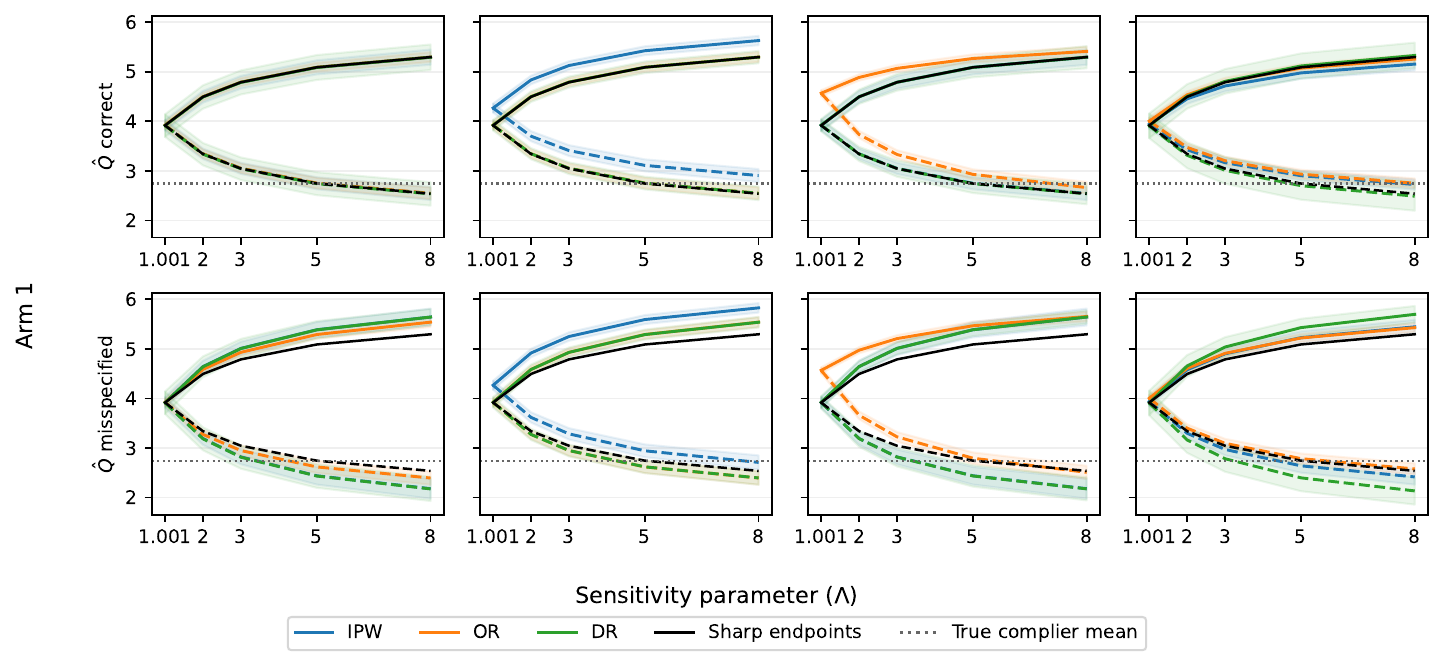}
\caption{Sensitivity bound estimation at $n=2000$ (arm 0 above, arm 1 below). Colored curves are mean endpoints (dashed lower, solid upper); ribbons extend one empirical standard error on either side. Black curves are reference sharp endpoints; the dotted horizontal line is the true complier mean. Columns show, from left to right, no misspecification (None), $\hat e_1$ misspecified, $\hat\mu$ misspecified, and $\hat p$ misspecified. Rows distinguish correctly specified and misspecified $\hat Q_{z\pm}$.
} \label{fig:sim_bounds_2000}
\end{figure}

Figure~\ref{fig:sim_bounds_2000} presents the estimated sensitivity bounds for $n=2000$. Additional results are provided in Figure~\ref{fig:sim_bounds_5000} and Tables~\ref{tab:sim_bounds_2000_0_0}--\ref{tab:sim_bounds_2000_0_1} and \ref{tab:sim_bounds_5000_0_0}--\ref{tab:sim_bounds_5000_0_1}. We report Monte Carlo means and empirical standard errors of the endpoint estimates.

When the quantile models were correctly specified, we assessed coverage of the sharp endpoints using two-sided 95\% percentile-bootstrap confidence intervals based on 500 bootstrap replicates per simulated data set. Under quantile model misspecification, we instead assessed one-sided validity, defined as the proportion of repetitions in which the 95\% lower (upper) confidence bound fell below (above) the corresponding sharp lower (upper) endpoint. Because misspecified quantile models can yield conservative limiting bounds, validity rates may exceed 95\%.

In each bootstrap replicate, we refitted the assignment propensity, principal-score, outcome-mean, and hinge regression models while holding the fitted quantile predictions fixed. Thus, the reported coverage and validity rates do not account for uncertainty from refitting the quantile models.

Under the baseline specifications and correctly specified quantile families, estimates were close to the reference endpoints and empirical standard errors decreased with sample size. At $n=2000$ and $\Lambda=5$, the reference intervals were $[2.869,4.506]$ for arm 0 and $[2.746,5.089]$ for arm 1, compared with the corresponding OR estimates of $[2.876,4.514]$ and $[2.754,5.090]$. With misspecified $\hat e_1$, the arm-0 IPW interval was $[2.592,4.344]$, and its lower- and upper-endpoint coverage rates fell to $19.8\%$ and $75.8\%$, respectively, whereas misspecified $\hat \mu$ reduced OR coverage. In contrast, DR estimates remained close to the reference endpoints under these settings. Misspecified $\hat p$ substantially reduced IPW and OR coverage, while DR estimates remained closer to the reference bounds. Baseline DR estimates were more variable than IPW in arm 1. This does not contradict efficiency in the nonparametric model, since IPW can exploit the correctly specified parametric nuisance models, while inverse propensity weights amplify the variability of the principal score corrections in this design.

Misspecified quantile models generally widened the estimated bounds, consistent with conservative limiting endpoints. Under the baseline specifications for the remaining nuisance models, the mean arm-1 intervals at $n=2000$ and $\Lambda=5$ were $[2.441,5.386]$ for IPW, $[2.620,5.290]$ for OR, and $[2.436,5.384]$ for DR. The corresponding lower- and upper-bound validity rates were $100\%/100\%$, $99.6\%/100\%$, and $99.4\%/100\%$, respectively. Validity deteriorated for IPW when $\hat e_1$ was also misspecified and for OR when $\hat\mu$ was also misspecified. In contrast, DR maintained high validity when either $\hat e_1$ or $\hat\mu$ alone was misspecified. These results agree with the alternative nuisance model conditions for DR. Near $\Lambda=1$, quantile misspecification provides little conservative widening, and finite-sample bootstrap error remains visible. The average bootstrap SE was smaller than the empirical SE in these settings, consistent with the undercoverage.

\subsection{Maximin policy learning}
\begin{figure}[!htbp]
\centering
\captionsetup{skip=4pt}
\includegraphics[width=\textwidth]{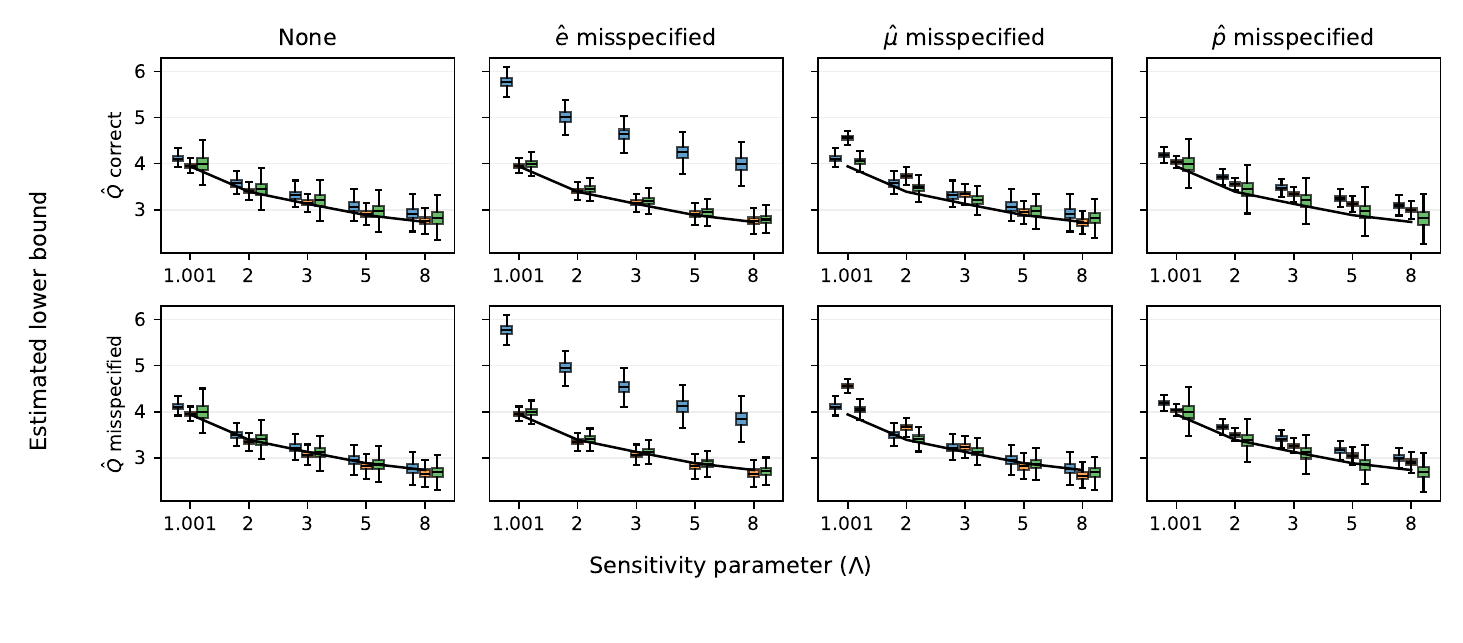}
\includegraphics[width=\textwidth]{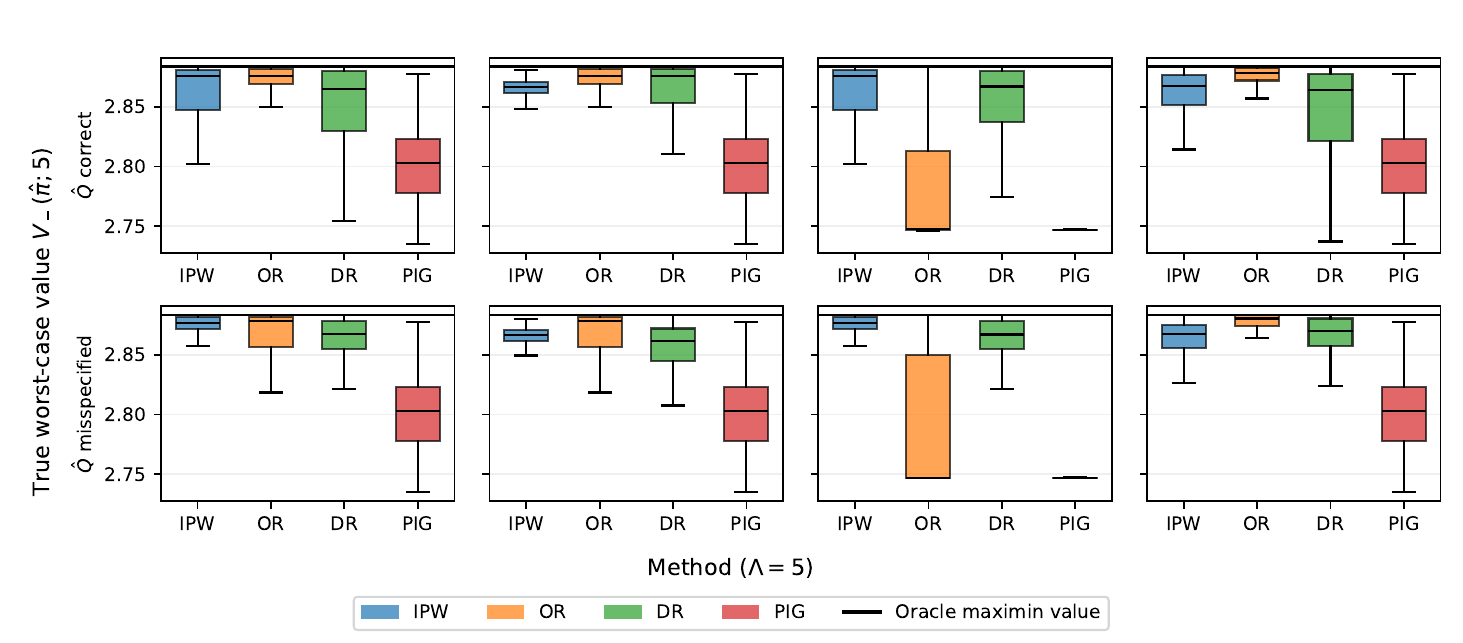}
\caption{Policy learning at $n=2000$. Upper panels show maximized estimated lower bounds for IPW, OR, and DR across sensitivity parameters. Solid black curves show the oracle true maximin value at each sensitivity parameter. Lower panels show true worst-case values at $\Lambda=5$ for all learned policies, including PIG, evaluated using the true sharp lower-bound functional. Columns show, from left to right, no misspecification (None), $\hat e_1$ misspecified, $\hat\mu$ misspecified, and $\hat p$ misspecified. Rows distinguish correctly specified and misspecified $\hat Q_{z\pm}$. Boxes show interquartile ranges and medians over 500 repetitions; whiskers extend to observations within 1.5 interquartile ranges, and outliers are hidden. The solid black line in the lower panels is the oracle maximum worst-case value over linear policies ($2.884$). Values closer to this line have smaller maximin regret.%
}    \label{fig:sim_policy_2000}
\end{figure}

We evaluated every learned policy using the same sharp lower-bound functional at $\Lambda=5$, whose oracle maximum over linear policies was 2.884. At $n=2000$, mean true worst-case values were 2.852 for IPW, 2.867 for OR, and 2.848 for DR, compared with 2.800 for PIG. Their corresponding mean regrets were 0.031, 0.017, 0.036, and 0.084. In this setting, all three methods that optimized an estimated lower bound had smaller mean regret than PIG. With correctly specified quantiles, DR had mean regrets of 0.024 and 0.033 when the assignment or outcome mean model alone was misspecified. IPW mean regret was 0.018 under misspecified $\hat e_1$, while OR and PIG regrets increased to 0.105 and 0.137 under misspecified $\hat\mu$ (Figure~\ref{fig:sim_policy_2000}; Table~\ref{tab:sim_oracle_2000}). The corresponding $n=5000$ results are presented in Figure~\ref{fig:sim_policy_5000} and Table~\ref{tab:sim_oracle_5000}.

Figure~\ref{fig:sim_policy_2000} also shows that the maximized estimated lower bounds decrease as \(\Lambda\) increases, reflecting greater uncertainty about principal ignorability. PIG instead maximizes a PI-based mean, so its estimated criterion is not directly comparable to a sensitivity lower bound. Results for all sensitivity levels are reported in Tables~\ref{tab:sim_estimated_2000_0}, \ref{tab:sim_estimated_2000_1} and \ref{tab:sim_estimated_5000_0}, \ref{tab:sim_estimated_5000_1}.

	\section{Oregon Health Insurance Experiment} \label{sec:oregon}

The Oregon Health Insurance Experiment (OHIE) used a 2008 lottery to offer low-income uninsured adults an opportunity to apply for Medicaid \citep{finkelstein2012oregon}. Selection did not guarantee enrollment, making the lottery a setting for sensitivity analysis among compliers. We use the public dataset accompanying \citet{hattab2024oregon}, distributed by \citet{hattab2023data}. After excluding 25 respondents with missing outcomes, the analysis includes 12,204 individuals in 10,921 households. The outcome is the SF-8 mental component score, with higher values indicating better mental-health-related quality of life. Survey weights account for the sampling and follow-up design. Lottery selection and enrollment define $Z$ and $S$; the policies assign an opportunity to apply, not insurance itself. Compliers are individuals whose enrollment would change from zero to one if their household were selected.

We use individual and household baseline covariates in nuisance models and restrict the policy to depth-one decision trees based on household-level covariates, so every observed member of a household receives the same assignment decision. The value averages over individual compliers using survey weights. We estimate propensity and principal scores with weighted logistic regression, outcome and hinge regressions with random forests, and quantiles with gradient boosting. We constrain fitted probabilities to maintain overlap and positive complier scores. We report projection and clipping diagnostics in Appendix~\ref{app:oregon_impl}.

For policy learning, 100 random splits assign 80\% of households to training and 20\% to testing; three-fold cross-fitting within training keeps households together. A DR policy is fitted for each $\Lambda\in\{1,1.1,1.3,1.5,2,3\}$ and evaluated using the held-out DR lower-bound estimator. We compare it with constant no-access and universal-access policies. We exhaustively search covariate cutoffs and leaf assignments, including constant policies, and independently verify the empirical maximum. Supplementary implementation details are provided in Appendix~\ref{app:oregon_impl}.

\subsection{Bounds on complier average treatment effect} \label{sec:oregon_effect}
We estimate DR bounds on $\psi_C$ with $\Lambda_0=\Lambda_1=\Lambda$, using all observations and household-level three-fold cross-fitting across 100 random fold assignments. Sampling standard errors (SEs) are estimated by resampling households while holding the fitted nuisance functions fixed. Under the error rate and regularity conditions for DR inference, nuisance estimation errors contribute only a negligible asymptotic remainder, and resampling the estimating-function contributions captures the leading sampling variation. 
%Appendix~\ref{app:oregon_impl} provides the asymptotic expansion and conditions for household-level inference.

Under PI ($\Lambda=1$), the median DR effect estimate was 0.294 SF-8 points, with a median household-bootstrap SE of 0.317, indicating substantial sampling uncertainty. Rather than assessing the robustness of a statistically significant effect, our sensitivity analysis examines how departures from PI change the range of complier effects compatible with the observed data. At $\Lambda=1.1$, the median estimated endpoints were $-0.363$ and $0.955$ SF-8 points; at $\Lambda=1.5$, they were $-2.434$ and $3.113$. Relaxing PI thus allows for increasingly substantial benefits and harms.

Table~\ref{tab:oregon_effect} reports the median estimated interval widths and endpoint SEs across cross-fitting repetitions. The median width increased from zero under PI to 1.317 at $\Lambda=1.1$ and 5.545 at $\Lambda=1.5$, while the endpoint SEs remained near 0.32. This contrast distinguishes two sources of uncertainty: sampling uncertainty in estimating the endpoints and identification uncertainty arising from departures from PI. For a fixed population law and sensitivity level, the latter persists even as the sample size increases.

The median estimated sensitivity interval included zero at $\Lambda=1.1$ and above, with the first inclusion occurring at $\Lambda=1.05$ on a finer grid with spacing 0.01 near the tipping point (Appendix~\ref{app:oregon_impl}). At this sensitivity level, the outcome-dependent odds of complier membership may range from $1/1.05$ to $1.05$ times the covariate-only odds. Using full-sample nuisance estimates and survey weights $w_i$, we compute the fitted complier-fraction summaries $\bar\gamma_z=\sum_i w_i\hat\pi_C(X_i)/\sum_i w_i\hat p_z(X_i)$. These give $\bar\gamma_0=0.274$ and $\bar\gamma_1=0.888$ (Appendix~\ref{app:oregon_impl}). At $\Lambda=1.05$, these correspond to membership probability ranges of approximately $[0.264,0.284]$ and $[0.883,0.893]$, respectively. Thus, even departures from PI that permit only small changes in complier membership probabilities can materially alter the range of compatible complier effects.

\subsection{Maximin policy learning}
Identification uncertainty can influence the choice of an assignment policy. Even when the average complier effect is imprecisely estimated, differences in conditional worst-case outcomes may favor particular access rules. We next examine which policies maximize the worst-case complier value under the specified departures from PI.

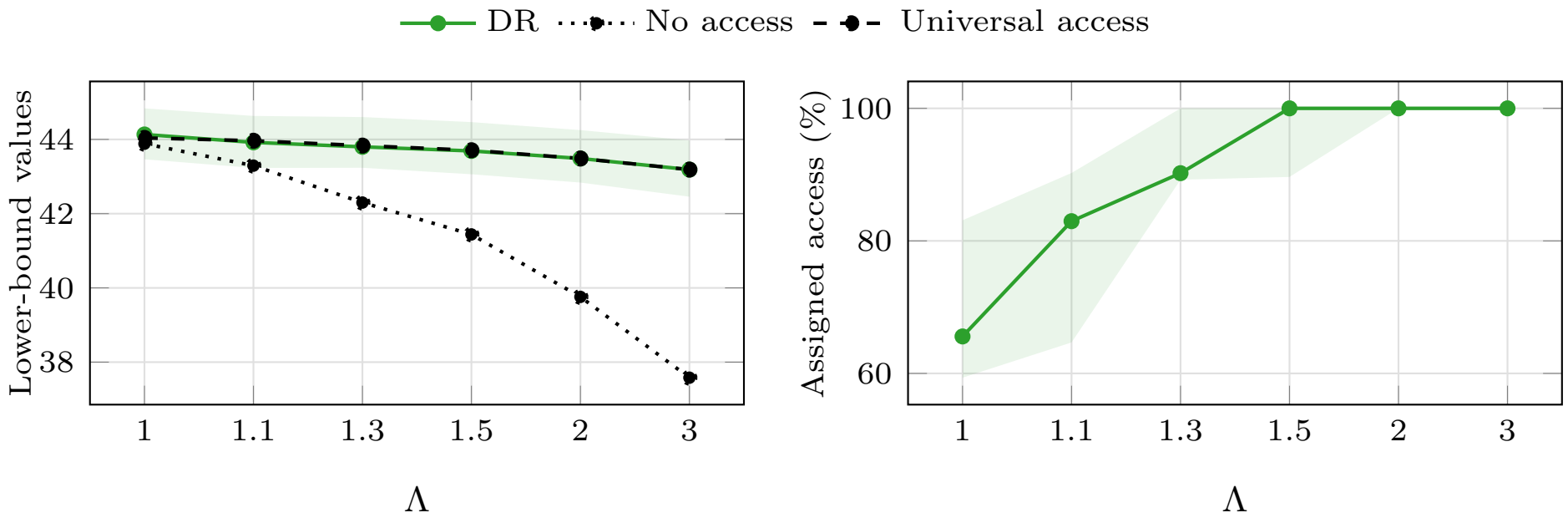
\begin{figure}[!htbp]\centering
{\small\ref{OHIElegend}}\par\smallskip
\begin{tikzpicture}
\begin{groupplot}[group style={group size=2 by 1,horizontal sep=13mm},width=.46\textwidth,height=4.14cm,symbolic x coords={1,1.1,1.3,1.5,2,3},xtick=data,xticklabels={1,1.1,1.3,1.5,2,3},xlabel={$\Lambda$},tick label style={font=\scriptsize},label style={font=\fontsize{8.1}{9.72}\selectfont},grid=major,grid style={black!12}, ylabel near ticks, ylabel style={font=\fontsize{7.29}{8.748}\selectfont}, ylabel shift=-5pt]
\nextgroupplot[ylabel={Lower-bound values},legend to name=OHIElegend,legend columns=3,legend style={font=\scriptsize,draw=none,/tikz/every even column/.append style={column sep=2pt}}]
\addplot[name path=ohie0DRlo,draw=none,forget plot] coordinates {(1,43.4644525) (1.1,43.2312162) (1.3,43.2356409) (1.5,43.0611708) (2,42.8373515) (3,42.4567878)};
\addplot[name path=ohie0DRhi,draw=none,forget plot] coordinates {(1,44.8365263) (1.1,44.6325866) (1.3,44.6023318) (1.5,44.4665299) (2,44.2510599) (3,43.9734595)};
\addplot[DRcolor,fill opacity=.10,draw=none,forget plot] fill between[of=ohie0DRlo and ohie0DRhi];
\addplot[DRcolor,thick,mark=*,mark size=1.4pt] coordinates {(1,44.1328118) (1.1,43.9198750) (1.3,43.7994042) (1.5,43.6893282) (2,43.4857498) (3,43.1869673)};\addlegendentry{DR}
\addplot[black,thick,mark=*,mark size=1.4pt,dotted] coordinates {(1,43.8839873) (1.1,43.2932380) (1.3,42.2986161) (1.5,41.4401367) (2,39.7581645) (3,37.5808669)};\addlegendentry{No access}
\addplot[black,thick,mark=*,mark size=1.4pt,dashed] coordinates {(1,44.0396358) (1.1,43.9643126) (1.3,43.8330547) (1.5,43.7083341) (2,43.4857498) (3,43.1869673)};\addlegendentry{Universal access}
\nextgroupplot[ylabel={Assigned access (\%)}]
\addplot[name path=ohie2DRlo,draw=none,forget plot] coordinates {(1,59.3540155) (1.1,64.6690207) (1.3,89.2281519) (1.5,89.6463740) (2,100.0000000) (3,100.0000000)};
\addplot[name path=ohie2DRhi,draw=none,forget plot] coordinates {(1,83.1080153) (1.1,90.2477605) (1.3,100.0000000) (1.5,100.0000000) (2,100.0000000) (3,100.0000000)};
\addplot[DRcolor,fill opacity=.10,draw=none,forget plot] fill between[of=ohie2DRlo and ohie2DRhi];
\addplot[DRcolor,thick,mark=*,mark size=1.4pt,forget plot] coordinates {(1,65.5783958) (1.1,82.9730282) (1.3,90.2199022) (1.5,100.0000000) (2,100.0000000) (3,100.0000000)};
\end{groupplot}
\end{tikzpicture}
\caption{OHIE DR depth-one policies. The left panel compares held-out lower-bound values with constant no-access and universal-access policies; the right panel shows survey-weighted access. Curves are medians, and shading shows descriptive 10th--90th percentiles across 100 household splits.}\label{fig:oregon_sensitivity}
\end{figure}

Figure~\ref{fig:oregon_sensitivity} presents the held-out lower-bound values and access shares of the learned DR policies, with constant no-access and universal-access policies serving as benchmarks under the same DR evaluation criterion. Table~\ref{tab:oregon_policy_se} reports the median split-specific bootstrap SEs of the held-out policy values. Conditional on the training data and learned policies, these SEs quantify sampling uncertainty within each test split, distinct from variation across the overlapping outer splits.

As the sensitivity level increased, the learned DR policies favored broader access. At $\Lambda=1.1$ and $1.3$, the median access shares were 83.0\% and 90.2\%, respectively, while the corresponding median held-out lower-bound values differed from that of universal access by less than 0.05 points. At $\Lambda=3$, the median access share reached 100\%, with a median lower-bound value of 43.19, matching that of universal access.

The increasing preference for access is consistent with the no-access lower bound decreasing more sharply than the universal-access bound as sensitivity increases. One contributing factor is the smaller fitted complier-fraction summary for nonselected nonenrollees ($\bar\gamma_0=0.274$) than for selected enrollees ($\bar\gamma_1=0.888$), which permits greater variation in the sensitivity weights; differences in outcome distributions also affect the bounds. These results should be interpreted in light of the maintained assumptions of conditional assignment exchangeability, monotonicity, overlap, and adequate nuisance specification. The analysis also does not impose costs or capacity constraints.

The selected DR policy tree varied across training splits and sensitivity levels. At $\Lambda=1$, the most frequent rule (44 of 100 splits) offered access when the household mean of the age group indicator \texttt{AG2} was at most $2/3$. At $\Lambda=1.3$, the most frequent rule (73 of 100 splits) withheld access from households with at least one analyzed respondent who requested non-English materials and offered access to all others. The excluded households accounted for 9.9\% of the survey-weighted analysis sample. Language preference may capture differences in access to care or unmet need, although the selected split alone does not establish this explanation. \citet{johnson2021heterogeneous} also identified language preference within heterogeneous effect subgroups in OHIE, but examined days without health-related activity limitations and considered language together with other characteristics. Their findings provide context for considering language in policy learning. The age-based splits are broadly consistent with the heterogeneity in lottery-induced enrollment across age groups reported by \citet{hattab2024oregon}, although their estimand differs from our worst-case policy value.

	\section{Discussion}\label{sec:discussion}
Our framework incorporates identified complier proportions and observed outcome distributions into sensitivity analysis for principal ignorability. Joint compatibility ensures that the bounds used to assess complier effects and evaluate assignment policies are attainable under a single full-data law. Retaining the full observed distribution is particularly valuable for nonbinary outcomes, as the mean already determines the distribution for binary outcomes. The identification characterization yields sharp bounds for effects and policy values, while their dependence on principal scores necessitates an additional correction in estimation.

Even with a common sensitivity parameter, the implications can differ across arms because complier fractions and outcome distributions differ. In OHIE, modest departures from PI along the common-parameter path allow the estimated effect interval to include zero, whereas larger departures favor universal access under the cost-free maximin objective within the considered policy class.

The theoretical guarantees rely on distinct regularity conditions. Conservative endpoints under quantile misspecification require the accompanying principal-score and assignment or outcome-regression conditions, while efficient inference additionally requires overlap, quantile regularity, and sufficient nuisance convergence rates. The simulations with working regressions illustrate finite-sample performance without necessarily satisfying every theorem assumption. Policy inference requires separation of the optimum, leaving near ties and nonunique optima for further investigation. Throughout, we assume conditional assignment exchangeability, monotonicity, and identification of the outcome distribution in the analysis population. Extending the framework to accommodate costs, capacity constraints, non-monotonic receipt, or informative response is a promising direction for future research.
\newpage
\bibliographystyle{abbrvnat}
  	\bibliography{ref}

@inproceedings{chu23c,
  title={Multiply Robust Off-policy Evaluation and Learning under Truncation by Death},
  author={Chu, Jianing and Yang, Shu and Lu, Wenbin},
  booktitle={Proceedings of the 40th International Conference on Machine Learning},
  year={2023},
  volume={202},
  pages={6195--6227}
}

@article{psog2,
  title={Principal stratification in causal inference},
  author={Frangakis, Constantine E and Rubin, Donald B},
  journal={Biometrics},
  volume={58},
  number={1},
  pages={21--29},
  year={2002},
  publisher={Oxford University Press}
}

@book{kosorok2008,
  title={Introduction to empirical processes and semiparametric inference},
  author={Kosorok, Michael R},
  year={2008},
  publisher={Springer}
}

@article{ps1,
  title={Identification and Estimation of Local Average Treatment Effects},
  author={Imbens, Guido W and Angrist, Joshua D},
  journal={Econometrica},
  volume={62},
  number={2},
  pages={467--475},
  year={1994},
  publisher={Econometric Society}
}

@article{ps2,
  title={The paired availability design: a proposal for evaluating epidural analgesia during labor},
  author={Baker, Stuart G and Lindeman, Karen S},
  journal={Statistics in Medicine},
  volume={13},
  number={21},
  pages={2269--2278},
  year={1994},
  publisher={Wiley Online Library}
}

@article{ps3,
  title={Multiply robust estimation of causal effects under principal ignorability},
  author={Jiang, Zhichao and Yang, Shu and Ding, Peng},
  journal={Journal of the Royal Statistical Society Series B: Statistical Methodology},
  volume={84},
  number={4},
  pages={1423--1445},
  year={2022},
  publisher={Oxford University Press}
}

@article{milgrom2002envelope,
  title={Envelope Theorems for Arbitrary Choice Sets},
  author={Milgrom, Paul and Segal, Ilya},
  journal={Econometrica},
  volume={70},
  number={2},
  pages={583--601},
  year={2002},
  doi={10.1111/1468-0262.00296}
}

@article{kitagawa2018welfare,
  title={Who Should Be Treated? {Empirical} Welfare Maximization Methods for Treatment Choice},
  author={Kitagawa, Toru and Tetenov, Aleksey},
  journal={Econometrica},
  volume={86}, number={2}, pages={591--616}, year={2018},
  doi={10.3982/ECTA13288}
}

@article{athey2021policy,
  title={Policy Learning With Observational Data},
  author={Athey, Susan and Wager, Stefan},
  journal={Econometrica},
  volume={89}, number={1}, pages={133--161}, year={2021},
  doi={10.3982/ECTA15732}
}

@article{sverdrup2020policytree,
  title={policytree: Policy learning via doubly robust empirical welfare maximization over trees},
  author={Sverdrup, Erik and Kanodia, Ayush and Zhou, Zhengyuan and Athey, Susan and Wager, Stefan},
  journal={Journal of Open Source Software},
  volume={5}, number={50}, pages={2232}, year={2020},
  doi={10.21105/joss.02232}
}

@inproceedings{leboeuf2020decision,
  title={Decision Trees as Partitioning Machines to Characterize Their Generalization Properties},
  author={Leboeuf, Jean-Samuel and LeBlanc, Fr\'ed\'eric and Marchand, Mario},
  booktitle={Advances in Neural Information Processing Systems},
  volume={33}, year={2020},
  pages={18135--18145},
  url={https://papers.nips.cc/paper/2020/hash/d2a10b0bd670e442b1d3caa3fbf9e695-Abstract.html}
}

@book{shalev2014understanding,
  title={Understanding machine learning: From theory to algorithms},
  author={Shalev-Shwartz, Shai and Ben-David, Shai},
  year={2014},
  publisher={Cambridge University Press}
}

@article{hattab2024oregon,
  author={Hattab, Zaid and Doherty, Edel and Ryan, Andrew M. and O'Neill, Stephen},
  title={Heterogeneity within the {Oregon} Health Insurance Experiment: An application of causal forests},
  journal={PLOS ONE}, year={2024}, volume={19}, number={1}, pages={e0297205},
  doi={10.1371/journal.pone.0297205}
}

@misc{hattab2023data,
  author={Hattab, Zaid}, title={Clean Data.csv}, year={2023},
  publisher={figshare}, doi={10.6084/m9.figshare.24718401},
  url={https://doi.org/10.6084/m9.figshare.24718401},
  note={Oregon Health Insurance Experiment respondent dataset; CC BY 4.0}
}

@article{finkelstein2012oregon,
  title={The {Oregon} health insurance experiment: evidence from the first year},
  author={Finkelstein, Amy and Taubman, Sarah and Wright, Bill and Bernstein, Mira and Gruber, Jonathan and Newhouse, Joseph P and Allen, Heidi and Baicker, Katherine and {the Oregon Health Study Group}},
  journal={The Quarterly Journal of Economics},
  volume={127},
  number={3},
  pages={1057--1106},
  year={2012},
  publisher={Oxford University Press}
}

@article{ding2017principal,
 author={Ding, Peng and Lu, Jiannan},
 title={Principal stratification analysis using principal scores},
 journal={Journal of the Royal Statistical Society Series B: Statistical Methodology},
 year={2017}, volume={79}, number={3}, pages={757--777},
 url={https://arxiv.org/abs/1602.01196}
}

@article{nguyen2024sensitivity,
 author={Nguyen, Trang Quynh and Stuart, Elizabeth A. and Scharfstein, Daniel O. and Ogburn, Elizabeth L.},
 title={Sensitivity analysis for principal ignorability violation in estimating complier and noncomplier average causal effects},
 journal={Statistics in Medicine}, year={2024}, volume={43}, number={19}, pages={3664--3688},
 url={https://arxiv.org/abs/2303.05032}
}

@article{chen2026beyond,
 author={Chen, Xinyuan and Harhay, Michael O. and Li, Fan},
 title={Beyond principal ignorability: Nonparametric sensitivity bounds for principal stratification},
 journal={arXiv preprint arXiv:2606.01669}, year={2026},
 doi={10.48550/arXiv.2606.01669}
}

@article{tan2006distributional,
 author={Tan, Zhiqiang},
 title={A Distributional Approach for Causal Inference Using Propensity Scores},
 journal={Journal of the American Statistical Association}, year={2006},
 volume={101}, number={476}, pages={1619--1637}, doi={10.1198/016214506000000023}
}

@article{zhao2019sensitivity,
 author={Zhao, Qingyuan and Small, Dylan S. and Bhattacharya, Bhaswar B.},
 title={Sensitivity analysis for inverse probability weighting estimators via the percentile bootstrap},
 journal={Journal of the Royal Statistical Society Series B: Statistical Methodology},
 year={2019}, volume={81}, number={4}, pages={735--761}, doi={10.1111/rssb.12327}
}

@article{dorn2023sharp,
 author={Dorn, Jacob and Guo, Kevin},
 title={Sharp Sensitivity Analysis for Inverse Propensity Weighting via Quantile Balancing},
 journal={Journal of the American Statistical Association},
 year={2023}, volume={118}, number={544}, pages={2645--2657}, doi={10.1080/01621459.2022.2069572}
}

@article{dorn2025doubly,
 author={Dorn, Jacob and Guo, Kevin and Kallus, Nathan},
 title={Doubly-Valid/Doubly-Sharp Sensitivity Analysis for Causal Inference with Unmeasured Confounding},
 journal={Journal of the American Statistical Association},
 year={2025}, volume={120}, number={549}, pages={331--342},
 doi={10.1080/01621459.2024.2335588}
}

@inproceedings{park2026evaluating,
 author = {Park, Sihyung and Lu, Wenbin and Yang, Shu},
 booktitle = {Advances in Neural Information Processing Systems},
 doi = {10.52202/085713-3655},
 editor = {D. Belgrave and C. Zhang and H. Lin and R. Pascanu and P. Koniusz and M. Ghassemi and N. Chen},
 pages = {109502--109538},
 publisher = {Curran Associates, Inc.},
 title = {Evaluating and Learning Optimal Dynamic Treatment Regimes under Truncation by Death},
 url = {https://proceedings.neurips.cc/paper_files/paper/2025/file/9deeb6f5be09657d9e6e6c0d2416af47-Paper-Conference.pdf},
 volume = {38, Main Conference},
 year = {2025}
}

@article{chen2026principal,
  title={Principal stratification with {U}-statistics under principal ignorability},
  author={Chen, Xinyuan and Li, Fan},
  journal={Journal of the Royal Statistical Society Series B: Statistical Methodology},
  volume={88},
  number={4},
  pages={1345--1365},
  year={2026},
  publisher={Oxford University Press, UK}
}

@article{lu2026principal,
  title={Principal stratification with continuous post-treatment variables: Nonparametric identification and semiparametric estimation},
  author={Lu, Sizhu and Jiang, Zhichao and Ding, Peng},
  journal={Journal of the Royal Statistical Society Series B: Statistical Methodology},
  volume={88},
  number={1},
  pages={239--260},
  year={2026},
  publisher={Oxford University Press, UK}
}

@article{rockafellar2000optimization,
  title={Optimization of conditional value-at-risk},
  author={Rockafellar, R Tyrrell and Uryasev, Stanislav},
  journal={Journal of Risk},
  volume={2},
  number={3},
  pages={21--41},
  year={2000},
  doi={10.21314/JOR.2000.038}
}

@article{johnson2021heterogeneous,
 author={Johnson, Michael and Cao, JiongYi and Kang, Hyunseung},
 title={Detecting heterogeneous treatment effects with instrumental variables and application to the {Oregon} health insurance experiment},
 journal={The Annals of Applied Statistics},
 year={2022}, volume={16}, number={2}, pages={1111--1129},
 doi={10.1214/21-AOAS1535}
}

@incollection{newey1994large,
  author = {Newey, Whitney K. and McFadden, Daniel},
  title = {Large Sample Estimation and Hypothesis Testing},
  booktitle = {Handbook of Econometrics},
  editor = {Engle, Robert F. and McFadden, Daniel},
  volume = {4},
  chapter = {36},
  pages = {2111--2245},
  publisher = {Elsevier},
  year = {1994},
  doi = {10.1016/S1573-4412(05)80005-4}
}

@article{luedtke2016statistical,
  author = {Luedtke, Alexander R. and van der Laan, Mark J.},
  title = {Statistical Inference for the Mean Outcome under a Possibly Non-Unique Optimal Treatment Strategy},
  journal = {The Annals of Statistics},
  year = {2016},
  volume = {44},
  number = {2},
  pages = {713--742},
  doi = {10.1214/15-AOS1384}
}

@article{stuart2015assessing,
 author={Stuart, Elizabeth A. and Jo, Booil},
 title={Assessing the sensitivity of methods for estimating principal causal effects},
 journal={Statistical Methods in Medical Research},
 year={2015}, volume={24}, number={6}, pages={657--674},
 doi={10.1177/0962280211421840}
}

@article{wang2023sensitivity,
 author={Wang, Craig and Zhang, Yufen and Mealli, Fabrizia and Bornkamp, Björn},
 title={Sensitivity analyses for the principal ignorability assumption using multiple imputation},
 journal={Pharmaceutical Statistics},
 year={2023}, volume={22}, number={1}, pages={64--78},
 doi={10.1002/pst.2260}
}

@article{lee2009training,
 author={Lee, David S.},
 title={Training, Wages, and Sample Selection: Estimating Sharp Bounds on Treatment Effects},
 journal={The Review of Economic Studies},
 year={2009}, volume={76}, number={3}, pages={1071--1102},
 doi={10.1111/j.1467-937X.2009.00536.x}
}

@inproceedings{kallus2018confounding,
 author={Kallus, Nathan and Zhou, Angela},
 title={Confounding-Robust Policy Improvement},
 booktitle={Advances in Neural Information Processing Systems},
 year={2018}, volume={31},
 url={https://papers.neurips.cc/paper_files/paper/2018/hash/3a09a524440d44d7f19870070a5ad42f-Abstract.html}
}

@misc{wu2024partial,
 author={Wu, Minxuan and Antonelli, Joseph},
 title={Partial identification of principal causal effects under violations of principal ignorability},
 year={2024},
 eprint={2412.06628},
 archivePrefix={arXiv},
 primaryClass={stat.ME},
 doi={10.48550/arXiv.2412.06628},
 note={Revised April 2026},
 url={https://arxiv.org/abs/2412.06628}
}

\newpage
\appendix
\section{Notations}
Throughout the Appendix, let $P$ denote the distribution of observed data $O$ and $P_X$ denote its marginal distribution of $X$. The nuisance functions are fitted on a training sample independent of the analysis observations $O_1,\ldots,O_n$, which are i.i.d. from $P$. For any function $f$ evaluated at either fitted or true nuisance functions, write $\E_n[f]=n^{-1}\sum_{i=1}^n f(O_i)$ for its mean over the analysis sample. Write $P_n=n^{-1}\sum_{i=1}^n\delta_{O_i}$ for the empirical distribution of the analysis observations, where $\delta_{O_i}$ places unit mass at $O_i$. Thus $\int f\,dP_n=\E_n[f]$ and $\|f\|_{L^2(P_n)}^2=n^{-1}\sum_{i=1}^n f(O_i)^2$. By contrast, $\E[f]=\E[f(O)]$ integrates over a new $O\sim P$ while holding any fitted functions fixed. Likewise, $\E[f(O)\mid X=x]$ integrates over the conditional law of a new observation given $X=x$, with fitted functions held fixed. Thus $\E[\hat f]$ can be random through the training sample, whereas $\E[f]$ is fixed when $f$ uses only true nuisance functions. In a cross-fitting implementation, the analysis observations correspond to a held-out fold and the nuisance functions are fitted on the other folds.

\section{Supporting identification representations}\label{app:supporting_representations}
This section presents the companion principal-stratum means, conditional IPW representation, and principal score weighting identities underlying the results in Section~\ref{sec:id}. These identities complement the sharp bounds and robust estimators developed in the main text.

The compatibility construction also determines the outcome distributions of the noncompliers in each observed mixture. Removing the complier component leaves the always-taker distribution in arm 1 and the never-taker distribution in arm 0. The resulting density ratios immediately give the conditional means in Corollary~\ref{cor:compat}.

\begin{cor}  \label{cor:compat}
	A distribution $P_{\full}$ constructed in Theorem \ref{thm:compat_plus} satisfies:
	\begin{align*}
		\mathbb{E}_{P_{\full}}[Y(1) \mid X, Z=1, U=11] = \mathbb{E}_P[W_{11} Y \mid X, Z=S=1], \\
		\mathbb{E}_{P_{\full}}[Y(0) \mid X, Z=0, U=00] = \mathbb{E}_P[W_{00} Y \mid X, Z=S=0],
	\end{align*}
	where $W_{11} = \{1 - \gamma_1(X)W_1\}/\{1 - \gamma_1(X)\}$ and $W_{00} = \{1 - \gamma_0(X)W_0\}/\{1 - \gamma_0(X)\}$.
\end{cor}
Under Assumptions \ref{A:consistency}, \ref{A:nuc}, and \ref{A:monoton}, the remaining stratum-specific potential outcome means are directly identified from the observed data:
\begin{align*}
\E_{P_{\full}}[Y(0)| X,U=11] = \E_P[Y| X,Z=0,S=1],\quad
\E_{P_{\full}}[Y(1)| X,U=00] = \E_P[Y| X,Z=1,S=0].
\end{align*}
Together, Theorem \ref{thm:compat_plus} and Corollary \ref{cor:compat} establish that a single full-data distribution simultaneously realizes all six stratum-specific conditional potential outcome means.

We next compare the marginal IPW representation in Section~\ref{sec:id} with the quantile representation used in marginal sensitivity analysis \citep{dorn2025doubly} which gives, after adapting the weights to the complier population,
\[
\frac{\E[\pi_C(X)\hat Q_{z\pm}(X)]
+\E\!\left[\frac{\mathbb{1}\{Z=S=z\}\gamma_z(X)}{e_z(X)}
\{\nu_{z\pm}(X,Y;\hat Q_{z\pm})-\hat Q_{z\pm}(X)\}\right]}{\pi_C}.
\]
With the true nuisance functions, this equals the marginal IPW representation for any candidate quantile. With true principal scores, propensity limit $e_z^*(X)$ and candidate quantile limit $Q^*_{z\pm}(X)$, the alternative minus the direct IPW limit equals
\[
\frac{1}{\pi_C}\E\!\left[\pi_C(X)
\left\{1-\frac{e_z(X)}{e_z^*(X)}\right\}Q^*_{z\pm}(X)\right].
\]
Thus the two fitted estimators need not agree under propensity misspecification.

We now return to bounding the complier mean. The conditional expectation of the transformed outcome $\nu_{z\pm}$ is $\omega_{z\pm}$, so the IPW bounds in Theorem~\ref{thm:ipw1x} also admit an outcome-regression representation. Theorem~\ref{thm:or1x} states this representation and its equality at the true quantiles.

\begin{theorem}[Outcome regression] \label{thm:or1x}
	For any candidate quantile functions $\hat Q_{z+}$ and $\hat Q_{z-}$, the conditional means $\omega_{z\pm}$ satisfy $\psi_{z+,X} \le \omega_{z+}(X;\hat{Q}_{z+})$ and $\psi_{z-,X} \ge \omega_{z-}(X;\hat{Q}_{z-})$ almost surely.
	Equalities hold when $\hat{Q}_{z\pm} = Q_{z\pm}$, respectively.
\end{theorem}

To obtain population bounds from the conditional expressions, we average over the covariate distribution among compliers. The following identities express this average using either principal score or inverse probability weighting of observed outcomes.

\begin{lemma}  \label{lemma:scoring}
	For any integrable $h(X)$ and $g(X,Y)$, and $z\in\{0,1\}$,
	\begin{gather*}
		\E[h(X) | U=01] = \E\left[ \frac{\pi_C(X)}{\pi_C} h(X) \right], \\
		\E[ \E\{ g(X,Y) | X, Z=S=z \} | U = 01 ] = \E\left[ \frac{\mathbb{1}\{Z=S=z\}\gamma_z(X)}{e_z(X)\pi_C} g(X,Y) \right].
	\end{gather*}
\end{lemma}

\begin{proof}
By iterated expectation and the definition of $\pi_C(X)$,
\[
\E[h(X)\mid U=01]
=\frac{\E[h(X)\mathbb{1}\{U=01\}]}{\pi_C}
=\frac{\E[h(X)\pi_C(X)]}{\pi_C}.
\]
Also, $\pr(Z=S=z\mid X)=e_z(X)p_z(X)$ and $\gamma_z(X)p_z(X)=\pi_C(X)$, so
\[
\E\!\left[\frac{\mathbb{1}\{Z=S=z\}\gamma_z(X)}{e_z(X)}g(X,Y)\right]
=\E\!\left[\pi_C(X)\E\{g(X,Y)\mid X,Z=S=z\}\right].
\]
Divide by $\pi_C$ and apply the first identity to the inner conditional mean.
\end{proof}

Applying Lemma~\ref{lemma:scoring} to the conditional bounds gives equivalent marginal OR and IPW representations. The next theorem states these representations for arm 1, including their conservativeness for candidate quantiles and sharpness at the true quantiles.

\begin{theorem}\label{thm:marginal_representations}
	Under the model \eqref{def:MSM2} and Assumptions \ref{A:consistency}--\ref{A:positivity_boundedmean},
	\begin{gather}
		\frac{\E\left[ \pi_C(X) \omega_{1\pm}(X;\hat{Q}_{1\pm}) \right]}{\pi_C}  = \frac{\E\left[ \frac{Z S \gamma_1(X)}{e_1(X)} \nu_{1\pm}(X,Y;\hat Q_{1\pm}) \right]}{\pi_C}  %{\E\left[ {Z S \gamma_1(X)}/{e_1(X)} \right]}.
	\end{gather}
	are valid bounds on $\psi_1$. The bounds are sharp when $\hat{Q}_{1\pm} = {Q}_{1\pm}$, respectively.
\end{theorem}
\begin{proof}
The result follows directly	from Theorems \ref{thm:ipw1x} and \ref{thm:or1x}, and Lemma \ref{lemma:scoring}.
\end{proof}

\section{Proofs of main results}\label{app:proofs}

\subsection{Proof of Proposition~\ref{prop:mixture_limits}}\label{app:mixture_limits}

Fix $z\in\{0,1\}$ and a covariate value $x$ satisfying the assumptions. Differentiating the weight limits gives
\[
\frac{\partial L_z(x)}{\partial\gamma_z(x)}=(\Lambda_z-1)L_z(x)^2\ge0,
\qquad
\frac{\partial U_z(x)}{\partial\gamma_z(x)}=(\Lambda_z^{-1}-1)U_z(x)^2\le0.
\]
With $F_z(\cdot\mid x)$ fixed, increasing $\gamma_z(x)$ narrows the weight interval and leaves normalization unchanged. The lower endpoint therefore increases and the upper endpoint decreases. For any admissible $W_z$,
\begin{align*}
&\left|\E[W_z(X,Y)Y\mid X=x,Z=S=z]-\mu_z(x)\right|\\
&\le\max\{1-L_z(x),U_z(x)-1\}\E[|Y|\mid X=x,Z=S=z].
\end{align*}
Since $L_z(x),U_z(x)\to1$ as $\gamma_z(x)\uparrow1$, conditional integrability implies that both endpoints converge to $\mu_z(x)$. Increasing $\Lambda_z$ widens the weight interval, so the feasible class expands. At $\Lambda_z=1$, the only admissible weight is one, and the displayed bound also gives convergence to $\mu_z(x)$ as $\Lambda_z\downarrow1$.

For $\Lambda_z>1$, write $W_z(x,y)=L_z(x)+\{U_z(x)-L_z(x)\}h_z(x,y)$, where $0\le h_z\le1$. Normalization requires
\[
\E[h_z(X,Y)\mid X=x,Z=S=z]=t_z(x)
:=\frac{1-L_z(x)}{U_z(x)-L_z(x)}
=\frac{1+(\Lambda_z-1)\gamma_z(x)}{\Lambda_z+1}.
\]
Set $h_{z-}(x,y)$ to one below $Q_{z,x}(t_z(x))$ and zero above it, choosing its value at the quantile to make its conditional expectation $t_z(x)$. For any feasible $h_z$,
\[
\{h_z(x,y)-h_{z-}(x,y)\}\{y-Q_{z,x}(t_z(x))\}\ge0.
\]
The conditional expectation of $h_z-h_{z-}$ is zero; thus this shows that $h_{z-}$ minimizes the weighted mean. Allocating the same mass to the largest outcomes maximizes it. Hence,
\begin{align*}
&\inf_{W_z\in\mathcal E_z(\Lambda_z)}\E[W_z(X,Y)Y\mid X=x,Z=S=z]\\
&\quad=L_z(x)\mu_z(x)+\{U_z(x)-L_z(x)\}\int_0^{t_z(x)}Q_{z,x}(u)\,du,\\
&\sup_{W_z\in\mathcal E_z(\Lambda_z)}\E[W_z(X,Y)Y\mid X=x,Z=S=z]\\
&\quad=L_z(x)\mu_z(x)+\{U_z(x)-L_z(x)\}\int_{1-t_z(x)}^1Q_{z,x}(u)\,du.
\end{align*}
As $\Lambda_z\to\infty$, $L_z(x)\to0$, $U_z(x)\to1/\gamma_z(x)$, and $t_z(x)\to\gamma_z(x)$. Since $\int_0^1|Q_{z,x}(u)|\,du<\infty$, these expressions converge to the stated limits.

To establish sharpness of these limits, note that the mixture identity holds exactly when the complier density ratio is normalized and satisfies $0\le W_z(x,y)\le1/\gamma_z(x)$. The component density ratios are then
\begin{align*}
\frac{dP_{Y(z)\mid X=x,U=01}}{dP_{Y\mid X=x,Z=S=z}}(y)=W_z(x,y), \qquad
\frac{dP_{Y(z)\mid X=x,U=zz}}{dP_{Y\mid X=x,Z=S=z}}(y)=\frac{1-\gamma_z(x)W_z(x,y)}{1-\gamma_z(x)}.
\end{align*}
These ratios are nonnegative, integrate to one, and reproduce the observed mixture. To attain the lower and upper limiting means, take
\begin{align*}
W_{z-}^{\infty}(x,y)
&=\frac{\mathbb{1}\{y<Q_{z,x}(\gamma_z(x))\}
+c_{z-}(x)\mathbb{1}\{y=Q_{z,x}(\gamma_z(x))\}}{\gamma_z(x)},\\
W_{z+}^{\infty}(x,y)
&=\frac{\mathbb{1}\{y>Q_{z,x}(1-\gamma_z(x))\}
+c_{z+}(x)\mathbb{1}\{y=Q_{z,x}(1-\gamma_z(x))\}}{\gamma_z(x)}.
\end{align*}
Choose $c_{z\pm}(x)\in[0,1]$ to normalize the weights, using fractional allocation at atoms when needed. Integrating these weights against $y$ gives the limits in the proposition. The allocation inequality above, applied with mass $\gamma_z(x)$, shows that no compatible weight can improve either endpoint. \qed

\subsection{Proof of Corollary~\ref{cor:information_relaxation}}
Fix an arm and a covariate value satisfying the assumptions. Under any conditional outcome law in the relaxed class, $0\le Y\le1$ and the conditional mean is $\mu_z(x)$. The weight limits imply
\begin{align*}
L_z(x)\mu_z(x)
&\le\E[W_z(X,Y)Y\mid X=x,Z=S=z]\le U_z(x)\mu_z(x),\\
L_z(x)\{1-\mu_z(x)\}
&\le\E[W_z(X,Y)(1-Y)\mid X=x,Z=S=z]\le U_z(x)\{1-\mu_z(x)\}.
\end{align*}
Normalization makes the two weighted expectations sum to one. The second inequality therefore gives another interval for the weighted outcome mean. Intersecting it with the first interval yields the stated bounds.

For attainability when $0<\mu_z(x)<1$, choose a Bernoulli outcome law with probability $\mu_z(x)$. The lower endpoint is attained by
\begin{align*}
W_{z-}(x,1)
=\max\left\{L_z(x),\frac{1-U_z(x)\{1-\mu_z(x)\}}{\mu_z(x)}\right\},\;\;
W_{z-}(x,0)
=\min\left\{U_z(x),\frac{1-L_z(x)\mu_z(x)}{1-\mu_z(x)}\right\}.
\end{align*}
Since $L_z(x) \le 1 \le U_z(x)$, the fraction in the maximum is at most $U_z(x)$, and the fraction in the minimum is at least $L_z(x)$. Thus, $L_z(x) \le W_{z-}(x,y) \le U_z(x)$. Direct substitution gives
\begin{align*}
\mu_z(x)W_{z-}(x,1)+\{1-\mu_z(x)\}W_{z-}(x,0)=1,\\
\mu_z(x)W_{z-}(x,1)
=\max\{L_z(x)\mu_z(x),\,1-U_z(x)\{1-\mu_z(x)\}\}.
\end{align*}
Thus, these weights are normalized and attain the lower bound. For the upper bound, take
\begin{align*}
W_{z+}(x,1)
=\min\left\{U_z(x),\frac{1-L_z(x)\{1-\mu_z(x)\}}{\mu_z(x)}\right\}, \;\;
W_{z+}(x,0)
=\max\left\{L_z(x),\frac{1-U_z(x)\mu_z(x)}{1-\mu_z(x)}\right\}.
\end{align*}
The same checks show that these weights lie in $[L_z(x),U_z(x)]$, are normalized, and attain $\min\{U_z(x)\mu_z(x),\,1-L_z(x)\{1-\mu_z(x)\}\}$. If $\mu_z(x)\in\{0,1\}$, the outcome is constant and both endpoints equal $\mu_z(x)$.
\qed

\subsection{Proof of Theorem \ref{thm:compat_plus}}

\paragraph{Necessity.}
Let $P_{\full}$ preserve $P$ and satisfy the theorem's structural and density ratio restrictions. Conditional exchangeability and monotonicity identify the stratum probabilities and imply the mixture identity in Section~\ref{sec:msm}. For any Borel set $B$, this identity gives
\[
\gamma_z(x)P_{\full}\{Y(z)\in B\mid X=x,U=01\}
\le P\{Y\in B\mid X=x,Z=S=z\}.
\]
Thus the complier law is absolutely continuous with respect to its observed mixture, and its derivative $W_z$ satisfies $0\le\gamma_zW_z\le1$ and conditional normalization. Bayes' rule gives posterior complier membership probability $\gamma_zW_z$. Substitution into \eqref{def:MSM1} yields $L_z\le W_z\le U_z$. Integrating $W_z$ against the observed conditional law therefore gives the complier law in $\mathcal G_z(P,\Lambda_z)$ and, on integrating $y$, its conditional mean.

\paragraph{Sufficiency.} Fix a pair of candidate laws and choose weights $W_0,W_1$ that generate them as in the definition of $\mathcal G_z(P,\Lambda_z)$ in Section~\ref{sec:datacompatible}. We retain the observed variables and construct the missing potential outcomes and receipt types as follows.

We work on a product extension of the observed data probability space supporting independent \(V_1,\ldots,V_4\sim\mathrm{Uniform}(0,1)\), independent of \((X,Z,S,Y,W_0,W_1)\).
Here $W_z = W_z(X,Y)$ is the given weight on the cell $Z=S=z$; extend it arbitrarily outside that cell. Define $\pi_S(X,Y) = W_0(X,Y)\gamma_0(X)$ on $Z=S=0$, and $\pi_S(X,Y) = 0$ outside $Z=S=0$. The normalization conditions are explicitly
\[
 E[W_z(X,Y)\mid X,Z=S=z]=1,\qquad z=0,1.
\]
The sensitivity bounds imply $0\le\gamma_z(X)W_z(X,Y)\le1$ on the corresponding cell, so these are valid allocation probabilities.

First generate the potential treatment indicators:
\begin{align*}
 S(1) &= ZS + (1-Z)\left[S + (1-S)\mathbb{1}\{V_1\le\pi_S(X,Y)\}\right],\\
 S(0) &= ZS\mathbb{1}\{V_2>W_1\gamma_1(X)\} + (1-Z)S.
\end{align*}
Now $(S(0),S(1))$ are defined, so we can derive the conditional distribution
$F(y\mid x,z,s_0,s_1)=P(Y\le y\mid X=x,Z=z,S(0)=s_0,S(1)=s_1)$ induced by this allocation. In particular,
\begin{align*}
 F(y\mid x,1,0,1)
 &= \frac{E[\mathbb{1}\{Y\le y\} \mathbb{1}\{V_2\le W_1\gamma_1(x)\} \mid X=x,Z=S=1]}{P(V_2\le W_1\gamma_1(x) \mid X=x,Z=S=1)}\\
 &= E[W_1\mathbb{1}\{Y\le y\} \mid X=x,Z=S=1],\\
 F(y\mid x,0,0,1)
 &= \frac{E[\mathbb{1}\{Y\le y\} \mathbb{1}\{V_1\le W_0\gamma_0(x)\} \mid X=x,Z=S=0]}{P(V_1\le W_0\gamma_0(x) \mid X=x,Z=S=0)}\\
 &= E[W_0\mathbb{1}\{Y\le y\} \mid X=x,Z=S=0].
\end{align*}
The denominators equal $\gamma_1(x)$ and $\gamma_0(x)$ by normalization. Conditioning on the complementary allocations gives
\begin{align*}
 F(y\mid x,1,1,1) &= \frac{E[(1-\gamma_1(x)W_1) \mathbb{1}\{Y\le y\} \mid X=x,Z=S=1]}{1-\gamma_1(x)},\\
 F(y\mid x,0,0,0) &= \frac{E[(1-\gamma_0(x)W_0)\mathbb{1}\{Y\le y\} \mid X=x,Z=S=0]}{1-\gamma_0(x)}.
\end{align*}
The remaining two distributions are directly observed:
$F(y\mid x,0,1,1)=P(Y\le y\mid X=x,Z=0,S=1)$, $F(y\mid x,1,0,0)=P(Y\le y\mid X=x,Z=1,S=0)$.
Each conditional distribution is well-defined: the allocation weights are nonnegative, and their normalization ensures that the corresponding conditional distributions have total mass one. Conditional distributions associated with zero-probability strata may be specified arbitrarily; e.g., the impossible defier stratum has probability zero, so its conditional law can be defined arbitrarily. 
These CDFs are derived from the observed data and the allocation above, before constructing the missing potential outcomes.

Let $J^{-1}(u)=\inf\{y\in\mathbb{R}:J(y)\ge u\}$ be the generalized inverse of a CDF. Complete the construction by
\begin{align*}
 Y(1)&=ZY+(1-Z)F^{-1}(V_3\mid X,1,S(0),S(1)),\\
 Y(0)&=ZF^{-1}(V_4\mid X,0,S(0),S(1))+(1-Z)Y.
\end{align*}
This leaves the original observed variables unchanged and hence preserves their joint law.

The construction ensures consistency and monotonicity. To check conditional exchangeability, observe that
\begin{align*}
	&P_{full}(S(0)=s_0, S(1)=s_1, Y(0)\le y_0, Y(1)\le y_1 | X,Z=1) \\
	&= P_{full}(S(0)=s_0, S(1)=s_1 \mid X, Z=1) \cdot P(Y(0) \leq y_0, Y(1) \leq y_1 \mid X, Z=1, S(0)=s_0, S(1)=s_1) \\
	&= P(S \mathbb{1}\{V_2 > W_1 \gamma_1(X)\} = s_0, S=s_1 \mid X, Z=1)  \\
	&\quad\cdot P(F^{-1}(V_4 \mid X, 0, s_0, s_1) \leq y_0, Y \leq y_1 \mid X, Z=1, S(0)=s_0, S(1)=s_1) \\
	&= P(S \mathbb{1}\{V_2 > W_1 \gamma_1(X)\} = s_0, S=s_1 \mid X, Z=1)  \\
	&\quad \cdot P(F^{-1}(V_4 \mid X, 0, s_0, s_1) \leq y_0 \mid X, Z=1, S(0)=s_0, S(1)=s_1) \\
	&\quad \cdot P(Y \leq y_1 \mid X, Z=1, S(0)=s_0, S(1)=s_1) \\
	&= P(S \mathbb{1}\{V_2 > W_1 \gamma_1(X)\} = s_0, S=s_1 \mid X, Z=1) \cdot F(y_0 \mid X, 0, s_0, s_1) \cdot F(y_1 \mid X, 1, s_0, s_1)
\end{align*}
and
\begin{align*}
	&P(S(0)=s_0, S(1)=s_1, Y(0) \leq y_0, Y(1) \leq y_1 \mid X, Z=0) \\
	&= P(S=s_0, S+(1-S)\mathbb{1}\{V_1\le\pi_S(X,Y)\}=s_1 \mid X, Z=0) \cdot F(y_0 \mid X, 0, s_0, s_1) \cdot F(y_1 \mid X, 1, s_0, s_1).
\end{align*}
We claim that
\begin{align*}
	&P(S \mathbb{1}\{V_2 > W_1 \gamma_1(X)\} = s_0, S=s_1 \mid X, Z=1) \\
	&= P(S=s_0, S+(1-S)\mathbb{1}\{V_1 \leq \pi_S(X, Y)\}=s_1 \mid X, Z=0).
\end{align*}
We can check that both probabilities have the same values in any of the four cases:
\begin{align*}
	\begin{cases}
		p_{0,1}(X), & s_1=1,~s_0=1 \\
		\pi_C(X), & s_1=1,~s_0=0 \\
		0, & s_1=0,~s_0=1 \\
		1-p_1(X), & s_1=0,~s_0=0
	\end{cases}
\end{align*}
Indeed, the complier probabilities in the two arms are $p_1(X)\gamma_1(X)=p_0(X)\gamma_0(X)=\pi_C(X)$ by normalization, and the complementary probabilities are $p_{0,1}(X)$ and $1-p_1(X)$. Independence of $V_3,V_4$ from the observed data and the stratum allocation justifies the factorizations above. Thus both arms have the same conditional joint law of $(S(0),S(1),Y(0),Y(1))$ given $X$.

Next, we verify the arm-1 mean identity. Note that
\begin{align}
	\E_{P_{full}}[Y(1) | X,Z=1,S(0)=0,S(1)=1] = \frac{\E_{P_{full}}[(1-S(0))Y | X,Z=S(1)=1]}{P_{full}(S(0)=0 | X,Z=S(1)=1)}
\end{align}
by consistency. The denominator equals $\gamma_1(X)$. By the law of iterated expectation, the numerator simplifies to
\begin{align*}
	&\E_{P_{full}}[Y(1) \E_{P_{full}}\{1-S(0) | X,Y(1),W_1,Z=S(1)=1\} | X,Z=S(1)=1] \\
	&= \E_{P_{full}}[Y(1) W_1\gamma_1(X) | X,Z=S(1)=1]
	= \E_{P}[Y W_1\gamma_1(X) | X,Z=S=1] \\
	&= \gamma_1(X) \E_P[W_1Y | X,Z=S=1].
\end{align*}
Hence, we have the desired result.

Finally, we show that, by construction,
\begin{align*}
	&\frac{ dP_{{full}}\!\left(Y(1)\in\cdot\mid X=x,Z=1,U=01\right) }{ dP_{{full}}\!\left(Y(1)\in\cdot\mid X=x,Z=1,S(1)=1\right) }(y) \\
	&= \frac{ P_{{full}}(U=01\mid X=x,Y(1)=y,Z=1,S(1)=1) }{ P_{{full}}(U=01\mid X=x,Z=1,S(1)=1) } = W_1(x,y).
\end{align*}
Since $L_1(X)\le W_1(X,y)\le U_1(X)$, the density-ratio bounds imply the arm-1 restriction in \eqref{def:MSM1}.

The construction already uses both $W_0$ and $W_1$ on the same probability space. The equality of stratum probabilities and the joint-law argument above therefore apply simultaneously. For the arm-0 mean identity, consistency and the allocation of $S(1)$ give
\begin{align*}
 &E_{P_{\full}}[Y(0)\mid X,Z=0,U=01]
 =\frac{E_{P_{full}}[S(1)Y\mid X,Z=S(0)=0]}{P_{full}(S(1)=1\mid X,Z=S(0)=0)}\\
 &=\frac{E_P[\gamma_0(X)W_0Y\mid X,Z=S=0]}{\gamma_0(X)}
 =E_P[W_0Y\mid X,Z=S=0].
\end{align*}
Similarly,
\[
 P_{\full}(S(1)=1\mid X,Y(0)=y,Z=S(0)=0)=\gamma_0(X)W_0(X,y),
\]
thus Bayes' rule identifies $W_0(X,y)$ as the complier-to-observed density ratio. Its bounds in \eqref{def:MSM2} imply that the odds of complier membership given $X$ and $Y=y$, relative to the odds given $X$ alone within $Z=S=0$, lie in $[\Lambda_0^{-1},\Lambda_0]$. This verifies \eqref{def:MSM1} for arm~0. The complier CDF identities derived from the allocations above hold simultaneously for both arms. Conditional exchangeability removes conditioning on $Z$, so $(P_{\full})_{Y(z)\mid X=x,U=01}(B)$ equals $\int_B W_z(x,y)\,dP_{Y\mid X=x,Z=S=z}(y)$ for both arms. The resulting law belongs to $\mathfrak P(P;\Lambda_0,\Lambda_1)$ and realizes that pair. \qed

\subsection{Proof of Theorem~\ref{thm:or1x}}
The argument compares weights separately for outcomes above and below the candidate quantile, as in \citet{dorn2025doubly}, and uses the quantile optimization characterization of conditional value-at-risk \citep{rockafellar2000optimization}.
Fix $z\in\{0,1\}$ and a covariate value $x$ for which the conditional outcome is integrable. For any $W_z\in\mathcal E_z(\Lambda_z)$ and any finite $q$, the pointwise bounds give
\begin{gather*}
L_z(x)(Y-q)-(U_z(x)-L_z(x))(q-Y)_+ \\
\le W_z(x,Y)(Y-q)
\le L_z(x)(Y-q)+(U_z(x)-L_z(x))(Y-q)_+.
\end{gather*}
The normalization constraint implies
\[
\E[W_z(X,Y)Y\mid X=x,Z=S=z]
=q+\E[W_z(X,Y)(Y-q)\mid X=x,Z=S=z].
\]
Taking conditional expectations in the pointwise inequalities therefore bounds this weighted mean from below by $\omega_{z-}(x;q)$ and from above by $\omega_{z+}(x;q)$, where a scalar $q$ denotes the value of the candidate quantile function at $x$. Taking the infimum and supremum over admissible $W_z$ proves the asserted inequalities.

For equality, first suppose $U_z(x)>L_z(x)$. The quantile definitions imply
\begin{align*}
\pr\{Y<Q_{z-}(x)\mid X=x,Z=S=z\}
&\le\tau_{z-}(x)\le
\pr\{Y\le Q_{z-}(x)\mid X=x,Z=S=z\},\\
\pr\{Y>Q_{z+}(x)\mid X=x,Z=S=z\}
&\le\tau_{z-}(x)\le
\pr\{Y\ge Q_{z+}(x)\mid X=x,Z=S=z\}.
\end{align*}
Consequently, choose $c_{z-}(x),c_{z+}(x)\in[0,1]$ so that each bracketed term below has conditional expectation $\tau_{z-}(x)$,
\begin{align*}
W_{z-}(x,y)&=L_z(x)+(U_z(x)-L_z(x))
\big[\mathbb{1}\{y<Q_{z-}(x)\}
+c_{z-}(x)\mathbb{1}\{y=Q_{z-}(x)\}\big],\\
W_{z+}(x,y)&=L_z(x)+(U_z(x)-L_z(x))
\big[\mathbb{1}\{y>Q_{z+}(x)\}
+c_{z+}(x)\mathbb{1}\{y=Q_{z+}(x)\}\big].
\end{align*}
Both weights satisfy the bounds and have conditional mean
$L_z(x)+(U_z(x)-L_z(x))\tau_{z-}(x)=1$.
The lower and upper pointwise inequalities become equalities at $q=Q_{z-}(x)$ and $q=Q_{z+}(x)$, respectively. Thus these admissible weights attain the claimed endpoints, including when the outcome law has atoms. Theorem~\ref{thm:compat_plus} ensures their full-data compatibility. If $L_z(x)=U_z(x)=1$, every expression reduces to $\mu_z(x)$.
\qed

\subsection{Proof of Theorem~\ref{thm:ipw1x}}
For any real $q$, the definition of $\nu_{z\pm}$ gives the pointwise identities
\begin{align*}
\nu_{z+}(x,y;q)
=q+\begin{cases}U_z(x)(y-q),&y>q\\L_z(x)(y-q),&y\le q\end{cases}, \quad
\nu_{z-}(x,y;q)
=q+\begin{cases}L_z(x)(y-q),&y>q\\U_z(x)(y-q)&y\le q\end{cases}.
\end{align*}
Choosing $L_z(x)$ or $U_z(x)$ according to whether the outcome lies above or below the candidate quantile is also used in the representations of \citet{dorn2025doubly}. Substituting the definitions of $L_z(x)$ and $U_z(x)$ yields
\[
\nu_{z\pm}(x,y;\hat Q_{z\pm})
=\hat Q_{z\pm}(x)
+\frac{y-\hat Q_{z\pm}(x)}
{\gamma_z(x)+\Lambda_z^{\mp\operatorname{sign}(y-\hat Q_{z\pm}(x))}
\{1-\gamma_z(x)\}}.
\]
At $y=\hat Q_{z\pm}(x)$ the residual is zero, so the identity also holds there.
Hold the candidate quantile functions fixed. Since $\pr(Z=S=z\mid X)=e_z(X)p_z(X)$, iterated expectation gives
\begin{align*}
&\E\!\left[\frac{\mathbb{1}\{Z=S=z\}}{e_z(X)p_z(X)}
\nu_{z\pm}(X,Y;\hat Q_{z\pm})\,\middle|\,X\right]\\
&= \frac{\pr(Z=S=z\mid X)}{e_z(X)p_z(X)}
\E[\nu_{z\pm}(X,Y;\hat Q_{z\pm})\mid X,Z=S=z]\\
&= \omega_{z\pm}(X;\hat Q_{z\pm}).
\end{align*}
Theorem~\ref{thm:or1x} then gives the leftmost and rightmost inequalities, with equality when the corresponding candidate quantile is true. The middle inequality follows because $\psi_{z-,X}$ and $\psi_{z+,X}$ are the infimum and supremum over the same nonempty class of admissible weights.

\subsection{Proof of Theorem \ref{thm:TR}}
We give the argument for arm 1.

\subsubsection{Conditional target and derivatives}  \label{AA:DR}
For a fixed quantile function $Q$, write $\rho_{1\pm}(x;Q)=\E[\{\pm(Y-Q(x))\}_+\mid X=x,Z=S=1]$ and $\omega_{1\pm}(Q)=L_1\mu_1+(1-L_1)Q\pm(U_1-L_1)\rho_{1\pm}(Q)$.
Here, the conditional outcome law and $Q$ are held fixed when differentiating in $(p_1,p_{0,1})$. Differentiating with respect to $\gamma_1$, we have $L_1' := (\Lambda_1-1)L_1^2$ and $U_1' := (\Lambda_1^{-1}-1)U_1^2$. Let $\omega'_{1\pm}(Q) = L_1'(\mu_1-Q)\pm(U_1'-L_1')\rho_{1\pm}(Q)$. We additionally have
\begin{align*}
 \partial_{p_1}\{\pi_C(X)\omega_{1\pm}(Q)\}
 &=\omega_{1\pm}(Q)+\gamma_1(1-\gamma_1)\omega'_{1\pm}(Q),\\
 \partial_{p_{0,1}}\{\pi_C(X)\omega_{1\pm}(Q)\}
 &=-\omega_{1\pm}(Q)-\gamma_1\omega'_{1\pm}(Q).
\end{align*}
These identities follow from $\pi_C(X)=p_1-p_{0,1}$, $\gamma_1=\pi_C(X)/p_1$, $\partial_{p_1}\gamma_1=(1-\gamma_1)/p_1$, and $\partial_{p_{0,1}}\gamma_1=-1/p_1$. 
The bounds in Assumptions \ref{A:positivity_boundedmean} and \ref{A:hat_positivity_boundedmean} keep $p_1$, the fitted $\hat p_1$, and the corresponding complier probabilities away from zero. Bounded candidate quantile functions and $\E[Y^2\mid X,Z=1,S=1]\le M$ then bound the first two derivatives of $\pi_C(X)\omega_{1\pm}(Q)$ uniformly over the admissible principal score values. In particular,
\[
 L_1-\hat L_1=(\Lambda_1-1)L_1\hat L_1(\gamma_1-\hat\gamma_1),\quad
 U_1-\hat U_1=(\Lambda_1^{-1}-1)U_1\hat U_1(\gamma_1-\hat\gamma_1).
\]

\subsubsection{Efficient influence function}\label{app:endpoint_eif}
Take $Q = Q_{1\pm}$ for the sharp endpoint. Under the conditional density and quantile uniqueness assumptions in Assumption \ref{A:rate}, the derivative of $\omega_{1\pm}(Q)$ with respect to $Q$ is zero at this quantile.
Thus the envelope argument \citep{milgrom2002envelope} removes the derivative of $Q$, including its dependence on $\gamma_1$.

Let $P_t$ be a regular observed-data submodel, and let dot notations denote derivatives at $t=0$. Factor its observed-data score as
\[
 s(O)=s(X)+s(Z\mid X)+s(S\mid X,Z)+s(Y\mid X,Z,S),
\]
where each conditional score has mean zero given the variables to its right. Holding $Q$ fixed as justified above, conditional differentiation gives
\begin{align*}
 \dot p_1&=\E[(S-p_1)s(S\mid X,Z=1)\mid X,Z=1],\\
 \dot p_{0,1}&=\E[(S-p_{0,1})s(S\mid X,Z=0)\mid X,Z=0],\\
 \dot\mu_1&=\E[(Y-\mu_1)s(Y\mid X,Z=S=1)\mid X,Z=S=1],\\
 \dot\rho_{1\pm}(Q)&=\E[(\{\pm(Y-Q)\}_+-\rho_{1\pm}(Q))s(Y\mid X,Z=S=1)\mid X,Z=S=1].
\end{align*}
For $\eta_{1\pm} := \E[\pi_C(X)\omega_{1\pm}(Q)]$, differentiating its outer expectation and then its inner components yields
\begin{align*}
 \dot\eta_{1\pm}={}&\E[\{\pi_C(X)\omega_{1\pm}(Q)-\eta_{1\pm} \}s(X)]\\
 &+\E[\{\omega_{1\pm}(Q)+\gamma_1(1-\gamma_1)\omega'_{1\pm}(Q)\}\dot p_1
   -\{\omega_{1\pm}(Q)+\gamma_1\omega'_{1\pm}(Q)\}\dot p_{0,1}]\\
 &+\E[\pi_C(X)\{L_1\dot\mu_1\pm(U_1-L_1)\dot\rho_{1\pm}(Q)\}],\\
 \dot\pi_C={}&\E[(\pi_C(X)-\pi_C)s(X)]+\E[\dot p_1-\dot p_{0,1}].
\end{align*}
By iterated expectation, the last term of $\dot\eta_{1\pm}$ becomes
\[
 \E\!\left[\frac{ZS\gamma_1}{e_1}\left\{L_1(Y-\mu_1)\pm(U_1-L_1)
 (\{\pm(Y-Q)\}_+-\rho_{1\pm}(Q))\right\}s(Y\mid X,Z,S)\right],
\]
because $\pi_C(X)/(e_1 p_1)=\gamma_1/e_1$. 
Each resulting residual has conditional mean zero given the variables preceding its score component, so it is orthogonal to the other score components. We may therefore replace each component score by $s(O)$ in its inner product. Thus, $\dot\eta_{1\pm}=\E[\varphi_{N_{1\pm}}s(O)]$ and $\dot\pi_C=\E[\varphi_Ds(O)]$, where
\begin{align*}
 \varphi_{N_{1\pm}}(O)={}&\pi_C(X)\omega_{1\pm}(Q)-\E[\pi_C(X)\omega_{1\pm}(Q)]\\
 &+\{\omega_{1\pm}(Q)+\gamma_1(1-\gamma_1)\omega'_{1\pm}(Q)\}\frac{Z}{e_1}(S-p_1)\\
 &-\{\omega_{1\pm}(Q)+\gamma_1\omega'_{1\pm}(Q)\}\frac{1-Z}{1-e_1}(S-p_{0,1})\\
 &+\frac{ZS\gamma_1}{e_1}\left[L_1(Y-\mu_1)\pm(U_1-L_1)
 \{\{\pm(Y-Q)\}_+-\rho_{1\pm}(Q)\}\right],\\
 \varphi_D(O)={}&\pi_C(X)-\pi_C+\frac{Z}{e_1}(S-p_1)-\frac{1-Z}{1-e_1}(S-p_{0,1}).
\end{align*}
Each is mean zero and square integrable. At an interior law the nonparametric tangent space is $L^2_0(P)$, identifying these as the canonical gradients. With the uncentered estimating functions $N_{1\pm} = \E[\pi_C(X)\omega_{1\pm}(Q)]+\varphi_{N_{1\pm}}$ and $D=\pi_C+\varphi_D$, the quotient rule gives
\[
 \varphi_{1\pm}=\frac{\varphi_{N_{1\pm}}-\psi_{1\pm}\varphi_D}{\pi_C}
 =\frac{N_{1\pm}-\psi_{1\pm}D}{\pi_C}.
\]
For arm 0, the same derivation applies after relabeling $(Z,S)$ as $(1-Z,1-S)$ and replacing $e_1$ by $1-e_1$. This provides the control-arm estimating function without an additional assumption. Therefore, at an interior law, $\varphi_{1\pm}$ is the efficient influence function (EIF) for the sharp endpoint \(\psi_{1\pm}\).

\subsubsection{Double validity and local principal score bias}\label{app:dr_local_bias}
We first establish an identity for fixed nuisance functions, then extend the result to their probability limits. Set $(\hat p_1,\hat p_{0,1})=(p_1,p_{0,1})$, so that the fitted sensitivity bounds equal the true bounds. The terms involving $S-\hat p_1$ and $S-\hat p_{0,1}$ then have conditional mean zero in their respective treatment groups. Taking the conditional expectation of the remaining terms gives
\[
 \E[\hat N_{1\pm}\mid X]
 =\pi_C(X)\hat\omega_{1\pm}(\hat Q)+\frac{e_1}{\hat e_1}\pi_C(X)
 \left[L_1(\mu_1-\hat\mu_1)\pm(U_1-L_1)
 \{\rho_{1\pm}(\hat Q)-\hat\rho_{1\pm}\}\right].
\]
In particular, the expression in square brackets equals $\omega_{1\pm}(\hat Q)-\hat\omega_{1\pm}(\hat Q)$. If $\hat e_1 = e_1$, adding this correction to $\hat\omega_{1\pm}(\hat Q)$ recovers $\omega_{1\pm}(\hat Q)$. If instead $\hat\mu_1=\mu_1$ and $\hat\rho_{1\pm}=\rho_{1\pm}(\hat Q)$, the correction is zero and $\hat\omega_{1\pm}(\hat Q)=\omega_{1\pm}(\hat Q)$, regardless of $\hat e_1$. Either case therefore gives
$\E[\hat N_{1\pm}\mid X]=\pi_C(X)\omega_{1\pm}(\hat Q)$.

For the denominator, taking conditional expectation gives
\[
 \E[\hat D\mid X]
 =\hat p_1-\hat p_{0,1}
 +\frac{e_1}{\hat e_1}(p_1-\hat p_1)
 -\frac{1-e_1}{1-\hat e_1}(p_{0,1}-\hat p_{0,1}).
\]
This reduces to $\pi_C(X)$ when either $\hat e_1=e_1$ or both fitted principal scores equal their true values. Thus $\E[\hat D]=\pi_C$ under either condition.

Now, let $\hat Q$ converge to $Q^*_{1\pm}$ and let the other nuisance estimators converge as in Assumption \ref{A:plim}. By Assumption \ref{A:hat_positivity_boundedmean}, $1/\hat e_1$, $1/(1-\hat e_1)$, and $1/\hat p_1$ are bounded; the bounded $\mu, \hat\mu$ and $Q, \hat Q$ also bound the factors multiplying the centered terms in $\hat N_{1\pm}$. Also observe that the hinge loss satisfies
\[
 |\{\pm(Y-\hat Q)\}_+-\{\pm(Y-Q^*_{1\pm})\}_+|
 \le|\hat Q-Q^*_{1\pm}|.
\]
The bound on $\E[Y^2 | X,Z=z,S=s]$ controls the terms containing $Y$ in $\hat N_{1\pm}$. Together with bounded inverse probabilities, bounded fitted regressions, and the hinge inequality, this convergence implies that replacing the fitted functions by their limits changes $\E[\hat N_{1\pm}]$ and $\E[\hat D]$ by $o_P(1)$. If $p_{z,s}^* = p_{z,s}$ and either $e_1^*=e_1$ or $(\mu_1^*,\rho_{1\pm}^*)=(\mu_1,\rho_{1\pm}(\cdot;Q^*_{1\pm}))$, the identities above imply
\[
 \E[\hat N_{1\pm}]\longrightarrow_P
 \E[\pi_C(X)\omega_{1\pm}(Q^*_{1\pm})],\qquad
 \E[\hat D] \xrightarrow{P} \pi_C.
\]
The outcome moment and overlap conditions in Assumption~\ref{A:positivity_boundedmean}, together with fitted overlap and boundedness in Assumption~\ref{A:hat_positivity_boundedmean}, imply that $\E[\hat N_{1\pm}^2]$ and $\E[\hat D^2]$ are bounded in probability. Conditional Chebyshev's inequality gives, for every $\varepsilon>0$,
\[
 P\left( \left|\E_n[\hat N_{1\pm}]-\E[\hat N_{1\pm}]\right|>\varepsilon \right)
 \le \frac{\E[\hat N_{1\pm}^2]}{n\varepsilon^2}=o_P(1).
\]
Thus, $\E_n[\hat N_{1\pm}] - \E[\hat N_{1\pm}] = o_P(1)$. The same calculation with $\hat D$ gives $\E_n[\hat D] - \E[\hat D] = o_P(1)$.
Since $\pi_C>0$, their ratio converges to
$\E[\pi_C(X)\omega_{1\pm}(Q^*_{1\pm})]/\pi_C$.

The conditional bound inequalities in Theorem \ref{thm:or1x} give
$\omega_{1-}(Q)\le\psi_{1-,X}$ and $\omega_{1+}(Q)\ge\psi_{1+,X}$ for every candidate quantile function $Q$. Averaging with the nonnegative weights $\pi_C(X)/\pi_C$ yields
\[
 \frac{\E[\pi_C(X)\omega_{1-}(Q^*_{1-})]}{\pi_C}\le\psi_{1-},
 \qquad
 \frac{\E[\pi_C(X)\omega_{1+}(Q^*_{1+})]}{\pi_C}\ge\psi_{1+}.
\]
These limits are consequently valid, possibly conservative bounds. The respective inequality becomes an equality when $Q^*_{1\pm}=Q_{1\pm}$, which proves sharp-bound consistency.

For the local principal score bias claim, keep the propensity and outcome regressions correct and hold one quantile function $Q$ fixed throughout the comparison. Consider perturbed principal scores $(\bar p_1,\bar p_{0,1})$. Set $\bar\pi_C(X)=\bar p_1-\bar p_{0,1}$ and $\bar\gamma_1=\bar\pi_C(X)/\bar p_1$; similarly, bars on $\omega_{1\pm}$ and $\omega'_{1\pm}$ mean that their bounds use these perturbed scores. The outcome residuals have zero conditional mean, whereas the principal score residuals have means $p_1-\bar p_1$ and $p_{0,1}-\bar p_{0,1}$. Hence
\begin{align*}
 &\E[N_{1\pm}(\bar p_1,\bar p_{0,1},e_1,\mu_1,\rho_{1\pm}(Q),Q)\mid X]\\
 &\quad=\bar\pi_C(X)\bar\omega_{1\pm}(Q)
 +\{\bar\omega_{1\pm}(Q)+\bar\gamma_1(1-\bar\gamma_1)\bar\omega'_{1\pm}(Q)\}(p_1-\bar p_1)\\
 &\qquad-\{\bar\omega_{1\pm}(Q)+\bar\gamma_1\bar\omega'_{1\pm}(Q)\}(p_{0,1}-\bar p_{0,1}).
\end{align*}
The right side is the first-order Taylor approximation to $\pi_C(X)\omega_{1\pm}(Q)$ about $(\bar p_1,\bar p_{0,1})$: its two correction coefficients are precisely the derivatives computed in Appendix \ref{AA:DR}. The bounded second derivatives therefore bound the error by
$C\{(p_1-\bar p_1)^2+(p_{0,1}-\bar p_{0,1})^2\}$.
After averaging over $X$, the numerator bias is at most
$C(\|\bar p_1-p_1\|_2^2+\|\bar p_{0,1}-p_{0,1}\|_2^2)$.
The denominator expectation is still exactly $\pi_C$, because the propensity is correct. Dividing by this positive constant proves the stated bound for the ratio.

\subsubsection{Remainder bound}
We now posit Assumptions \ref{A:positivity_boundedmean}-\ref{A:hat_positivity_boundedmean} and \ref{A:rate}, and bound the error in $\E[\hat N_{1\pm}]$. First, we compare it with the true conditional target at the fitted quantile function $\hat Q$. If the propensity and outcome regressions were replaced by $e_1$, $\mu_1$, and $\rho_{1\pm}(\cdot;\hat Q)$, the preceding Taylor argument would leave only a principal score error of order $r_p^2$.
To analyze the errors in the fitted outcome regressions, write
\[
 \rho_{1\pm}(\hat Q)
 =\E[\{\pm(Y-\hat Q(X))\}_+\mid X,Z=S=1],
 \qquad
 \delta_m=|\hat\mu_1-\mu_1|+|\hat\rho_{1\pm}-\rho_{1\pm}(\hat Q)|.
\]
Thus the hinge regression is compared with its true mean at the fitted quantile function, rather than at the true quantile. When the principal scores and propensity are correct, the preceding cancellation eliminates both regression errors exactly. Smoothness of the sensitivity bounds and fitted overlap imply that, away from those true values, the absolute contribution of these errors is bounded by
\[
 C\bigl(|\hat p_1-p_1|+|\hat p_{0,1}-p_{0,1}|+|\hat e_1-e_1|\bigr)\delta_m.
\]
Cauchy--Schwarz bounds its expectation by $C(r_p+r_e)r_m$.

Replacing the true propensity by $\hat e_1$ also changes the weights of the principal-score residuals. For example,
\[
 \frac{1-e_1}{1-\hat e_1}-1=\frac{\hat e_1-e_1}{1-\hat e_1},
 \qquad
 \E[S-\hat p_{0,1}\mid X,Z=0]=p_{0,1}-\hat p_{0,1}.
\]
Their product is bounded by a constant multiple of
$|\hat e_1-e_1||\hat p_{0,1}-p_{0,1}|$; the treatment-arm term has the same form with $p_1$. Hence, their expectations are of order $r_e r_p$. For the denominator, the conditional identity above gives the error
\[
 \E[\hat D\mid X]-\pi_C(X)
 =(\hat e_1-e_1)\left(
 \frac{\hat p_1-p_1}{\hat e_1}
 +\frac{\hat p_{0,1}-p_{0,1}}{1-\hat e_1}\right),
\]
whose expected absolute value is also bounded by $C r_e r_p$. Combining these bounds yields
\begin{align*}
 \left|\E[\hat N_{1\pm}]
 -\E[\pi_C(X)\omega_{1\pm}(\hat Q)]\right|
 &\le C\{r_p^2+(r_e+r_p)r_m+r_e r_p\},\\
 |\E[\hat D]-\pi_C|&\le C r_e r_p.
\end{align*}
We note that the argument bounds the expectation of the absolute conditional error. Thus, multiplying the conditional errors by any policy $0\le\pi(X)\le1$ preserves these bounds; this will be used in the maximin policy learning results.

It remains to compare the target at $\hat Q$ with the sharp target at $Q_{1\pm}$. The derivative of $\omega_{1\pm}(Q)$ is zero at this quantile. The bounded conditional density in Assumption \ref{A:rate} implies
$|\partial_{Q}\omega_{1\pm}(Q)-\partial_{Q}\omega_{1\pm}(Q_{1\pm})|
 \le C|Q-Q_{1\pm}|$
between the true and fitted quantiles. Integrating this inequality gives
\[
 |\omega_{1\pm}(\hat Q)-\omega_{1\pm}(Q_{1\pm})|
 \le C|\hat Q-Q_{1\pm}|^2.
\]
Multiplying by $\pi_C(X)$ and averaging adds at most $C r_Q^2$ to the numerator error. Thus,
\begin{align}
 \left|\E[\hat N_{1\pm}]
 -\E[\pi_C(X)\omega_{1\pm}(Q_{1\pm})]\right|
 +|\E[\hat D]-\pi_C|
 \le C R_n=o_P(n^{-1/2}).  \label{eq:NandDbound}
\end{align}
This establishes that nuisance estimation contributes negligible bias on the $n^{-1/2}$ scale. When $\Lambda_1=1$, $L_1=U_1=1$ and $\omega_{1\pm}(Q)=\mu_1$, so no quantile estimation error enters this calculation.

\subsubsection{Asymptotic linearity and efficiency}
Let $N_{1\pm}$ and $D$ be the estimating functions evaluated at the true nuisance functions and quantiles. To control the difference between fitted and true estimating functions, first note that overlap bounds the inverse propensity weights, while bounded regressions and quantiles control terms that do not involve $Y$. For the remaining terms, the conditional second-moment assumption ensures that any coefficient difference $c(X)$ satisfies $\E[c(X)^2Y^2]\le M\E[c(X)^2]$. Moreover, the hinge function changes by at most the change in its quantile argument. Combining these bounds with nuisance consistency in Assumption \ref{A:rate} gives
\[
 \|\hat N_{1\pm}-N_{1\pm}\|_{L^2(P)}
 +\|\hat D-D\|_{L^2(P)}=o_P(1).
\]
The conditional variance satisfies
\begin{align}  \label{eq:eff_var_N}
 \operatorname{Var}\!\left(
 \sqrt n(\E_n-\E)(\hat N_{1\pm}-N_{1\pm})
 \right)
 \le\|\hat N_{1\pm}-N_{1\pm}\|_{L^2(P)}^2=o_P(1).
\end{align}
For every $\varepsilon>0$, the conditional probability that the absolute quantity inside this variance exceeds $\varepsilon$ is at most its variance divided by $\varepsilon^2$. Since that variance is $o_P(1)$ and the conditional probability is at most one, taking its expectation shows that the unconditional probability tends to zero. The same reasoning applies to $D$. Consequently,
\[
 (\E_n-\E)(\hat N_{1\pm}-N_{1\pm})=o_P(n^{-1/2}),
 \qquad
 (\E_n-\E)(\hat D-D)=o_P(n^{-1/2}).
\]
For example, the numerator error decomposes as
\begin{align*}
 &\E_n[\hat N_{1\pm}]-\E[N_{1\pm}] \\
 &= (\E_n-\E)N_{1\pm}
   + (\E_n-\E)(\hat N_{1\pm}-N_{1\pm})
   + \E[\hat N_{1\pm}]-\E[N_{1\pm}].
\end{align*}
The second term is $o_P(n^{-1/2})$ by \eqref{eq:eff_var_N}. The third is $o_P(n^{-1/2})$ by \eqref{eq:NandDbound}. For the denominator, the identical decomposition is
\[
 \E_n[\hat D]-\pi_C
 =(\E_n-\E)D+(\E_n-\E)(\hat D-D)
   +\{\E[\hat D]-\pi_C\}.
\]
Its last two terms are also $o_P(n^{-1/2})$. The first-order terms are therefore the centered finite-sample means of the true $N_{1\pm}$ and $D$.

Since $\E[N_{1\pm}]=\pi_C\psi_{1\pm}$ and $\E[D]=\pi_C>0$, we have $\E_n[\hat D]\longrightarrow_P\pi_C$. The exact ratio identity and these two decompositions give
\begin{align}
 \hat\psi_{1\pm}-\psi_{1\pm}
 &=\frac{\E_n[\hat N_{1\pm}]-\psi_{1\pm}\E_n[\hat D]}{\E_n[\hat D]}\nonumber\\
 &=\frac{(\E_n-\E)N_{1\pm}-\psi_{1\pm}(\E_n-\E)D}{\pi_C}
   +o_P(n^{-1/2})\nonumber\\
 &=\E_n[\varphi_{1\pm}]+o_P(n^{-1/2}).\label{eq:clt-form}
\end{align}
The influence function $\varphi_{1\pm}$ has mean zero and finite variance. The i.i.d. central limit theorem therefore gives
$\sqrt n(\hat\psi_{1\pm}-\psi_{1\pm})
 \xrightarrow{d}\mathcal N(0,\E[\varphi_{1\pm}^2])$.
At the interior laws considered in Appendix~\ref{app:endpoint_eif}, $\varphi_{1\pm}$ is the canonical gradient of the sharp endpoint. Its variance is the semiparametric efficiency bound, which the estimator attains. \qed

\subsection{Proof of Theorem \ref{thm:opl_param}}
A depth-$H$ tree has at most $2^H-1$ internal nodes. The VC-dimension bound for binary decision trees with axis-aligned splits on $p$ real-valued features \citep{leboeuf2020decision} therefore shows that $\Pi_H^{\mathrm{tree}}$ has finite VC dimension when $p$ and $H$ are fixed.

Let $N^*_{z-}$ and $D^*$ be the estimating functions obtained by replacing each fitted nuisance with its deterministic probability limit. These limits may differ from the true nuisance functions. 
The conditional expectation identities proved for Theorem \ref{thm:TR} give
\[
 \E[N^*_{z-}\mid X]=\pi_C(X)\omega_{z-}(X;Q^*_{z-}),\qquad
 \E[D^*\mid X]=\pi_C(X).
\]
Here, $\omega_{z-}(X;Q)$ %$\omega_{z-}(X;Q)=L_z(X)\mu_z(X)+\{1-L_z(X)\}Q(X)-\{U_z(X)-L_z(X)\}\rho_{z-}(X;Q)$ 
is the conditional lower bound for arm $z$ evaluated at $Q$. Thus the population value formed from $N^*_{z-}$ and $D^*$ equals $V_{Q^*}$ in the theorem.

We next show that replacing the probability limits by fitted functions has vanishing effect. 
Specifically, we make the bound on the estimating functions explicit for arm 1; the formula for arm 0 follows by the relabeling described in the proof of Theorem \ref{thm:TR}. On the event in Assumption \ref{A:hat_positivity_boundedmean}, all fitted inverse probabilities are at most $1/\epsilon$. Since $\Lambda_1$ is fixed and finite, the functions $L_1,U_1$ and their derivatives with respect to $\gamma_1$ are bounded on $0\le\gamma_1\le1$. Therefore, $|\hat\omega_{1-}|+|\hat\omega'_{1-}|\le C$. In addition, $|S-\hat p_{z,s}(X)|\le1$ and
\[
 |Y-\hat\mu_1(X)|\le |Y|+C,\qquad
 |\{\hat Q_{1-}(X)-Y\}_+-\hat\rho_{1-}(X)|\le |Y|+C.
\]
Using \eqref{eq:N1}--\eqref{eq:N4} and taking absolute values gives
\begin{align*}
 |\hat N_{1-}|\le{}&|\hat\pi_C\hat\omega_{1-}|
 +\frac{Z}{\hat e_1}|\hat\omega_{1-}+\hat\gamma_1(1-\hat\gamma_1)\hat\omega'_{1-}|
   |S-\hat p_1|\\
 &+\frac{1-Z}{1-\hat e_1}|\hat\omega_{1-}+\hat\gamma_1\hat\omega'_{1-}|
   |S-\hat p_{0,1}|\\
 &+\frac{ZS\hat\gamma_1}{\hat e_1}
   \left[\hat L_1|Y-\hat\mu_1|+(\hat U_1-\hat L_1)
   |\{\hat Q_{1-}-Y\}_+-\hat\rho_{1-}|\right]
 \le C(1+|Y|).
\end{align*}
The arm 0 formula has the same structure with index $0$ and hence the same bound. Directly from the denominator formula, we have
\[
 |\hat D|\le |\hat\pi_C(X)|+
 \frac{Z}{\hat e_1(X)}|S-\hat p_1(X)|+
 \frac{1-Z}{1-\hat e_1(X)}|S-\hat p_{0,1}(X)|
 \le 1+\frac{2}{\epsilon}.
\]
The deterministic probability limits obey the same bounds, so these inequalities also hold for $N^*_{z-}$ and $D^*$. The conditional second moment assumption implies $\E[Y^2]\le M$, making $C(1+|Y|)$ square integrable.

We now show convergence of estimating functions. Write $\hat\theta$ for the vector of fitted propensity, principal scores, quantile, and outcome regressions, and $\theta^*$ for its deterministic limit. For each index $j$, $\hat\theta_{j}(X)$ is one fitted nuisance function in that vector, such as $\hat e_1(X)$ or $\hat Q_{z-}(X)$, and $\theta^*_j(X)$ is the probability limit of that same function. For bounded components $u,v$ and limits $u^*,v^*$, $|uv-u^*v^*|\le C(|u-u^*|+|v-v^*|)$. Overlap likewise gives $|1/\hat e_1-1/e_1^*|\le\epsilon^{-2}|\hat e_1-e_1^*|$, with the same bound for $1/(1-\hat e_1)$. Applying these inequalities to the formulas for $\gamma_z$, $L_z$, $U_z$, $\omega_{z-}$, and $\omega'_{z-}$ shows that each coefficient changes by at most $C\sum_j|\hat\theta_{j}-\theta^*_j|$. For the hinge residual,
\[
 |\{\hat Q_{z-}(X)-Y\}_+-\{Q^*_{z-}(X)-Y\}_+|
 \le |\hat Q_{z-}(X)-Q^*_{z-}(X)|.
\]
After subtracting $N^*_{z-}$ from $\hat N_{z-}$, terms without $Y$ are bounded by $C\sum_j|\hat\theta_{j}(X)-\theta^*_j(X)|$. Terms involving $Y$ have the form $(\hat c(X)-c^*(X))Y$, for some $\hat{c}, c^*$, times an indicator bounded by $1$. For any such coefficient, the conditional moment bound gives
\[
 \E\!\left[(\hat c(X)-c^*(X))^2Y^2\right]
 \le M\|\hat c-c^*\|_{L^2(P_X)}^2.
\]
Now, the denominator difference contains only bounded residual terms and is handled in the same way without $Y$. Assumption \ref{A:plim} makes every component error in $L^2(P_X)$ tend to zero; therefore, the bounds give
\[
 \|\hat N_{z-}-N^*_{z-}\|_{L^2(P)}=o_P(1),\qquad
 \|\hat D-D^*\|_{L^2(P)}=o_P(1),\quad z\in\{0,1\}.
\]

We now prove the uniform law used below. Let $N^*_\pi=\pi(X)N^*_{1-}+\{1-\pi(X)\}N^*_{0-}$ and $\hat N_\pi=\pi(X)\hat N_{1-}+\{1-\pi(X)\}\hat N_{0-}$. Since every tree policy takes values in $\{0,1\}$,
\begin{align*}
 &\sup_{\pi\in\Pi}\big|\E_n[\hat N_\pi]-\E[N^*_\pi]\big|\\
 &\quad\le \E_n[|\hat N_{1-}-N^*_{1-}|+|\hat N_{0-}-N^*_{0-}|]
 +\sup_{\pi\in\Pi}\big|(\E_n-\E)N^*_\pi\big|.
\end{align*}
The first term has conditional expectation $\E[|\hat N_{1-}-N^*_{1-}|+|\hat N_{0-}-N^*_{0-}|]$. Cauchy--Schwarz bounds this expectation by $\|\hat N_{1-}-N^*_{1-}\|_{L^2(P)}+\|\hat N_{0-}-N^*_{0-}\|_{L^2(P)}=o_P(1)$. For any $\varepsilon,\delta>0$, conditional Markov's inequality gives
\[
 P\left(\E_n[|\hat N_{1-}-N^*_{1-}|+|\hat N_{0-}-N^*_{0-}|]>\varepsilon\right)
 \le P\left(\|\hat N_{1-}-N^*_{1-}\|_{L^2(P)}+
            \|\hat N_{0-}-N^*_{0-}\|_{L^2(P)}>\delta\right)
     + \frac{\delta}{\varepsilon}.
\]
The probability term on the right tends to zero for every fixed $\delta$ by the $L^2(P)$ convergence above. Letting $\delta$ decrease to zero proves that the first term is $o_P(1)$.

For the second term, write $g=N^*_{1-}-N^*_{0-}$, so that $N^*_\pi=N^*_{0-}+\pi(X)g(O)$. The ordinary law of large numbers handles $N^*_{0-}$. To handle the supremum over $\pi$, fix $K>0$ and set $g_K=\operatorname{sign}(g)\min(|g|,K)$. Then
\begin{align*}
 \sup_{\pi\in\Pi}|(\E_n-\E)\{\pi(X)g(O)\}|
 &\le \sup_{\pi\in\Pi}|(\E_n-\E)\{\pi(X)g_K(O)\}|\\
 &\quad+\E_n[|g|\mathbf 1\{|g|>K\}]
      +\E[|g|\mathbf 1\{|g|>K\}].
\end{align*}
For fixed $K$, the first term tends to zero by the uniform law for a bounded function multiplying a fixed VC class. By the ordinary law of large numbers, the second term converges to its expectation. Because $|g|$ is integrable, that expectation can be made arbitrarily small by increasing $K$. Thus, $\sup_{\pi\in\Pi}|(\E_n-\E)N^*_\pi|=o_P(1)$. Together with the bound on the first term, this establishes the uniform law $\sup_{\pi\in\Pi} | \E_n[\hat{N}_\pi] - \E[N_\pi^*] | = o_P(1)$. A similar argument applied to $\hat D-D^*$ and $D^*$ gives $\E_n[\hat D] \xrightarrow{p} \E[D^*]=\pi_C>0$.

Since $\E[N_\pi^*]$ is uniformly bounded over $\Pi$, because $0\le\pi\le1$ and the $N_{z-}^*$, $z=0,1$ are integrable, dividing by the common denominator, which converges to the positive number $\pi_C$, gives
\[
 b_n:=\sup_{\pi\in\Pi}|\hat V_-(\pi)-V_{Q^*}(\pi)|=o_P(1).
\]
Finally, the definition of $\hat\pi$ gives $\hat V_-(\pi)-\hat V_-(\hat\pi)\le0$ for every $\pi\in\Pi$. Add and subtract the same terms to obtain, for all $\pi\in\Pi$,
\begin{align*}
 &V_{Q^*}(\pi)-V_{Q^*}(\hat\pi) \\
 &=\{V_{Q^*}(\pi)-\hat V_-(\pi)\}
    + \{\hat V_-(\pi)-\hat V_-(\hat\pi)\} + \{\hat V_-(\hat\pi)-V_{Q^*}(\hat\pi)\} \\
 &\le b_n+0+b_n=2b_n.
\end{align*}
Taking the supremum over $\pi$ yields $0 \le \sup_{\pi\in\Pi} V_{Q^*}(\pi) - V_{Q^*}(\hat\pi) \le 2b_n = o_P(1)$. \qed

\subsection{Proof of Lemma \ref{lem:policy_convergence}}

\subsubsection{Population separation and policy convergence}
Define $z_n:=\sup_{\pi\in\Pi}|\hat V_-(\pi)-V_-(\pi)|$. Empirical maximization gives $\hat V_-(\hat\pi)\ge\hat V_-(\pi^*)$, thus
\begin{align*}
 0\le R(\hat\pi)
 &=V_-(\pi^*)-V_-(\hat\pi)\\
 &\le\{V_-(\pi^*)-\hat V_-(\pi^*)\}
      +\{\hat V_-(\hat\pi)-V_-(\hat\pi)\}
 \le2z_n=o_P(1).
\end{align*}
For each $\eta>0$, separation in Assumption~\ref{A:opl} gives $c_\eta>0$ such that $R(\pi)\ge c_\eta$ whenever $\delta(\pi,\pi^*)\ge\eta$. Therefore,
\[
 P\{\delta(\hat\pi,\pi^*)\ge\eta\}
 \le P\{R(\hat\pi)\ge c_\eta\}
 \le P\{2z_n\ge c_\eta\} \to 0.
\]
If no policy is at distance at least $\eta$, the event on the left is empty. Thus $\delta(\hat\pi,\pi^*)=o_P(1)$.

\subsubsection{Uniform expansion of the estimated value}
Write $V(\pi)=V_-(\pi)$ and $\hat V(\pi)=\hat V_-(\pi)$. We derive an expansion valid simultaneously for all $\pi\in\Pi$. 
%The nuisance functions are fitted independently of $O_1,\ldots,O_n$. In the calculations below, expectations, variances, and probability bounds involving fitted functions condition on the training sample. In particular, $\E[\hat N_\pi]$ integrates over a new $O\sim P$ while holding the fitted functions fixed, so it remains random through the training sample. 
Set $\mathbb G_n f=\sqrt n(\E_n-\E)f$.
The ratio estimator satisfies
\begin{align}
 \hat V(\pi)-V(\pi)
 =\frac{\E_n[\hat N_\pi]}{\E_n[\hat D]}
   -\frac{\E[N_\pi]}{\pi_C} =\frac{\{\E_n[\hat N_\pi]-\E[N_\pi]\}
       -V(\pi)\{\E_n[\hat D]-\pi_C\}}{\E_n[\hat D]}.  \label{eq:err_decomp}
\end{align}
Decompose the numerator and denominator errors as
\begin{align}
 \E_n[\hat N_\pi]-\E[N_\pi]
 &=(\E_n-\E)N_\pi
   +(\E_n-\E)(\hat N_\pi-N_\pi)
   +\E[\hat N_\pi-N_\pi], \label{eq:err_decomp_N}  \\
 \E_n[\hat D]-\pi_C
 &=(\E_n-\E)D
   +(\E_n-\E)(\hat D-D)
   +\E[\hat D-D].  \label{eq:err_decomp_D}
\end{align}
We bound the final two terms in each decomposition, uniformly over $\pi$ for the numerator.

For the conditional mean errors, \eqref{eq:NandDbound} gives an armwise bound of $C R_n$ after averaging absolute errors over $X$. Since $0\le\pi(X)\le1$,
\begin{align*}
 |\E[\hat N_\pi-N_\pi]|
 \le \E\left[\pi(X)|\E[\hat N_{1-}-N_{1-}\mid X]|
       +\{1-\pi(X)\}|\E[\hat N_{0-}-N_{0-}\mid X]|\right] 
 \le C R_n.
\end{align*}
Also, $|\E[\hat D-D]|\le C R_n$. These bounds hold for every policy, and thus Assumption~\ref{A:rate} gives $\sup_{\pi\in\Pi}|\E[\hat N_\pi-N_\pi]|+|\E[\hat D-D]|=o_P(n^{-1/2})$.

For the centered empirical errors, write $h_z:=\hat N_{z-}-N_{z-}$, $z=0,1$. Since $\hat N_\pi-N_\pi=h_0+\pi(X)(h_1-h_0)$,
\begin{align}
 \sup_{\pi\in\Pi}|\mathbb G_n(\hat N_\pi-N_\pi)|
 \le |\mathbb G_nh_0|+\sup_{\pi\in\Pi}|\mathbb G_n\{\pi(X)(h_1-h_0)\}|.  \label{eq:lemma2_middle}
\end{align}
The proof of Theorem~\ref{thm:TR} gives $\|h_0\|_{L^2(P)}+\|h_1\|_{L^2(P)}=o_P(1)$. $\mathbb G_nh_0$ has mean zero and variance
\[
 \operatorname{Var}(\mathbb G_nh_0)
 =\operatorname{Var}_P\{h_0(O)\}
 \le\E_P[h_0(O)^2]=\|h_0\|_{L^2(P)}^2.
\]
For any $\varepsilon>0$, Chebyshev's inequality therefore gives $P\{|\mathbb G_nh_0|>\varepsilon\} \le \min \{1,{\|h_0\|_{L^2(P)}^2}/{\varepsilon^2}\}$.
This conditional probability bound tends to zero in probability and is bounded by one. Averaging over the training sample therefore gives $\mathbb G_nh_0=o_P(1)$ unconditionally. The same argument applied to $\hat D-D$ gives $\mathbb G_n(\hat D-D)=o_P(1)$.

To control the supremum in \eqref{eq:lemma2_middle}, use the finite VC dimension of $\Pi$ established in the proof of Theorem~\ref{thm:opl_param}. Let $h:=h_1-h_0$. Symmetrization with independent Rademacher variables $\xi_1,\ldots,\xi_n$, each taking values $1$ and $-1$ with equal probability, gives \citep[Section~8.2]{kosorok2008}
\[
 \E\left[\sup_{\pi\in\Pi}|\mathbb G_n\{\pi(X)h(O)\}|\right]
 \le 2\E\left[\sup_{\pi\in\Pi}
       \left|n^{-1/2}\sum_{i=1}^n\xi_i\pi(X_i)h(O_i)\right|\right].
\]
For a probability measure $\mathsf M$ with $\int h^2\,d\mathsf M>0$, define
\[
d\mathsf M_h(o)=\frac{h(o)^2}{\int h^2\,d\mathsf M}\,d\mathsf M(o).
\]
Then $\|(\pi-\tilde\pi)h\|_{L^2(\mathsf M)}=\|h\|_{L^2(\mathsf M)}\|\pi-\tilde\pi\|_{L^2(\mathsf M_h)}$. The finite VC dimension bounds the covering number of $\Pi$ by $(A/\varepsilon)^v$ under every probability measure, with $A,v$ depending only on the class \citep[Theorem~9.3]{kosorok2008}. Applying this bound under $\mathsf M_h$ gives a cover of $\{\pi h:\pi\in\Pi\}$ under $\mathsf M$, with radius multiplied by $\|h\|_{L^2(\mathsf M)}$. If this norm is zero, the multiplied class is zero in $L^2(\mathsf M)$.
Thus,
\begin{align}
 \mathbf{N}\left(\varepsilon\|h\|_{L^2(\mathsf M)},
     \{\pi h:\pi\in\Pi\}, L^2(\mathsf M)\right)
 \le (A/\varepsilon)^v, \qquad 0 < \varepsilon < 1,   \label{eq:covering_bound}
\end{align}
where $ \mathbf{N}(\cdot)$ is the number of $\varepsilon\|h\|_{L^2(\mathsf M)}$-balls needed to cover $\{\pi h:\pi\in\Pi\}$. This gives the finite entropy integral $J := \int_0^1\sqrt{v\log(A/\varepsilon)}\,d\varepsilon<\infty$.
Let $\E_\xi$ average only over the Rademacher variables. For two policies $\pi,\tilde\pi$,
\begin{align}
 &\E_\xi\!\left[\left\{n^{-1/2}\sum_{i=1}^n
 \xi_i\bigl(\pi(X_i)-\tilde\pi(X_i)\bigr)h(O_i)\right\}^{\!2}\right]  \nonumber \\
 &=\frac1n\sum_{i=1}^n\bigl(\pi(X_i)-\tilde\pi(X_i)\bigr)^2h(O_i)^2
 =\|(\pi-\tilde\pi)h\|_{L^2(P_n)}^2.  \label{eq:maximal_ineq_rademacher}
\end{align}
Thus the Rademacher process has increments controlled by the empirical $L^2$ distance. \eqref{eq:covering_bound} with $\mathsf M = P_n$, \eqref{eq:maximal_ineq_rademacher}, and the sub-Gaussian maximal inequality \citep[Lemma~8.7 and Corollary~8.5]{kosorok2008} give
$
 \E_\xi [\sup_{\pi\in\Pi}
 |n^{-1/2}\sum_{i=1}^n\xi_i\pi(X_i)h(O_i)|]
 \le C J\|h\|_{L^2(P_n)}.
$
Taking expectation over the analysis sample and applying Jensen's inequality gives
 $\E[\|h\|_{L^2(P_n)}]\le\{\E[\|h\|_{L^2(P_n)}^2]\}^{1/2}
 =\{\E[h(O)^2]\}^{1/2}=\|h\|_{L^2(P)}$. Hence,
\[
 \E\!\left[\sup_{\pi\in\Pi}|\mathbb G_n\{\pi(X)h(O)\}|\right]
 \le C\|h\|_{L^2(P)}
 \le C\{\|h_1\|_{L^2(P)}+\|h_0\|_{L^2(P)}\}=o_P(1),
\]
Conditional Markov's inequality, followed by averaging its bounded probability bound over training, makes the supremum $o_P(1)$. Together with \eqref{eq:lemma2_middle}, this proves $\sup_{\pi\in\Pi}|\mathbb G_n(\hat N_\pi-N_\pi)|=o_P(1)$. The middle terms in \eqref{eq:err_decomp_N}--\eqref{eq:err_decomp_D} are therefore $o_P(n^{-1/2})$.

Combining these bounds in \eqref{eq:err_decomp_N}--\eqref{eq:err_decomp_D} gives
\begin{align}
 \sup_{\pi\in\Pi}\left| \E_n[\hat N_\pi]-\E[N_\pi] - n^{-1/2} \mathbb G_nN_\pi \right| = o_P(n^{-1/2}),\\
 \E_n[\hat D] - \pi_C = n^{-1/2} \mathbb G_nD + o_P(n^{-1/2}).  \label{eq:expansionD}
\end{align}
The finite second moment of $D$ gives $\mathbb G_nD=O_P(1)$, thus \eqref{eq:expansionD} implies $\E_n[\hat D]\xrightarrow{P}\pi_C>0$. Also, $N_\pi=\{1-\pi(X)\}N_{0-}+\pi(X)N_{1-}$ and $0\le\pi(X)\le1$ imply
\[
 \sup_{\pi\in\Pi}|V(\pi)|
 =\sup_{\pi\in\Pi}\frac{|\E[N_\pi]|}{\pi_C}
 \le\frac{\E[|N_{0-}|+|N_{1-}|]}{\pi_C}<\infty.
\]
The same VC covering argument, now with the fixed square-integrable multiplier $N_{1-}-N_{0-}$, gives $\sup_{\pi\in\Pi}|\mathbb G_nN_\pi|=O_P(1)$. Hence $\sup_{\pi\in\Pi}|\mathbb G_nN_\pi-V(\pi)\mathbb G_nD|=O_P(1)$. Substitution into \eqref{eq:err_decomp} yields
\[
 \sup_{\pi\in\Pi} \left| \sqrt n\{\hat V(\pi)-V(\pi)\}
   - \frac{ \mathbb{G}_nN_\pi - V(\pi) \mathbb{G}_nD }{ \E_n[\hat{D}] } \right| = o_P(1).
\]
Moreover, $\sup_{\pi\in\Pi}|\mathbb G_nN_\pi-V(\pi)\mathbb G_nD| |{1}/{\E_n[\hat D]} - {1}/{\pi_C}| = O_P(1)o_P(1)=o_P(1).$
Therefore $\E_n[\hat D]$ can be replaced by $\pi_C$ in the leading term, giving
\[
 \sup_{\pi\in\Pi}\left|\sqrt n\{\hat V(\pi)-V(\pi)\}
   -\mathbb G_n\!\left\{\frac{N_\pi-V(\pi)D}{\pi_C}\right\}\right|=o_P(1).
\]
The function inside $\mathbb G_n$ is $\varphi_\pi$. Uniformity allows substitution of $\hat\pi$. Approximating split cutoffs from above by rational values makes the policy suprema measurable, and deterministic tie-breaking makes $\hat\pi$ measurable.

\subsubsection{Negligible regret}
The uniform expansion gives $z_n=O_P(n^{-1/2})$, and separation then gives $\delta(\hat\pi,\pi^*)=o_P(1)$. The identity $N_\pi-N_{\pi^*}=\{\pi(X)-\pi^*(X)\}(N_{1-}-N_{0-})$ and the bounded conditional second moments of the armwise estimating functions imply
\[
 \|N_\pi-N_{\pi^*}\|_{L^2(P)}^2
 \le C P_X\{\pi(X)\ne\pi^*(X)\}=C\delta(\pi,\pi^*)^2,
\]
and thus $|V(\pi)-V(\pi^*)|\le C\delta(\pi,\pi^*)^2$ as well. Using $\varphi_\pi = \{N_\pi-V(\pi)D\}/\pi_C$, we obtain
\[
 \|\varphi_\pi-\varphi_{\pi^*}\|_{L^2(P)}
 \le \left\{\|N_\pi-N_{\pi^*}\|_{L^2(P)}
       +|V(\pi)-V(\pi^*)|\|D\|_{L^2(P)}\right\} / \pi_C.
\]
Thus, $\|\varphi_{\hat\pi}-\varphi_{\pi^*}\|_{L^2(P)}=o_P(1)$. The influence function class has a square-integrable envelope, a finite uniform entropy integral, and the countable approximation established above. It is therefore $P$-Donsker \citep[Theorem~8.19]{kosorok2008}, with a Gaussian limit continuous in $L^2(P)$. Asymptotic equicontinuity \citep[Theorem~7.19]{kosorok2008} then gives $\mathbb G_n(\varphi_{\hat\pi}-\varphi_{\pi^*})=o_P(1)$.

Finally, empirical maximization and the uniform expansion give
\begin{align*}
 0\le R(\hat\pi)
 &\le\{\hat V(\hat\pi)-V(\hat\pi)\}
       -\{\hat V(\pi^*)-V(\pi^*)\}\\
 &=n^{-1/2}\mathbb G_n(\varphi_{\hat\pi}-\varphi_{\pi^*})
       +o_P(n^{-1/2})
 =o_P(n^{-1/2}).  \qed
\end{align*}

\subsection{Proof of Theorem \ref{thm:opl}}
The uniform expansion in Lemma \ref{lem:policy_convergence} holds simultaneously for every $\pi\in\Pi$, so it remains valid at $\pi=\hat\pi$. We begin with the exact decomposition
\[
 \hat V_-(\hat\pi)-V_-(\pi^*)
 =\{\hat V_-(\hat\pi)-V_-(\hat\pi)\}
  +\{V_-(\hat\pi)-V_-(\pi^*)\}.
\]
For the first term, the lemma gives $\sqrt n\{\hat V_-(\hat\pi)-V_-(\hat\pi)\}=\mathbb G_n\varphi_{\hat\pi}+o_P(1)$ and $\delta(\hat\pi,\pi^*)=o_P(1)$, and therefore $\mathbb G_n(\varphi_{\hat\pi}-\varphi_{\pi^*}) = o_P(1)$ and $\sqrt n\{V_-(\hat\pi)-V_-(\pi^*)\} = o_P(1)$. Hence,
\begin{align*}
 \sqrt n\{\hat V_-(\hat\pi)-V_-(\pi^*)\}
 &=\mathbb G_n\varphi_{\hat\pi}
   +\sqrt n\{V_-(\hat\pi)-V_-(\pi^*)\}+o_P(1)\\
 &=\mathbb G_n\varphi_{\pi^*}
   +\mathbb G_n(\varphi_{\hat\pi}-\varphi_{\pi^*})+o_P(1)\\
 &=\mathbb G_n\varphi_{\pi^*}+o_P(1).
\end{align*}
Since the oracle policy $\pi^*$ is fixed, and $\varphi_{\pi^*}(O_i)$ are i.i.d. with mean zero and finite variance $\Upsilon(\pi^*)=\E[\varphi_{\pi^*}^2]$, the central limit theorem and Slutsky's theorem give the desired normal limit.   \qed

\subsubsection{Efficiency of the optimal value}
Theorems~\ref{thm:opl} and \ref{thm:rkhs_opl} establish asymptotic normality for fixed-depth trees and fixed linear-feature policies, respectively. Establishing semiparametric efficiency for the optimal value requires showing that the supremum over policies is pathwise differentiable. The following result states this requirement for either class; related results are given by \citet{luedtke2016statistical}.

\begin{prop}[Efficiency of the optimal value]\label{prop:optimal_value_efficiency}
Let $\Pi$ be a fixed policy class satisfying Assumption~\ref{A:opl}, and define $\mathcal V_-^*(P)=\sup_{\pi\in\Pi}V_-(\pi;P)$. Suppose the fixed-policy canonical gradients satisfy $\sup_{\pi\in\Pi}\|\varphi_\pi\|_{L^2(P)}<\infty$ and are continuous in $L^2(P)$ at $\pi^*$ under $\delta$. At an interior observed-data law with tangent space $L^2_0(P)$, assume that every regular submodel $P_t$ through $P$, with score $s(O)$, satisfies
\[
\sup_{\pi\in\Pi}
\left|V_-(\pi;P_t)-V_-(\pi;P)
-t\E_P[\varphi_\pi(O)s(O)]\right|=o(|t|).
\]
Then $\mathcal V_-^*$ has canonical gradient $\varphi_{\pi^*}$ and efficiency bound $\Upsilon(\pi^*)$. An estimator with asymptotically linear expansion
$\sqrt n\{\hat V_-(\hat\pi)-\mathcal V_-^*(P)\}
 = \mathbb G_n\varphi_{\pi^*}+o_P(1)$
is regular and efficient at $P$.
\end{prop}

\begin{proof}
Fix a regular submodel. The uniform pathwise expansion and gradient bound give $\sup_{\pi\in\Pi}|V_-(\pi;P_t)-V_-(\pi;P)|=O(|t|)$. Choose a policy $\pi_t$ within $t^2$ of the supremum under $P_t$. Its regret under $P$ is at most twice this uniform value difference plus $t^2$. Separation therefore implies $\delta(\pi_t,\pi^*)\to0$, and gradient continuity gives $\E_P[(\varphi_{\pi_t}-\varphi_{\pi^*})s]\to0$. Since
\begin{align*}
V_-(\pi^*;P_t)-V_-(\pi^*;P)
&\le\mathcal V_-^*(P_t)-\mathcal V_-^*(P)\\
&\le V_-(\pi_t;P_t)-V_-(\pi_t;P)+t^2,
\end{align*}
the uniform expansion yields
\[
\mathcal V_-^*(P_t)-\mathcal V_-^*(P)
=t\E_P[\varphi_{\pi^*}s]+o(|t|).
\]
The full tangent space identifies $\varphi_{\pi^*}$ as the canonical gradient. For a fixed $c\in\mathbb R$, let $P_{c/\sqrt n}$ denote the distribution on this submodel at $t=c/\sqrt n$, and let $P_{c/\sqrt n}^{\otimes n}$ denote the joint law of $n$ independent observations from that distribution. These local alternatives approach $P$ at rate $n^{-1/2}$. Under these joint laws, contiguity transfers the negligible remainder in the estimator expansion, and Le Cam's third lemma shifts its leading term by $c\E_P[\varphi_{\pi^*}s]$. The target has the same shift by the displayed derivative, establishing regularity and attainment of the efficiency bound \citep{kosorok2008}.
\end{proof}

\section{Alternative policy classes}  \label{AA:alternative_policy}
We consider policies defined by the sign of a normalized linear combination of fixed features in a reproducing kernel Hilbert space (RKHS). This extends ordinary linear policies by allowing nonlinear features of $X$. The learning criterion is the estimated lower-bound value, as in Theorem \ref{thm:opl}. The target throughout this section is the best policy within the specified feature class.

\subsection{Normalized policies with RKHS features}  \label{AA:rkhs_learning}
Let $\mathcal H_{\mathsf K}$ be an RKHS with measurable kernel $\mathsf K$. Choose fixed measurable features $\phi_1,\ldots,\phi_m\in\mathcal H_{\mathsf K}$, with $m$ independent of sample size, and write $\Phi(x)=(\phi_1(x),\ldots,\phi_m(x))^\top$. Define
\[
 \pi_\beta(x)=\mathbf1\{\beta^\top\Phi(x)>0\},\qquad
 \Pi^\Phi=\{\pi_\beta:\beta\in\mathbb R^m,\ \|\beta\|_2=1\}.
\]
Positive rescaling of a nonzero coefficient vector leaves its treatment decisions unchanged, so normalization removes this scale ambiguity. Different normalized coefficient vectors can still represent the same policy. The decision function $\beta^\top\Phi(x)$ need not equal the conditional contrast $B(x)=\psi_{1-,X=x}-\psi_{0-,X=x}$. Each decision function belongs to the fixed span of the RKHS features; the results below concern this finite-dimensional span.

Choose a measurable coefficient vector $\hat\beta$ attaining the maximum of $\hat V_-(\pi_\beta)$ over $\|\beta\|_2=1$, and set $\hat\pi=\pi_{\hat\beta}$. On the event $\E_n[\hat D]>0$, the optimization is equivalently
\[
 \hat\beta\in\argmax_{\|\beta\|_2=1}
 \E_n[\pi_\beta(X)\{\hat N_{1-}(O)-\hat N_{0-}(O)\}].
\]
For a fixed sample, only finitely many vectors $(\pi_\beta(X_1),\ldots,\pi_\beta(X_n))$ occur, so the empirical maximum is attained. We assume a measurable selection among maximizers and measurable empirical-process suprema over $\Pi^\Phi$. On the event $\E_n[\hat D]\le0$, define $\hat\pi$ to be any fixed member of $\Pi^\Phi$; this event has probability tending to zero under the inference conditions below and is thus negligible.

Throughout this section, we use Assumption \ref{A:opl} with $\Pi=\Pi^\Phi$, so $\pi^*$ maximizes the population value within this feature class.
Sufficient conditions for Assumption \ref{A:opl} are that $P_X\{\beta^\top\Phi(X)=0\}=0$ for every unit vector $\beta$ and that every maximizing policy makes the same decisions as $\pi^*$ almost surely. If $\beta_j\to\beta$, then $\pi_{\beta_j}(X)\to\pi_\beta(X)$ outside the zero-probability boundary $\beta^\top\Phi(X)=0$, so dominated convergence gives $\delta(\pi_{\beta_j},\pi_\beta)\to0$. The unit sphere is compact, hence its image $\Pi^\Phi$, identifying policies that agree almost surely, is compact under $\delta$. Moreover, the conditional moment bounds give
\[
 |V_-(\pi)-V_-(\tilde\pi)|
 =\frac{|\E[(\pi(X)-\tilde\pi(X))(N_{1-}-N_{0-})]|}{\pi_C}
 \le C\delta(\pi,\tilde\pi)^2.
\]
Thus, the value is continuous. Uniqueness of the maximizing treatment decisions then makes the maximum over any closed set at distance at least $\eta$ strictly smaller than $V_-(\pi^*)$. This is the separation argument based on compactness, continuity and uniqueness in \citet[proof of Theorem~2.1]{newey1994large}, applied to policies identified up to $P_X$-null sets.

Lemma \ref{lem:policy_convergence} also holds with $\Pi=\Pi^\Phi$, under its same separation and estimation conditions. This class has VC dimension at most $m$ \citep[Theorem 9.2]{shalev2014understanding}. In the proof of the lemma, finite VC dimension supplies the covering bounds used to control empirical errors simultaneously over all policies. Combined with the moment and nuisance rate conditions, these bounds give the uniform value expansion and the empirical equicontinuity needed for negligible regret. These arguments therefore apply to $\Pi^\Phi$ as well.

\begin{theorem}[Inference for normalized policies with RKHS features]\label{thm:rkhs_opl}
Under the conditions in Lemma \ref{lem:policy_convergence}, with $\Pi = \Pi^\Phi$,
\[
 \sqrt n\{\hat V_-(\hat\pi)-V_-(\pi^*)\}
 =\mathbb G_n\varphi_{\pi^*} + o_P(1)
  \overset d\to \mathcal N(0,\Upsilon(\pi^*)).
\]
\end{theorem}

\begin{proof}
The lemma gives negligible regret, and its proof gives $\mathbb G_n(\varphi_{\hat\pi}-\varphi_{\pi^*})=o_P(1)$. Hence,
\begin{align*}
 \sqrt n\{\hat V_-(\hat\pi)-V_-(\pi^*)\}
 &=\mathbb G_n\varphi_{\hat\pi}
   +\sqrt n\{V_-(\hat\pi)-V_-(\pi^*)\}+o_P(1)\\
 &=\mathbb G_n\varphi_{\pi^*}+o_P(1).
\end{align*}
The oracle influence function is centered and square-integrable, and thus the central limit theorem and Slutsky's theorem give the desired result.
\end{proof}

\begin{example}[Ordinary linear policies]\label{ex:rkhs_linear}
Take $\Phi(x)=(1,x^\top)^\top$, with the dimension of $X$ fixed. The kernel $\mathsf K(x,z)=\Phi(x)^\top\Phi(z)=1+x^\top z$ has an RKHS containing the intercept and coordinate features. Thus,
\[
 \Pi^\Phi = \{x \mapsto \mathbb{1}\{\beta_0+\beta_X^\top x>0\}: \|\beta\|_2=1\}
\]
is a special case of the class above, and Lemma \ref{lem:policy_convergence} (with $\Pi=\Pi^\Phi$) and Theorem \ref{thm:rkhs_opl} apply. For example, if $X$ has a density with respect to Lebesgue measure, every nonconstant affine decision boundary has probability zero; a unit vector with zero slope has a nonzero intercept and no decision boundary. Together with uniqueness of the maximizing treatment decisions, this verifies the sufficient separation conditions above.
\end{example}

\newpage

\section{Supplementary simulation design and evaluation}\label{app:sim_impl}
Section~\ref{sec:sim} describes the simulation design and analysis. Here, we specify the outcome distributions underlying the data-generating mechanism. Let $f_z(y\mid x)$ denote the normal density with mean $\mu_z(x)$ and standard deviation $\sigma=1.5$ in the observed group $(Z,S)=(z,z)$. The conditional complier fractions in these groups are $\gamma_0=\pi_C/(1-p_{0,1})$ and $\gamma_1=\pi_C/p_1$. At $\Lambda_0=\Lambda_1=5$, we specify the density ratios as
\begin{gather*}
	 W_0(x,y)=L_0(x)+\{U_0(x)-L_0(x)\}\mathbb{1}\{y>Q_{0+}(x)\},\\
	 W_1(x,y)=L_1(x)+\{U_1(x)-L_1(x)\}\mathbb{1}\{y<Q_{1-}(x)\}.
\end{gather*}
The resulting conditional complier density is $W_zf_z$, while the companion principal-stratum density is $(1-\gamma_z W_z)f_z/(1-\gamma_z)$. Both densities integrate to one, and their mixture with weights $\gamma_z$ and $1-\gamma_z$ recovers $f_z$. Outcomes outside $S(z)=z$ follow the corresponding baseline normal law. The true quantiles are given by $Q_{z\pm}(x)=\mu_z(x)+\sigma\Phi^{-1}\{\tau_{z\pm}(x)\}$, where $\tau_{z-}=(1-L_z)/(U_z-L_z)$ and $\tau_{z+}=(U_z-1)/(U_z-L_z)$.

The propensity and principal score models are fitted by logistic regression. The former regresses $Z$ on an intercept and $X$, while the latter regresses $S$ on an intercept, $X$, and $Z$. Under misspecification, the propensity model includes only an intercept, whereas the principal score model includes an intercept and $Z$ but omits $X$. Within each observed $(Z,S)=(z,z)$ group, the outcome mean is estimated by linear regression on an intercept and $X$, or on an intercept alone under misspecification. Fitted probabilities are projected to $0.01\le\hat e_1\le0.99$, $0.01\le\hat p_{0,1}\le0.98$, and $\hat p_{0,1}+0.01\le\hat p_1\le0.99$.

Within each arm, we fit quantile regressions at 50 levels between 0.001 and 0.999 and interpolate to obtain estimates at $\hat\tau_{z\pm}(X)$. The correctly specified quantile family includes an intercept and $X$, whereas the misspecified family includes only an intercept. Even under the latter, the interpolated $\hat Q_{z\pm}(X)$ may vary with $X$ because the estimated quantile levels depend on $X$. Hinge regressions use linear working models for the corresponding hinge responses and may remain misspecified even in the baseline setting.

\subsection{Sensitivity bound estimation}
We use an independent data set of one million observations to compute reference endpoint values. 

In each repetition, we construct percentile-bootstrap confidence intervals for the endpoints using 500 bootstrap samples. The propensity, principal-score, outcome, and hinge models are refitted in each bootstrap sample, while the fitted quantile predictions are held fixed. Under correctly specified quantile models, endpoint coverage is the proportion of repetitions in which the two-sided 95\% confidence interval contains the corresponding sharp endpoint. Under misspecified quantile models, we instead assess the validity of the one-sided confidence bounds through
\[
\frac{1}{500}\sum_{m=1}^{500}\mathbb{1}\{L_{z-}^{(m)}\le\psi_{z-}\},\qquad
\frac{1}{500}\sum_{m=1}^{500}\mathbb{1}\{U_{z+}^{(m)}\ge\psi_{z+}\},
\]
where $L_{z-}^{(m)}$ and $U_{z+}^{(m)}$ denote the one-sided 95\% lower and upper percentile-bootstrap confidence bounds, corresponding to the fifth and 95th percentiles, respectively. These rates assess the two endpoints separately and do not represent simultaneous coverage of the identification interval.

\begin{table}[tbp]
\centering
\caption{Sensitivity bound estimation over 500 repetitions: $n=2000$, $\hat Q_{z\pm}$ correctly specified. At $\Lambda=5$, the reference sharp endpoints are [2.869, 4.506] for arm 0 and [2.746, 5.089] for arm 1. Lower and Upper report mean endpoint estimates. At $\Lambda=5$, bold entries highlight selected results with correctly specified principal score models and assignment propensity (IPW), outcome mean models (OR), or either (DR). Hinge regressions remain working approximations. Misspec.\ identifies the misspecified model among $\hat e_1$, $\hat p$, and $\hat\mu$; None indicates that all three are correctly specified.%
  % Draft return reference; comment out this line for camera-ready.
}
\label{tab:sim_bounds_2000_0_0}
\setlength{\tabcolsep}{2.5pt}
\scriptsize
\textbf{Arm 0}

\;

\resizebox{\textwidth}{!}{%
\begin{tabular}{llrrrrrrrrr}
\toprule
\multirow{2}{*}{Misspec.} & \multirow{2}{*}{$\Lambda$} & \multicolumn{3}{c}{IPW} & \multicolumn{3}{c}{OR} & \multicolumn{3}{c}{DR} \\
\cmidrule(lr){3-5}\cmidrule(lr){6-8}\cmidrule(lr){9-11}
 &  & Lower & Upper & CI cov.\ (\%) & Lower & Upper & CI cov.\ (\%) & Lower & Upper & CI cov.\ (\%) \\
\midrule
\multirow{5}{*}{None} & 1.001 & 3.698 (0.222) & 3.700 (0.222) & 92.4 / 92.4 & 3.694 (0.076) & 3.696 (0.076) & 96.8 / 96.8 & 3.691 (0.187) & 3.692 (0.187) & 94.8 / 94.8 \\
 & 2 & 3.291 (0.216) & 4.107 (0.235) & 93.6 / 93.0 & 3.286 (0.086) & 4.104 (0.084) & 96.2 / 97.4 & 3.282 (0.192) & 4.100 (0.194) & 94.6 / 92.8 \\
 & 3 & 3.087 (0.216) & 4.311 (0.244) & 94.2 / 93.8 & 3.081 (0.096) & 4.308 (0.093) & 95.6 / 96.0 & 3.078 (0.198) & 4.305 (0.201) & 95.8 / 92.6 \\
 & 5 & 2.883 (0.217) & 4.515 (0.253) & 94.6 / 94.4 & \boldwidth{2.876} (0.109) & \boldwidth{4.514} (0.105) & \boldwidth{95.2 / 94.8} & \boldwidth{2.873} (0.206) & \boldwidth{4.510} (0.211) & \boldwidth{95.4 / 93.2} \\
 & 8 & 2.746 (0.219) & 4.651 (0.261) & 94.2 / 94.6 & 2.740 (0.118) & 4.651 (0.114) & 94.8 / 95.0 & 2.736 (0.212) & 4.647 (0.218) & 94.8 / 93.2 \\
\addlinespace
\multirow{5}{*}{$\hat e_1$} & 1.001 & 3.467 (0.074) & 3.469 (0.074) & 22.2 / 22.2 & 3.694 (0.076) & 3.696 (0.076) & 96.8 / 96.8 & 3.693 (0.096) & 3.694 (0.096) & 96.0 / 96.0 \\
 & 2 & 3.031 (0.077) & 3.906 (0.092) & 13.6 / 50.0 & 3.286 (0.086) & 4.104 (0.084) & 96.2 / 97.4 & 3.285 (0.106) & 4.103 (0.100) & 94.6 / 97.0 \\
 & 3 & 2.812 (0.086) & 4.125 (0.106) & 15.2 / 66.0 & 3.081 (0.096) & 4.308 (0.093) & 95.6 / 96.0 & 3.080 (0.115) & 4.307 (0.107) & 94.6 / 96.6 \\
 & 5 & 2.592 (0.099) & 4.344 (0.123) & 19.8 / 75.8 & \boldwidth{2.876} (0.109) & \boldwidth{4.514} (0.105) & \boldwidth{95.2 / 94.8} & \boldwidth{2.875} (0.127) & \boldwidth{4.512} (0.117) & \boldwidth{94.4 / 96.8} \\
 & 8 & 2.446 (0.109) & 4.491 (0.135) & 23.4 / 81.4 & 2.740 (0.118) & 4.651 (0.114) & 94.8 / 95.0 & 2.738 (0.135) & 4.649 (0.125) & 94.4 / 96.4 \\
\addlinespace
\multirow{5}{*}{$\hat\mu$} & 1.001 & 3.698 (0.222) & 3.700 (0.222) & 92.4 / 92.4 & 3.289 (0.069) & 3.290 (0.069) & 0.0 / 0.0 & 3.695 (0.115) & 3.697 (0.115) & 95.0 / 95.0 \\
 & 2 & 3.291 (0.216) & 4.107 (0.235) & 93.6 / 93.0 & 3.009 (0.077) & 3.827 (0.077) & 7.4 / 9.4 & 3.286 (0.129) & 4.104 (0.127) & 93.8 / 93.8 \\
 & 3 & 3.087 (0.216) & 4.311 (0.244) & 94.2 / 93.8 & 2.871 (0.087) & 4.098 (0.090) & 37.4 / 44.8 & 3.081 (0.145) & 4.308 (0.145) & 94.6 / 94.2 \\
 & 5 & 2.883 (0.217) & 4.515 (0.253) & 94.6 / 94.4 & 2.735 (0.101) & 4.372 (0.104) & 77.0 / 76.0 & \boldwidth{2.875} (0.168) & \boldwidth{4.512} (0.170) & \boldwidth{95.0 / 94.2} \\
 & 8 & 2.746 (0.219) & 4.651 (0.261) & 94.2 / 94.6 & 2.644 (0.112) & 4.555 (0.114) & 90.4 / 88.8 & 2.738 (0.186) & 4.648 (0.190) & 95.0 / 93.8 \\
\addlinespace
\multirow{5}{*}{$\hat p$} & 1.001 & 3.406 (0.204) & 3.407 (0.204) & 56.8 / 56.6 & 3.755 (0.079) & 3.756 (0.079) & 90.8 / 91.0 & 3.686 (0.204) & 3.688 (0.204) & 94.2 / 94.2 \\
 & 2 & 3.088 (0.197) & 3.725 (0.215) & 70.0 / 46.4 & 3.413 (0.083) & 4.099 (0.085) & 71.4 / 96.8 & 3.263 (0.203) & 4.111 (0.222) & 95.4 / 93.0 \\
 & 3 & 2.928 (0.194) & 3.885 (0.222) & 77.6 / 43.4 & 3.241 (0.088) & 4.270 (0.091) & 61.2 / 94.4 & 3.052 (0.210) & 4.322 (0.236) & 95.4 / 93.8 \\
 & 5 & 2.768 (0.193) & 4.045 (0.230) & 84.8 / 40.0 & 3.069 (0.095) & 4.442 (0.099) & 54.6 / 91.4 & 2.841 (0.220) & 4.533 (0.253) & 95.2 / 94.6 \\
 & 8 & 2.661 (0.193) & 4.151 (0.235) & 88.8 / 38.0 & 2.954 (0.100) & 4.557 (0.104) & 50.0 / 89.0 & 2.700 (0.228) & 4.674 (0.266) & 95.2 / 94.4 \\
\addlinespace
\bottomrule
\end{tabular}}

\;

\;

\;

\;

\textbf{Arm 1}

\;

\resizebox{\textwidth}{!}{%
\begin{tabular}{llrrrrrrrrr}
\toprule
\multirow{2}{*}{Misspec.} & \multirow{2}{*}{$\Lambda$} & \multicolumn{3}{c}{IPW} & \multicolumn{3}{c}{OR} & \multicolumn{3}{c}{DR} \\
\cmidrule(lr){3-5}\cmidrule(lr){6-8}\cmidrule(lr){9-11}
 &  & Lower & Upper & CI cov.\ (\%) & Lower & Upper & CI cov.\ (\%) & Lower & Upper & CI cov.\ (\%) \\
\midrule
\multirow{5}{*}{None} & 1.001 & 3.918 (0.106) & 3.920 (0.106) & 94.0 / 94.2 & 3.921 (0.074) & 3.923 (0.074) & 94.4 / 94.4 & 3.913 (0.228) & 3.914 (0.228) & 93.8 / 93.8 \\
 & 2 & 3.344 (0.107) & 4.494 (0.119) & 94.8 / 94.2 & 3.347 (0.085) & 4.498 (0.075) & 93.0 / 95.2 & 3.337 (0.226) & 4.490 (0.236) & 93.6 / 93.4 \\
 & 3 & 3.051 (0.112) & 4.788 (0.129) & 95.2 / 93.4 & 3.054 (0.094) & 4.791 (0.081) & 92.2 / 94.4 & 3.043 (0.228) & 4.784 (0.243) & 93.4 / 93.0 \\
 & 5 & 2.751 (0.122) & 5.088 (0.143) & 94.8 / 92.8 & \boldwidth{2.754} (0.107) & \boldwidth{5.090} (0.090) & \boldwidth{91.6 / 94.6} & \boldwidth{2.742} (0.232) & \boldwidth{5.085} (0.251) & \boldwidth{93.4 / 93.0} \\
 & 8 & 2.545 (0.132) & 5.294 (0.154) & 94.4 / 92.4 & 2.549 (0.118) & 5.295 (0.098) & 92.2 / 94.8 & 2.536 (0.238) & 5.292 (0.258) & 93.8 / 93.0 \\
\addlinespace
\multirow{5}{*}{$\hat e_1$} & 1.001 & 4.266 (0.091) & 4.268 (0.091) & 2.8 / 2.8 & 3.921 (0.074) & 3.923 (0.074) & 94.4 / 94.4 & 3.921 (0.106) & 3.922 (0.106) & 95.8 / 95.8 \\
 & 2 & 3.700 (0.101) & 4.835 (0.087) & 5.0 / 2.2 & 3.347 (0.085) & 4.498 (0.075) & 93.0 / 95.2 & 3.346 (0.112) & 4.497 (0.108) & 94.6 / 96.8 \\
 & 3 & 3.410 (0.110) & 5.125 (0.087) & 8.0 / 2.6 & 3.054 (0.094) & 4.791 (0.081) & 92.2 / 94.4 & 3.053 (0.119) & 4.790 (0.112) & 94.4 / 97.4 \\
 & 5 & 3.111 (0.121) & 5.423 (0.089) & 11.2 / 2.8 & \boldwidth{2.754} (0.107) & \boldwidth{5.090} (0.090) & \boldwidth{91.6 / 94.6} & \boldwidth{2.754} (0.129) & \boldwidth{5.089} (0.119) & \boldwidth{93.0 / 97.4} \\
 & 8 & 2.906 (0.131) & 5.628 (0.093) & 18.0 / 4.0 & 2.549 (0.118) & 5.295 (0.098) & 92.2 / 94.8 & 2.549 (0.137) & 5.294 (0.125) & 93.0 / 97.4 \\
\addlinespace
\multirow{5}{*}{$\hat\mu$} & 1.001 & 3.918 (0.106) & 3.920 (0.106) & 94.0 / 94.2 & 4.565 (0.061) & 4.567 (0.061) & 0.0 / 0.0 & 3.921 (0.123) & 3.923 (0.123) & 92.0 / 92.0 \\
 & 2 & 3.344 (0.107) & 4.494 (0.119) & 94.8 / 94.2 & 3.734 (0.073) & 4.885 (0.064) & 0.0 / 0.0 & 3.342 (0.143) & 4.495 (0.145) & 93.2 / 94.2 \\
 & 3 & 3.051 (0.112) & 4.788 (0.129) & 95.2 / 93.4 & 3.331 (0.086) & 5.068 (0.072) & 8.8 / 3.0 & 3.046 (0.164) & 4.787 (0.170) & 93.4 / 94.6 \\
 & 5 & 2.751 (0.122) & 5.088 (0.143) & 94.8 / 92.8 & 2.930 (0.103) & 5.266 (0.083) & 52.0 / 40.0 & \boldwidth{2.744} (0.190) & \boldwidth{5.087} (0.201) & \boldwidth{93.8 / 94.2} \\
 & 8 & 2.545 (0.132) & 5.294 (0.154) & 94.4 / 92.4 & 2.663 (0.115) & 5.409 (0.094) & 77.2 / 76.8 & 2.537 (0.210) & 5.293 (0.224) & 94.8 / 94.0 \\
\addlinespace
\multirow{5}{*}{$\hat p$} & 1.001 & 3.938 (0.088) & 3.940 (0.088) & 94.6 / 94.6 & 4.002 (0.067) & 4.004 (0.067) & 74.0 / 74.2 & 3.908 (0.249) & 3.909 (0.249) & 93.0 / 93.0 \\
 & 2 & 3.425 (0.086) & 4.453 (0.100) & 82.0 / 89.2 & 3.473 (0.073) & 4.533 (0.071) & 55.6 / 91.4 & 3.314 (0.260) & 4.503 (0.246) & 93.0 / 93.4 \\
 & 3 & 3.166 (0.090) & 4.713 (0.109) & 68.8 / 83.2 & 3.206 (0.080) & 4.801 (0.077) & 49.2 / 94.0 & 3.011 (0.270) & 4.806 (0.248) & 93.0 / 93.4 \\
 & 5 & 2.903 (0.096) & 4.976 (0.120) & 57.4 / 76.6 & 2.935 (0.088) & 5.071 (0.085) & 43.4 / 95.4 & 2.702 (0.283) & 5.116 (0.254) & 93.6 / 94.0 \\
 & 8 & 2.726 (0.102) & 5.153 (0.128) & 51.6 / 72.4 & 2.752 (0.096) & 5.255 (0.091) & 39.8 / 92.8 & 2.490 (0.293) & 5.328 (0.259) & 93.2 / 94.6 \\
\addlinespace
\bottomrule
\end{tabular}}
\par\smallskip\begin{minipage}{\textwidth}\scriptsize Entries are Monte Carlo means (empirical SEs) over 500 repetitions. CI cov.\ reports separate marginal coverage rates (percent) for the lower and upper sharp endpoints using two-sided 95\% percentile bootstrap intervals with 500 bootstrap draws per repetition. \end{minipage}
\end{table}

\clearpage

\begin{table}[tbp]
\centering
\caption{Sensitivity bound estimation over 500 repetitions: $n=2000$, $\hat Q_{z\pm}$ misspecified. At $\Lambda=5$, the reference sharp endpoints are [2.869, 4.506] for arm 0 and [2.746, 5.089] for arm 1. See the caption of Table~\ref{tab:sim_bounds_2000_0_0} for a description of the shared elements.%
  % Draft return reference; comment out this line for camera-ready.
}
\label{tab:sim_bounds_2000_0_1}
\setlength{\tabcolsep}{2.5pt}
\scriptsize
\textbf{Arm 0}

\;

\resizebox{\textwidth}{!}{%
\begin{tabular}{llrrrrrrrrr}
\toprule
\multirow{2}{*}{Misspec.} & \multirow{2}{*}{$\Lambda$} & \multicolumn{3}{c}{IPW} & \multicolumn{3}{c}{OR} & \multicolumn{3}{c}{DR} \\
\cmidrule(lr){3-5}\cmidrule(lr){6-8}\cmidrule(lr){9-11}
 &  & Lower & Upper & Valid.\ (\%) & Lower & Upper & Valid.\ (\%) & Lower & Upper & Valid.\ (\%) \\
\midrule
\multirow{5}{*}{None} & 1.001 & 3.698 (0.222) & 3.700 (0.222) & 99.2 / 90.6 & 3.694 (0.076) & 3.696 (0.076) & 96.2 / 97.0 & 3.690 (0.187) & 3.692 (0.187) & 97.0 / 92.4 \\
 & 2 & 3.231 (0.199) & 4.167 (0.254) & 99.4 / 94.0 & 3.249 (0.085) & 4.136 (0.088) & 97.6 / 98.8 & 3.223 (0.171) & 4.160 (0.198) & 98.8 / 96.0 \\
 & 3 & 2.997 (0.192) & 4.401 (0.273) & 99.6 / 95.6 & 3.024 (0.097) & 4.355 (0.100) & 98.0 / 98.8 & 2.988 (0.162) & 4.395 (0.201) & 99.2 / 97.2 \\
 & 5 & 2.762 (0.188) & 4.635 (0.292) & \boldwidth{99.8 / 96.6} & 2.797 (0.111) & 4.573 (0.114) & \boldwidth{98.6 / 98.8} & 2.754 (0.155) & 4.630 (0.204) & \boldwidth{99.4 / 98.2} \\
 & 8 & 2.605 (0.187) & 4.791 (0.306) & 99.8 / 97.0 & 2.645 (0.122) & 4.717 (0.124) & 98.4 / 99.0 & 2.597 (0.154) & 4.787 (0.206) & 99.2 / 98.6 \\
\addlinespace
\multirow{5}{*}{$\hat e_1$} & 1.001 & 3.467 (0.074) & 3.469 (0.074) & 100.0 / 12.2 & 3.694 (0.076) & 3.696 (0.076) & 96.2 / 97.0 & 3.693 (0.096) & 3.694 (0.096) & 94.8 / 96.0 \\
 & 2 & 2.993 (0.078) & 3.943 (0.094) & 100.0 / 53.0 & 3.249 (0.085) & 4.136 (0.088) & 97.6 / 98.8 & 3.248 (0.103) & 4.134 (0.100) & 96.2 / 99.0 \\
 & 3 & 2.755 (0.090) & 4.182 (0.110) & 100.0 / 74.2 & 3.024 (0.097) & 4.355 (0.100) & 98.0 / 98.8 & 3.023 (0.111) & 4.353 (0.107) & 96.4 / 99.6 \\
 & 5 & 2.516 (0.105) & 4.420 (0.129) & 100.0 / 83.6 & 2.797 (0.111) & 4.573 (0.114) & \boldwidth{98.6 / 98.8} & 2.797 (0.121) & 4.572 (0.118) & \boldwidth{98.2 / 99.0} \\
 & 8 & 2.357 (0.117) & 4.579 (0.143) & 100.0 / 89.4 & 2.645 (0.122) & 4.717 (0.124) & 98.4 / 99.0 & 2.644 (0.129) & 4.717 (0.126) & 98.6 / 99.0 \\
\addlinespace
\multirow{5}{*}{$\hat\mu$} & 1.001 & 3.698 (0.222) & 3.700 (0.222) & 99.2 / 90.6 & 3.289 (0.069) & 3.290 (0.069) & 100.0 / 0.0 & 3.695 (0.115) & 3.697 (0.115) & 96.0 / 94.6 \\
 & 2 & 3.231 (0.199) & 4.167 (0.254) & 99.4 / 94.0 & 2.972 (0.079) & 3.859 (0.079) & 100.0 / 14.6 & 3.227 (0.110) & 4.164 (0.137) & 99.4 / 96.4 \\
 & 3 & 2.997 (0.192) & 4.401 (0.273) & 99.6 / 95.6 & 2.814 (0.090) & 4.144 (0.094) & 100.0 / 56.0 & 2.992 (0.117) & 4.398 (0.154) & 99.6 / 98.0 \\
 & 5 & 2.762 (0.188) & 4.635 (0.292) & \boldwidth{99.8 / 96.6} & 2.656 (0.105) & 4.431 (0.112) & 100.0 / 85.8 & 2.756 (0.128) & 4.632 (0.173) & \boldwidth{99.8 / 98.8} \\
 & 8 & 2.605 (0.187) & 4.791 (0.306) & 99.8 / 97.0 & 2.550 (0.118) & 4.622 (0.123) & 99.8 / 93.6 & 2.599 (0.139) & 4.789 (0.186) & 99.8 / 98.8 \\
\addlinespace
\multirow{5}{*}{$\hat p$} & 1.001 & 3.406 (0.204) & 3.407 (0.204) & 100.0 / 48.0 & 3.755 (0.079) & 3.756 (0.079) & 82.4 / 99.2 & 3.686 (0.204) & 3.688 (0.204) & 95.8 / 92.6 \\
 & 2 & 3.041 (0.184) & 3.772 (0.229) & 100.0 / 46.6 & 3.385 (0.081) & 4.127 (0.089) & 67.4 / 98.4 & 3.205 (0.182) & 4.174 (0.224) & 98.8 / 95.8 \\
 & 3 & 2.857 (0.176) & 3.955 (0.243) & 100.0 / 46.4 & 3.199 (0.086) & 4.313 (0.097) & 61.4 / 96.6 & 2.967 (0.175) & 4.420 (0.232) & 99.2 / 97.6 \\
 & 5 & 2.674 (0.169) & 4.139 (0.257) & 100.0 / 46.6 & 3.012 (0.093) & 4.499 (0.107) & 59.6 / 94.8 & 2.731 (0.173) & 4.666 (0.239) & 99.4 / 98.6 \\
 & 8 & 2.551 (0.166) & 4.261 (0.267) & 100.0 / 45.6 & 2.888 (0.099) & 4.624 (0.113) & 56.8 / 93.8 & 2.574 (0.176) & 4.831 (0.244) & 99.6 / 99.0 \\
\addlinespace
\bottomrule
\end{tabular}}

\;

\;

\;

\;

\textbf{Arm 1}

\;

\resizebox{\textwidth}{!}{%
\begin{tabular}{llrrrrrrrrr}
\toprule
\multirow{2}{*}{Misspec.} & \multirow{2}{*}{$\Lambda$} & \multicolumn{3}{c}{IPW} & \multicolumn{3}{c}{OR} & \multicolumn{3}{c}{DR} \\
\cmidrule(lr){3-5}\cmidrule(lr){6-8}\cmidrule(lr){9-11}
 &  & Lower & Upper & Valid.\ (\%) & Lower & Upper & Valid.\ (\%) & Lower & Upper & Valid.\ (\%) \\
\midrule
\multirow{5}{*}{None} & 1.001 & 3.918 (0.106) & 3.920 (0.106) & 96.8 / 92.0 & 3.921 (0.074) & 3.923 (0.074) & 94.8 / 94.4 & 3.913 (0.228) & 3.915 (0.228) & 95.6 / 94.2 \\
 & 2 & 3.193 (0.127) & 4.643 (0.118) & 99.6 / 99.4 & 3.274 (0.092) & 4.586 (0.070) & 98.8 / 99.6 & 3.186 (0.232) & 4.639 (0.210) & 98.8 / 98.6 \\
 & 3 & 2.822 (0.149) & 5.010 (0.134) & 100.0 / 100.0 & 2.949 (0.105) & 4.930 (0.075) & 99.4 / 99.8 & 2.816 (0.233) & 5.007 (0.195) & 99.2 / 100.0 \\
 & 5 & 2.441 (0.178) & 5.386 (0.155) & \boldwidth{100.0 / 100.0} & 2.620 (0.122) & 5.290 (0.086) & \boldwidth{99.6 / 100.0} & 2.436 (0.237) & 5.384 (0.176) & \boldwidth{99.4 / 100.0} \\
 & 8 & 2.181 (0.200) & 5.642 (0.171) & 100.0 / 100.0 & 2.396 (0.135) & 5.539 (0.098) & 99.6 / 100.0 & 2.176 (0.244) & 5.642 (0.162) & 99.8 / 100.0 \\
\addlinespace
\multirow{5}{*}{$\hat e_1$} & 1.001 & 4.266 (0.091) & 4.268 (0.091) & 1.8 / 100.0 & 3.921 (0.074) & 3.923 (0.074) & 94.8 / 94.4 & 3.921 (0.106) & 3.922 (0.106) & 94.8 / 95.8 \\
 & 2 & 3.618 (0.104) & 4.917 (0.085) & 13.2 / 100.0 & 3.274 (0.092) & 4.586 (0.070) & 98.8 / 99.6 & 3.274 (0.113) & 4.585 (0.100) & 98.6 / 99.8 \\
 & 3 & 3.286 (0.115) & 5.249 (0.086) & 28.4 / 100.0 & 2.949 (0.105) & 4.930 (0.075) & 99.4 / 99.8 & 2.950 (0.119) & 4.929 (0.099) & 98.8 / 100.0 \\
 & 5 & 2.945 (0.129) & 5.591 (0.091) & 48.2 / 100.0 & 2.620 (0.122) & 5.290 (0.086) & \boldwidth{99.6 / 100.0} & 2.621 (0.129) & 5.288 (0.102) & \boldwidth{99.4 / 100.0} \\
 & 8 & 2.710 (0.142) & 5.827 (0.096) & 60.0 / 100.0 & 2.396 (0.135) & 5.539 (0.098) & 99.6 / 100.0 & 2.398 (0.138) & 5.538 (0.107) & 99.8 / 100.0 \\
\addlinespace
\multirow{5}{*}{$\hat\mu$} & 1.001 & 3.918 (0.106) & 3.920 (0.106) & 96.8 / 92.0 & 4.565 (0.061) & 4.567 (0.061) & 0.0 / 100.0 & 3.921 (0.123) & 3.923 (0.123) & 91.6 / 96.6 \\
 & 2 & 3.193 (0.127) & 4.643 (0.118) & 99.6 / 99.4 & 3.662 (0.078) & 4.973 (0.063) & 1.0 / 100.0 & 3.191 (0.165) & 4.643 (0.114) & 99.0 / 99.8 \\
 & 3 & 2.822 (0.149) & 5.010 (0.134) & 100.0 / 100.0 & 3.226 (0.096) & 5.207 (0.070) & 42.4 / 100.0 & 2.819 (0.190) & 5.010 (0.121) & 99.6 / 100.0 \\
 & 5 & 2.441 (0.178) & 5.386 (0.155) & \boldwidth{100.0 / 100.0} & 2.796 (0.117) & 5.466 (0.084) & 88.4 / 100.0 & 2.438 (0.216) & 5.386 (0.128) & \boldwidth{99.6 / 100.0} \\
 & 8 & 2.181 (0.200) & 5.642 (0.171) & 100.0 / 100.0 & 2.510 (0.132) & 5.653 (0.097) & 95.8 / 100.0 & 2.177 (0.233) & 5.643 (0.133) & 99.8 / 100.0 \\
\addlinespace
\multirow{5}{*}{$\hat p$} & 1.001 & 3.938 (0.088) & 3.940 (0.088) & 92.6 / 95.8 & 4.002 (0.067) & 4.004 (0.067) & 64.2 / 99.8 & 3.907 (0.249) & 3.910 (0.249) & 95.6 / 92.4 \\
 & 2 & 3.297 (0.101) & 4.579 (0.100) & 99.0 / 98.4 & 3.400 (0.080) & 4.607 (0.067) & 82.0 / 99.8 & 3.163 (0.261) & 4.656 (0.220) & 98.8 / 99.4 \\
 & 3 & 2.972 (0.117) & 4.900 (0.113) & 99.4 / 99.4 & 3.095 (0.090) & 4.911 (0.073) & 87.4 / 99.8 & 2.785 (0.264) & 5.038 (0.199) & 99.2 / 100.0 \\
 & 5 & 2.642 (0.138) & 5.226 (0.129) & 99.8 / 99.6 & 2.787 (0.103) & 5.219 (0.082) & 90.0 / 99.8 & 2.399 (0.270) & 5.429 (0.178) & 99.4 / 100.0 \\
 & 8 & 2.418 (0.155) & 5.445 (0.141) & 99.6 / 99.6 & 2.579 (0.113) & 5.428 (0.090) & 90.4 / 99.8 & 2.136 (0.277) & 5.696 (0.168) & 99.8 / 100.0 \\
\addlinespace
\bottomrule
\end{tabular}}
\par\smallskip\begin{minipage}{\textwidth}\scriptsize {Entries are Monte Carlo means (empirical SEs) over 500 repetitions. Valid.\ gives lower and upper confidence bound validity rates (percent). Strictly conservative limiting bounds can yield rates above 95\%.}  \end{minipage}
\end{table}

\clearpage
\begin{table}[tbp]
\centering
\caption{Sensitivity bound estimation over 500 repetitions: $n=5000$, $\hat Q_{z\pm}$ correctly specified. At $\Lambda=5$, the reference sharp endpoints are [2.869, 4.506] for arm 0 and [2.746, 5.089] for arm 1. See the caption of Table~\ref{tab:sim_bounds_2000_0_0} for a description of the shared elements.%
  % Draft return reference; comment out this line for camera-ready.
}
\label{tab:sim_bounds_5000_0_0}\label{tab:sim_bounds_5000_1_0}
\setlength{\tabcolsep}{2.5pt}
\scriptsize
\textbf{Arm 0}

\;

\resizebox{\textwidth}{!}{%
\begin{tabular}{llrrrrrrrrr}
\toprule
\multirow{2}{*}{Misspec.} & \multirow{2}{*}{$\Lambda$} & \multicolumn{3}{c}{IPW} & \multicolumn{3}{c}{OR} & \multicolumn{3}{c}{DR} \\
\cmidrule(lr){3-5}\cmidrule(lr){6-8}\cmidrule(lr){9-11}
 &  & Lower & Upper & CI cov.\ (\%) & Lower & Upper & CI cov.\ (\%) & Lower & Upper & CI cov.\ (\%) \\
\midrule
\multirow{5}{*}{None} & 1.001 & 3.693 (0.137) & 3.694 (0.137) & 92.6 / 92.6 & 3.690 (0.052) & 3.691 (0.052) & 95.4 / 95.4 & 3.694 (0.114) & 3.695 (0.114) & 95.6 / 95.6 \\
 & 2 & 3.283 (0.134) & 4.104 (0.145) & 92.6 / 93.8 & 3.280 (0.060) & 4.101 (0.055) & 95.0 / 95.8 & 3.283 (0.119) & 4.106 (0.117) & 94.8 / 95.0 \\
 & 3 & 3.078 (0.134) & 4.310 (0.150) & 92.6 / 93.2 & 3.074 (0.066) & 4.307 (0.060) & 94.6 / 95.2 & 3.077 (0.123) & 4.312 (0.121) & 94.4 / 94.4 \\
 & 5 & \boldwidth{2.872} (0.135) & \boldwidth{4.515} (0.157) & \boldwidth{92.8 / 93.2} & \boldwidth{2.868} (0.074) & \boldwidth{4.513} (0.067) & \boldwidth{94.2 / 96.2} & \boldwidth{2.871} (0.128) & \boldwidth{4.518} (0.126) & \boldwidth{94.8 / 94.6} \\
 & 8 & 2.735 (0.137) & 4.652 (0.161) & 92.4 / 93.2 & 2.730 (0.080) & 4.651 (0.072) & 94.8 / 96.4 & 2.733 (0.133) & 4.656 (0.131) & 94.6 / 94.4 \\
\addlinespace
\multirow{5}{*}{$\hat e_1$} & 1.001 & 3.461 (0.050) & 3.462 (0.050) & 0.2 / 0.2 & 3.690 (0.052) & 3.691 (0.052) & 95.4 / 95.4 & 3.690 (0.063) & 3.691 (0.063) & 95.8 / 95.8 \\
 & 2 & 3.022 (0.051) & 3.902 (0.061) & 0.0 / 12.4 & 3.280 (0.060) & 4.101 (0.055) & 95.0 / 95.8 & 3.279 (0.069) & 4.101 (0.065) & 96.4 / 95.2 \\
 & 3 & 2.802 (0.057) & 4.122 (0.070) & 0.0 / 30.0 & 3.074 (0.066) & 4.307 (0.060) & 94.6 / 95.2 & 3.074 (0.075) & 4.307 (0.069) & 95.6 / 95.4 \\
 & 5 & 2.581 (0.065) & 4.342 (0.080) & 0.2 / 50.4 & \boldwidth{2.868} (0.074) & \boldwidth{4.513} (0.067) & \boldwidth{94.2 / 96.2} & \boldwidth{2.867} (0.082) & \boldwidth{4.513} (0.075) & \boldwidth{95.6 / 95.6} \\
 & 8 & 2.434 (0.072) & 4.489 (0.087) & 0.6 / 59.6 & 2.730 (0.080) & 4.651 (0.072) & 94.8 / 96.4 & 2.730 (0.088) & 4.651 (0.080) & 95.6 / 95.8 \\
\addlinespace
\multirow{5}{*}{$\hat\mu$} & 1.001 & 3.693 (0.137) & 3.694 (0.137) & 92.6 / 92.6 & 3.283 (0.046) & 3.285 (0.046) & 0.0 / 0.0 & 3.692 (0.074) & 3.693 (0.074) & 93.2 / 93.2 \\
 & 2 & 3.283 (0.134) & 4.104 (0.145) & 92.6 / 93.8 & 3.002 (0.052) & 3.824 (0.050) & 0.0 / 0.0 & 3.281 (0.082) & 4.105 (0.078) & 94.8 / 94.6 \\
 & 3 & 3.078 (0.134) & 4.310 (0.150) & 92.6 / 93.2 & 2.863 (0.059) & 4.097 (0.057) & 3.6 / 7.0 & 3.076 (0.091) & 4.311 (0.087) & 95.0 / 94.8 \\
 & 5 & \boldwidth{2.872} (0.135) & \boldwidth{4.515} (0.157) & \boldwidth{92.8 / 93.2} & 2.726 (0.068) & 4.372 (0.066) & 42.8 / 51.2 & \boldwidth{2.870} (0.105) & \boldwidth{4.518} (0.101) & \boldwidth{94.4 / 95.0} \\
 & 8 & 2.735 (0.137) & 4.652 (0.161) & 92.4 / 93.2 & 2.635 (0.076) & 4.556 (0.072) & 73.8 / 76.6 & 2.733 (0.116) & 4.656 (0.113) & 94.6 / 95.0 \\
\addlinespace
\multirow{5}{*}{$\hat p$} & 1.001 & 3.403 (0.131) & 3.404 (0.131) & 39.2 / 39.2 & 3.751 (0.054) & 3.752 (0.054) & 76.8 / 76.8 & 3.693 (0.122) & 3.694 (0.122) & 95.6 / 95.6 \\
 & 2 & 3.082 (0.126) & 3.725 (0.139) & 54.8 / 27.2 & 3.406 (0.057) & 4.096 (0.056) & 41.0 / 95.4 & 3.268 (0.123) & 4.119 (0.130) & 94.6 / 94.8 \\
 & 3 & 2.921 (0.124) & 3.886 (0.144) & 65.0 / 23.0 & 3.234 (0.060) & 4.269 (0.059) & 26.2 / 93.4 & 3.055 (0.127) & 4.331 (0.138) & 94.8 / 94.8 \\
 & 5 & 2.759 (0.122) & 4.047 (0.148) & 76.6 / 19.6 & 3.061 (0.065) & 4.442 (0.063) & 19.4 / 82.8 & 2.844 (0.134) & 4.543 (0.147) & 94.4 / 95.4 \\
 & 8 & 2.651 (0.122) & 4.155 (0.152) & 82.6 / 17.4 & 2.945 (0.068) & 4.558 (0.067) & 14.2 / 75.4 & 2.703 (0.139) & 4.684 (0.154) & 95.0 / 96.2 \\
\addlinespace
\bottomrule
\end{tabular}}

\;

\;

\;

\;

\textbf{Arm 1}

\;

\resizebox{\textwidth}{!}{%
\begin{tabular}{llrrrrrrrrr}
\toprule
\multirow{2}{*}{Misspec.} & \multirow{2}{*}{$\Lambda$} & \multicolumn{3}{c}{IPW} & \multicolumn{3}{c}{OR} & \multicolumn{3}{c}{DR} \\
\cmidrule(lr){3-5}\cmidrule(lr){6-8}\cmidrule(lr){9-11}
 &  & Lower & Upper & CI cov.\ (\%) & Lower & Upper & CI cov.\ (\%) & Lower & Upper & CI cov.\ (\%) \\
\midrule
\multirow{5}{*}{None} & 1.001 & 3.913 (0.067) & 3.914 (0.067) & 95.2 / 95.2 & 3.915 (0.045) & 3.917 (0.045) & 95.0 / 95.0 & 3.919 (0.148) & 3.921 (0.148) & 93.2 / 93.2 \\
 & 2 & 3.337 (0.066) & 4.490 (0.076) & 94.0 / 94.2 & 3.340 (0.049) & 4.493 (0.047) & 95.4 / 94.4 & 3.342 (0.148) & 4.498 (0.152) & 94.4 / 93.0 \\
 & 3 & 3.044 (0.068) & 4.783 (0.083) & 94.2 / 93.6 & 3.047 (0.055) & 4.786 (0.052) & 95.0 / 94.8 & 3.048 (0.149) & 4.793 (0.156) & 94.8 / 94.0 \\
 & 5 & \boldwidth{2.744} (0.073) & \boldwidth{5.084} (0.092) & \boldwidth{93.4 / 93.4} & \boldwidth{2.747} (0.062) & \boldwidth{5.086} (0.058) & \boldwidth{95.2 / 94.4} & \boldwidth{2.746} (0.152) & \boldwidth{5.094} (0.160) & \boldwidth{93.8 / 94.0} \\
 & 8 & 2.538 (0.079) & 5.290 (0.098) & 93.8 / 93.6 & 2.542 (0.068) & 5.291 (0.064) & 95.6 / 94.0 & 2.540 (0.155) & 5.301 (0.164) & 93.6 / 94.2 \\
\addlinespace
\multirow{5}{*}{$\hat e_1$} & 1.001 & 4.265 (0.055) & 4.266 (0.055) & 0.0 / 0.0 & 3.915 (0.045) & 3.917 (0.045) & 95.0 / 95.0 & 3.915 (0.066) & 3.917 (0.066) & 96.8 / 96.8 \\
 & 2 & 3.698 (0.060) & 4.834 (0.053) & 0.0 / 0.0 & 3.340 (0.049) & 4.493 (0.047) & 95.4 / 94.4 & 3.340 (0.068) & 4.493 (0.070) & 96.8 / 95.2 \\
 & 3 & 3.407 (0.065) & 5.124 (0.054) & 0.0 / 0.0 & 3.047 (0.055) & 4.786 (0.052) & 95.0 / 94.8 & 3.047 (0.071) & 4.786 (0.074) & 96.6 / 94.6 \\
 & 5 & 3.109 (0.072) & 5.423 (0.055) & 0.0 / 0.0 & \boldwidth{2.747} (0.062) & \boldwidth{5.086} (0.058) & \boldwidth{95.2 / 94.4} & \boldwidth{2.747} (0.076) & \boldwidth{5.086} (0.079) & \boldwidth{95.8 / 94.2} \\
 & 8 & 2.903 (0.078) & 5.629 (0.058) & 0.0 / 0.0 & 2.542 (0.068) & 5.291 (0.064) & 95.6 / 94.0 & 2.542 (0.080) & 5.291 (0.084) & 95.8 / 94.0 \\
\addlinespace
\multirow{5}{*}{$\hat\mu$} & 1.001 & 3.913 (0.067) & 3.914 (0.067) & 95.2 / 95.2 & 4.563 (0.037) & 4.564 (0.037) & 0.0 / 0.0 & 3.918 (0.078) & 3.920 (0.078) & 93.0 / 93.0 \\
 & 2 & 3.337 (0.066) & 4.490 (0.076) & 94.0 / 94.2 & 3.729 (0.041) & 4.882 (0.041) & 0.0 / 0.0 & 3.341 (0.093) & 4.496 (0.093) & 93.6 / 95.2 \\
 & 3 & 3.044 (0.068) & 4.783 (0.083) & 94.2 / 93.6 & 3.325 (0.049) & 5.064 (0.047) & 0.0 / 0.0 & 3.046 (0.107) & 4.791 (0.109) & 93.6 / 94.8 \\
 & 5 & \boldwidth{2.744} (0.073) & \boldwidth{5.084} (0.092) & \boldwidth{93.4 / 93.4} & 2.924 (0.059) & 5.263 (0.054) & 17.8 / 7.2 & \boldwidth{2.745} (0.124) & \boldwidth{5.093} (0.128) & \boldwidth{93.2 / 94.2} \\
 & 8 & 2.538 (0.079) & 5.290 (0.098) & 93.8 / 93.6 & 2.657 (0.066) & 5.406 (0.061) & 59.8 / 52.6 & 2.539 (0.137) & 5.300 (0.142) & 93.0 / 94.0 \\
\addlinespace
\multirow{5}{*}{$\hat p$} & 1.001 & 3.934 (0.055) & 3.935 (0.055) & 94.0 / 94.0 & 3.997 (0.041) & 3.998 (0.041) & 54.0 / 54.2 & 3.917 (0.157) & 3.919 (0.157) & 95.2 / 95.2 \\
 & 2 & 3.420 (0.052) & 4.450 (0.063) & 67.8 / 86.0 & 3.467 (0.043) & 4.528 (0.045) & 22.0 / 86.6 & 3.324 (0.165) & 4.512 (0.153) & 93.4 / 94.6 \\
 & 3 & 3.160 (0.054) & 4.710 (0.068) & 45.4 / 74.4 & 3.199 (0.046) & 4.796 (0.049) & 13.4 / 94.2 & 3.022 (0.172) & 4.814 (0.154) & 92.8 / 95.0 \\
 & 5 & 2.896 (0.057) & 4.974 (0.075) & 27.2 / 61.8 & 2.928 (0.051) & 5.068 (0.055) & 7.8 / 92.8 & 2.714 (0.180) & 5.124 (0.156) & 93.0 / 94.8 \\
 & 8 & 2.719 (0.060) & 5.152 (0.080) & 19.4 / 49.8 & 2.744 (0.055) & 5.251 (0.059) & 5.4 / 88.2 & 2.502 (0.187) & 5.335 (0.159) & 92.8 / 94.6 \\
\addlinespace
\bottomrule
\end{tabular}}
\par\smallskip\begin{minipage}{\textwidth}\scriptsize Entries are Monte Carlo means (empirical SEs) over 500 repetitions. CI cov.\ reports separate marginal coverage rates (percent) for the lower and upper sharp endpoints using two-sided 95\% percentile bootstrap intervals with 500 bootstrap draws per repetition. \end{minipage}
\end{table}

\clearpage
\begin{table}[tbp]
\centering
\caption{Sensitivity bound estimation over 500 repetitions: $n=5000$, $\hat Q_{z\pm}$ misspecified. At $\Lambda=5$, the reference sharp endpoints are [2.869, 4.506] for arm 0 and [2.746, 5.089] for arm 1. See the caption of Table~\ref{tab:sim_bounds_2000_0_0} for a description of the shared elements.%
  % Draft return reference; comment out this line for camera-ready.
}
\label{tab:sim_bounds_5000_0_1}  \label{tab:sim_bounds_5000_1_1}
\setlength{\tabcolsep}{2.5pt}
\scriptsize
\textbf{Arm 0}

\;

\resizebox{\textwidth}{!}{%
\begin{tabular}{llrrrrrrrrr}
\toprule
\multirow{2}{*}{Misspec.} & \multirow{2}{*}{$\Lambda$} & \multicolumn{3}{c}{IPW} & \multicolumn{3}{c}{OR} & \multicolumn{3}{c}{DR} \\
\cmidrule(lr){3-5}\cmidrule(lr){6-8}\cmidrule(lr){9-11}
 &  & Lower & Upper & Valid.\ (\%) & Lower & Upper & Valid.\ (\%) & Lower & Upper & Valid.\ (\%) \\
\midrule
\multirow{5}{*}{None} & 1.001 & 3.693 (0.137) & 3.694 (0.137) & 96.4 / 91.4 & 3.690 (0.052) & 3.691 (0.052) & 95.2 / 95.8 & 3.694 (0.114) & 3.695 (0.115) & 97.0 / 93.8 \\
 & 2 & 3.224 (0.126) & 4.163 (0.154) & 98.8 / 96.0 & 3.243 (0.059) & 4.133 (0.058) & 99.2 / 99.0 & 3.223 (0.107) & 4.165 (0.118) & 98.6 / 98.4 \\
 & 3 & 2.989 (0.123) & 4.398 (0.164) & 98.8 / 98.4 & 3.017 (0.066) & 4.353 (0.064) & 99.6 / 99.8 & 2.987 (0.103) & 4.399 (0.119) & 99.0 / 99.4 \\
 & 5 & 2.753 (0.121) & 4.633 (0.175) & \boldwidth{99.4 / 99.0} & 2.789 (0.075) & 4.571 (0.072) & \boldwidth{99.6 / 99.6} & 2.751 (0.100) & 4.634 (0.120) & \boldwidth{100.0 / 99.6} \\
 & 8 & 2.596 (0.122) & 4.789 (0.183) & 99.8 / 99.2 & 2.637 (0.082) & 4.717 (0.078) & 99.6 / 99.6 & 2.593 (0.100) & 4.791 (0.121) & 100.0 / 99.8 \\
\addlinespace
\multirow{5}{*}{$\hat e_1$} & 1.001 & 3.461 (0.050) & 3.462 (0.050) & 100.0 / 0.0 & 3.690 (0.052) & 3.691 (0.052) & 95.2 / 95.8 & 3.690 (0.063) & 3.691 (0.063) & 95.6 / 95.0 \\
 & 2 & 2.984 (0.052) & 3.939 (0.062) & 100.0 / 21.2 & 3.243 (0.059) & 4.133 (0.058) & 99.2 / 99.0 & 3.243 (0.067) & 4.133 (0.066) & 98.8 / 98.2 \\
 & 3 & 2.745 (0.059) & 4.178 (0.072) & 100.0 / 50.4 & 3.017 (0.066) & 4.353 (0.064) & 99.6 / 99.8 & 3.017 (0.072) & 4.352 (0.070) & 99.8 / 99.4 \\
 & 5 & 2.506 (0.068) & 4.417 (0.083) & 100.0 / 70.2 & 2.789 (0.075) & 4.571 (0.072) & \boldwidth{99.6 / 99.6} & 2.790 (0.078) & 4.571 (0.075) & \boldwidth{99.8 / 99.4} \\
 & 8 & 2.346 (0.075) & 4.577 (0.091) & 100.0 / 80.6 & 2.637 (0.082) & 4.717 (0.078) & 99.6 / 99.6 & 2.637 (0.084) & 4.716 (0.080) & 99.6 / 99.6 \\
\addlinespace
\multirow{5}{*}{$\hat\mu$} & 1.001 & 3.693 (0.137) & 3.694 (0.137) & 96.4 / 91.4 & 3.283 (0.046) & 3.285 (0.046) & 100.0 / 0.0 & 3.692 (0.074) & 3.693 (0.074) & 96.6 / 93.6 \\
 & 2 & 3.224 (0.126) & 4.163 (0.154) & 98.8 / 96.0 & 2.966 (0.053) & 3.856 (0.051) & 100.0 / 0.0 & 3.222 (0.072) & 4.163 (0.084) & 99.2 / 99.0 \\
 & 3 & 2.989 (0.123) & 4.398 (0.164) & 98.8 / 98.4 & 2.807 (0.060) & 4.142 (0.060) & 100.0 / 21.8 & 2.986 (0.076) & 4.399 (0.093) & 99.6 / 99.8 \\
 & 5 & 2.753 (0.121) & 4.633 (0.175) & \boldwidth{99.4 / 99.0} & 2.648 (0.070) & 4.430 (0.070) & 100.0 / 73.2 & 2.750 (0.084) & 4.634 (0.103) & \boldwidth{100.0 / 100.0} \\
 & 8 & 2.596 (0.122) & 4.789 (0.183) & 99.8 / 99.2 & 2.542 (0.078) & 4.621 (0.077) & 100.0 / 92.0 & 2.593 (0.090) & 4.791 (0.111) & 100.0 / 100.0 \\
\addlinespace
\multirow{5}{*}{$\hat p$} & 1.001 & 3.403 (0.131) & 3.404 (0.131) & 100.0 / 29.6 & 3.751 (0.054) & 3.752 (0.054) & 68.4 / 99.8 & 3.692 (0.122) & 3.694 (0.122) & 96.2 / 94.8 \\
 & 2 & 3.035 (0.119) & 3.771 (0.146) & 100.0 / 29.2 & 3.378 (0.055) & 4.125 (0.059) & 42.2 / 98.2 & 3.210 (0.112) & 4.181 (0.130) & 99.0 / 98.2 \\
 & 3 & 2.851 (0.114) & 3.955 (0.154) & 100.0 / 28.6 & 3.191 (0.058) & 4.312 (0.063) & 33.8 / 97.0 & 2.970 (0.109) & 4.425 (0.134) & 99.8 / 99.4 \\
 & 5 & 2.666 (0.110) & 4.139 (0.162) & 100.0 / 28.6 & 3.004 (0.063) & 4.499 (0.069) & 29.4 / 95.8 & 2.732 (0.109) & 4.670 (0.138) & 100.0 / 99.6 \\
 & 8 & 2.542 (0.108) & 4.262 (0.168) & 100.0 / 28.0 & 2.879 (0.066) & 4.624 (0.072) & 27.6 / 93.8 & 2.574 (0.111) & 4.834 (0.141) & 100.0 / 99.8 \\
\addlinespace
\bottomrule
\end{tabular}}

\;

\;

\;

\;

\textbf{Arm 1}

\;

\resizebox{\textwidth}{!}{%
\begin{tabular}{llrrrrrrrrr}
\toprule
\multirow{2}{*}{Misspec.} & \multirow{2}{*}{$\Lambda$} & \multicolumn{3}{c}{IPW} & \multicolumn{3}{c}{OR} & \multicolumn{3}{c}{DR} \\
\cmidrule(lr){3-5}\cmidrule(lr){6-8}\cmidrule(lr){9-11}
 &  & Lower & Upper & Valid.\ (\%) & Lower & Upper & Valid.\ (\%) & Lower & Upper & Valid.\ (\%) \\
\midrule
\multirow{5}{*}{None} & 1.001 & 3.912 (0.067) & 3.915 (0.067) & 95.8 / 94.0 & 3.915 (0.045) & 3.917 (0.045) & 94.8 / 94.6 & 3.919 (0.148) & 3.921 (0.148) & 95.0 / 94.6 \\
 & 2 & 3.186 (0.076) & 4.638 (0.078) & 100.0 / 100.0 & 3.267 (0.053) & 4.581 (0.045) & 100.0 / 100.0 & 3.191 (0.150) & 4.646 (0.136) & 99.6 / 99.4 \\
 & 3 & 2.815 (0.088) & 5.006 (0.089) & 100.0 / 100.0 & 2.941 (0.060) & 4.928 (0.050) & 100.0 / 100.0 & 2.818 (0.150) & 5.014 (0.125) & 100.0 / 100.0 \\
 & 5 & 2.435 (0.105) & 5.382 (0.102) & \boldwidth{100.0 / 100.0} & 2.611 (0.070) & 5.288 (0.058) & \boldwidth{100.0 / 100.0} & 2.435 (0.151) & 5.390 (0.111) & \boldwidth{100.0 / 100.0} \\
 & 8 & 2.174 (0.119) & 5.638 (0.112) & 100.0 / 100.0 & 2.386 (0.078) & 5.538 (0.065) & 100.0 / 100.0 & 2.173 (0.153) & 5.648 (0.101) & 100.0 / 100.0 \\
\addlinespace
\multirow{5}{*}{$\hat e_1$} & 1.001 & 4.265 (0.055) & 4.267 (0.055) & 0.0 / 100.0 & 3.915 (0.045) & 3.917 (0.045) & 94.8 / 94.6 & 3.915 (0.066) & 3.917 (0.066) & 95.8 / 94.8 \\
 & 2 & 3.615 (0.062) & 4.917 (0.052) & 0.0 / 100.0 & 3.267 (0.053) & 4.581 (0.045) & 100.0 / 100.0 & 3.267 (0.068) & 4.581 (0.066) & 99.8 / 100.0 \\
 & 3 & 3.283 (0.068) & 5.250 (0.053) & 3.8 / 100.0 & 2.941 (0.060) & 4.928 (0.050) & 100.0 / 100.0 & 2.942 (0.070) & 4.927 (0.066) & 100.0 / 100.0 \\
 & 5 & 2.941 (0.076) & 5.593 (0.055) & 18.8 / 100.0 & 2.611 (0.070) & 5.288 (0.058) & \boldwidth{100.0 / 100.0} & 2.612 (0.074) & 5.287 (0.068) & \boldwidth{100.0 / 100.0} \\
 & 8 & 2.705 (0.083) & 5.830 (0.058) & 36.2 / 100.0 & 2.386 (0.078) & 5.538 (0.065) & 100.0 / 100.0 & 2.388 (0.080) & 5.537 (0.072) & 100.0 / 100.0 \\
\addlinespace
\multirow{5}{*}{$\hat\mu$} & 1.001 & 3.912 (0.067) & 3.915 (0.067) & 95.8 / 94.0 & 4.562 (0.037) & 4.564 (0.037) & 0.0 / 100.0 & 3.918 (0.078) & 3.920 (0.078) & 92.0 / 95.6 \\
 & 2 & 3.186 (0.076) & 4.638 (0.078) & 100.0 / 100.0 & 3.656 (0.044) & 4.971 (0.041) & 0.0 / 100.0 & 3.189 (0.105) & 4.644 (0.075) & 99.8 / 100.0 \\
 & 3 & 2.815 (0.088) & 5.006 (0.089) & 100.0 / 100.0 & 3.219 (0.054) & 5.206 (0.047) & 10.4 / 100.0 & 2.816 (0.120) & 5.013 (0.078) & 100.0 / 100.0 \\
 & 5 & 2.435 (0.105) & 5.382 (0.102) & \boldwidth{100.0 / 100.0} & 2.788 (0.066) & 5.465 (0.056) & 87.4 / 100.0 & 2.434 (0.135) & 5.389 (0.081) & \boldwidth{100.0 / 100.0} \\
 & 8 & 2.174 (0.119) & 5.638 (0.112) & 100.0 / 100.0 & 2.501 (0.076) & 5.652 (0.064) & 99.0 / 100.0 & 2.172 (0.145) & 5.647 (0.083) & 100.0 / 100.0 \\
\addlinespace
\multirow{5}{*}{$\hat p$} & 1.001 & 3.934 (0.055) & 3.936 (0.055) & 91.6 / 97.4 & 3.997 (0.041) & 3.998 (0.041) & 39.6 / 100.0 & 3.917 (0.157) & 3.919 (0.157) & 94.6 / 94.2 \\
 & 2 & 3.291 (0.060) & 4.575 (0.064) & 99.8 / 99.2 & 3.393 (0.046) & 4.603 (0.044) & 75.4 / 100.0 & 3.173 (0.165) & 4.664 (0.137) & 99.4 / 99.8 \\
 & 3 & 2.965 (0.069) & 4.898 (0.072) & 100.0 / 99.6 & 3.087 (0.052) & 4.908 (0.048) & 85.4 / 100.0 & 2.794 (0.166) & 5.044 (0.123) & 100.0 / 100.0 \\
 & 5 & 2.635 (0.081) & 5.224 (0.082) & 100.0 / 99.8 & 2.778 (0.059) & 5.218 (0.054) & 90.4 / 100.0 & 2.407 (0.168) & 5.433 (0.109) & 100.0 / 100.0 \\
 & 8 & 2.410 (0.091) & 5.444 (0.090) & 100.0 / 99.8 & 2.569 (0.064) & 5.427 (0.059) & 91.8 / 100.0 & 2.143 (0.171) & 5.699 (0.102) & 100.0 / 100.0 \\
\addlinespace
\bottomrule
\end{tabular}}
\par\smallskip\begin{minipage}{\textwidth}\scriptsize {Entries are Monte Carlo means (empirical SEs) over 500 repetitions. Valid.\ gives lower and upper confidence bound validity rates (percent). Strictly conservative limiting bounds can yield rates above 95\%.}  \end{minipage}
\end{table}

\clearpage
\begin{figure}[!htbp]
\centering
\captionsetup{skip=4pt}
\includegraphics[width=\textwidth]{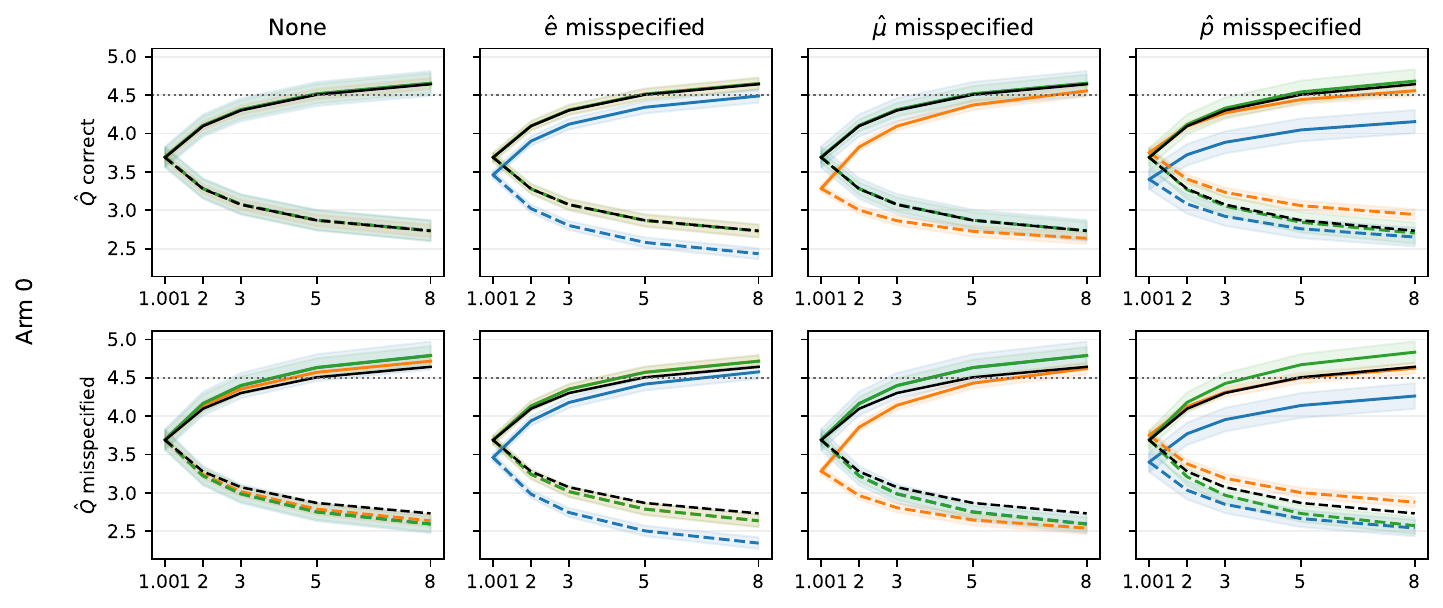}
\includegraphics[width=\textwidth]{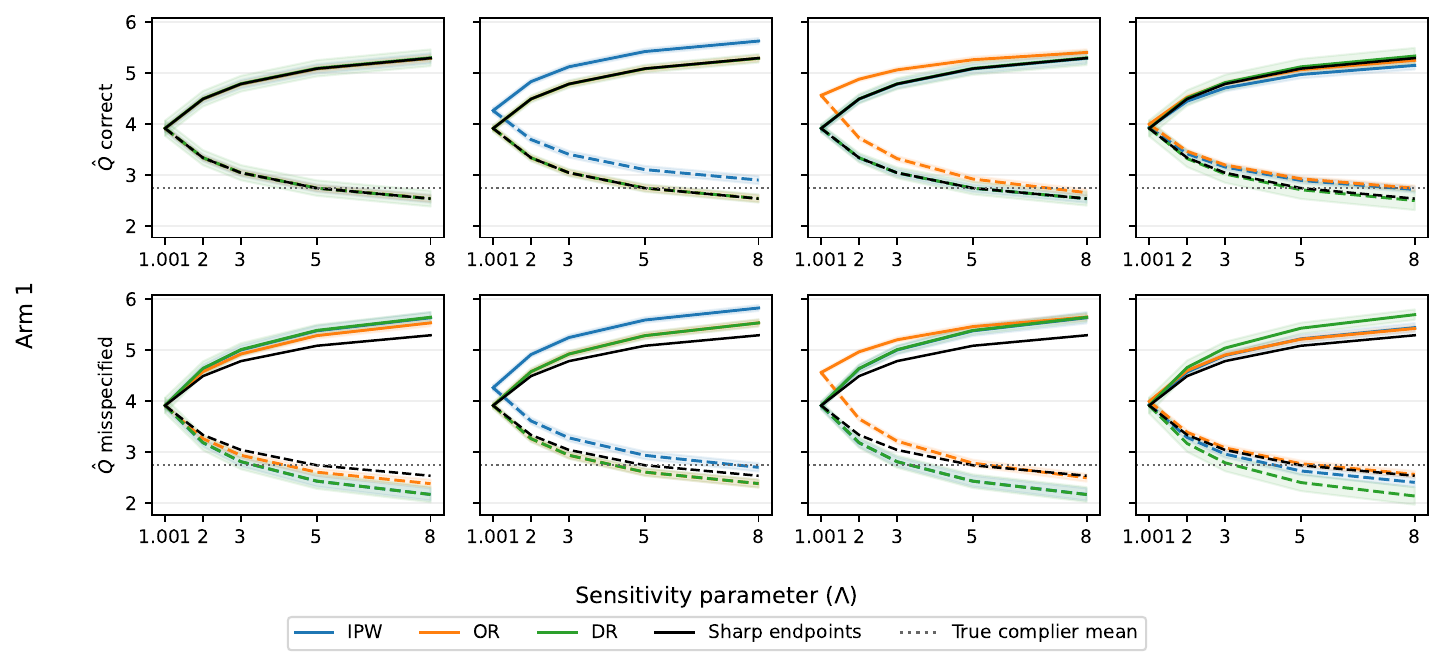}
\caption{Sensitivity bound estimation at $n=5000$ (arm 0 above, arm 1 below). See the caption of Figure~\ref{fig:sim_bounds_2000} for a description of the shared elements.%
  % Draft return reference; comment out this line for camera-ready.
}  \label{fig:sim_bounds_5000}
\end{figure}
\clearpage

\subsection{Maximin policy learning}
As a benchmark, the principal-ignorability-guided outcome regression criterion (PIG) uses
\[
\hat\psi_z^{\mathrm{PIG}}=
\frac{\E_n[\hat\pi_C(X)\hat\mu_z(X)]}{\E_n[\hat\pi_C(X)]},\qquad
\hat V_{\mathrm{PIG}}(\pi)=
\frac{\E_n[\hat\pi_C(X)\{\pi(X)\hat\mu_1(X)+(1-\pi(X))\hat\mu_0(X)\}]}{\E_n[\hat\pi_C(X)]}.
\]
This criterion does not depend on $\Lambda$ or the quantile regressions. The reported true worst-case policy values and oracle maxima are instead computed by numerical integration of the known conditional outcome laws and principal scores. For policies learned at $\Lambda=5$, the true worst-case value and oracle maximum are
\[
V_-(\pi;5)=\frac{\E[\pi_C(X)\{\pi(X)\omega_{1-}(X;5)+(1-\pi(X))\omega_{0-}(X;5)\}]}{\E[\pi_C(X)]},\qquad
V_-^*(5)=\sup_{\pi\in\Pi^{\mathrm{linear}}}V_-(\pi;5).
\]
The reported regret is $V_-^*(5)-V_-(\hat\pi;5)$. All learned policies, including PIG, are evaluated using the same sharp lower-bound functional.

\begin{table}[tbp]
\centering
\caption{True worst-case policy performance at $\Lambda=5$, $n=2000$, over 500 repetitions. The oracle maximum over linear policies is 2.88356. All learned policies, including PIG, are evaluated using the known outcome laws and principal scores. WCV is the worst-case value, i.e., true sharp lower bound on the learned policy's value. Gap is mean maximin regret, i.e., the oracle maximum minus mean WCV. Bold entries identify the largest mean WCV in each scenario (ties at the displayed precision). Misspec.\ identifies the misspecified model among $\hat e_1$, $\hat p$, and $\hat\mu$; None indicates that all three are correctly specified. 
%The first column gives the quantile specification status.
%
  % Draft return reference; comment out this line for camera-ready.
}
\label{tab:sim_oracle_2000}
\setlength{\tabcolsep}{2.5pt}
\scriptsize
\resizebox{\textwidth}{!}{%
\begin{tabular}{llrrrrrrrr}
\toprule
\multirow{2}{*}{$\hat Q_{z\pm}$} & \multirow{2}{*}{Misspec.} & \multicolumn{2}{c}{IPW} & \multicolumn{2}{c}{OR} & \multicolumn{2}{c}{DR} & \multicolumn{2}{c}{PIG} \\
\cmidrule(lr){3-4}\cmidrule(lr){5-6}\cmidrule(lr){7-8}\cmidrule(lr){9-10}
 &  & WCV & Gap & WCV & Gap & WCV & Gap & WCV & Gap \\
\midrule
\multirow{4}{*}{Specified} & None & 2.852 (0.048) & 0.031 & \boldwidth{2.867} (0.028) & \boldwidth{0.017} & 2.848 (0.042) & 0.036 & 2.800 (0.031) & 0.084 \\
 & $\hat e_1$ & 2.866 (0.006) & 0.018 & \boldwidth{2.867} (0.028) & \boldwidth{0.017} & 2.859 (0.036) & 0.024 & 2.800 (0.031) & 0.084 \\
 & $\hat\mu$ & \boldwidth{2.852} (0.048) & \boldwidth{0.031} & 2.779 (0.042) & 0.105 & 2.851 (0.040) & 0.033 & 2.747 (0.000) & 0.137 \\
 & $\hat p$ & 2.857 (0.031) & 0.026 & \boldwidth{2.875} (0.014) & \boldwidth{0.009} & 2.844 (0.045) & 0.040 & 2.800 (0.031) & 0.084 \\
\addlinespace
\multirow{4}{*}{Misspecified} & None & \boldwidth{2.871} (0.022) & \boldwidth{0.013} & 2.863 (0.031) & 0.020 & 2.864 (0.018) & 0.020 & 2.800 (0.031) & 0.084 \\
 & $\hat e_1$ & \boldwidth{2.866} (0.006) & \boldwidth{0.017} & 2.863 (0.031) & 0.020 & 2.856 (0.025) & 0.028 & 2.800 (0.031) & 0.084 \\
 & $\hat\mu$ & \boldwidth{2.871} (0.022) & \boldwidth{0.013} & 2.785 (0.056) & 0.099 & 2.864 (0.018) & 0.020 & 2.747 (0.000) & 0.137 \\
 & $\hat p$ & 2.864 (0.016) & 0.020 & \boldwidth{2.877} (0.011) & \boldwidth{0.006} & 2.866 (0.019) & 0.018 & 2.800 (0.031) & 0.084 \\
\addlinespace
\bottomrule
\end{tabular}}
\par\smallskip\begin{minipage}{\textwidth}\scriptsize {WCV entries are Monte Carlo means (empirical SEs across 500 repetitions); Gap entries are mean regrets. Numerical integration uses the true conditional normal laws and $\pi_C(X)$.
%; its agreement with adaptive quadrature is within $10^{-8}$. 
%PIG is trained under principal ignorability, but evaluated under the same $\Lambda=5$ sensitivity model.
}\end{minipage}
\end{table}

\clearpage
\begin{table}[tbp]
\centering
\caption{Method-specific maximized value estimates across sensitivity parameters, $n=2000$, $\hat Q_{z\pm}$ correctly specified, over 500 repetitions. Each IPW, OR, or DR estimate uses that method's learned policy at the displayed $\Lambda$. 
Bold means highlight OR's invariance to $\hat e_1$ misspecification, IPW's invariance to $\hat\mu$ misspecification, and double robustness of DR. Misspec.\ identifies the misspecified model among $\hat e_1$, $\hat p$, and $\hat\mu$; None indicates that all three are correctly specified. LB denotes the estimated lower bound on policy value.%
  % Draft return reference; comment out this line for camera-ready.
}
\label{tab:sim_estimated_2000_0}
\setlength{\tabcolsep}{2.5pt}
\scriptsize
\resizebox{\textwidth}{!}{%
\begin{tabular}{llrrrr}
\toprule
\multirow{2}{*}{Misspec.} & \multirow{2}{*}{$\Lambda$} & IPW & OR & DR & PIG \\
\cmidrule(lr){3-3}\cmidrule(lr){4-4}\cmidrule(lr){5-5}\cmidrule(lr){6-6}
 & & Estimated LB & Estimated LB & Estimated LB & Estimated PI value \\
\midrule
\multirow{5}{*}{None} & 1.001 & 4.138 (0.156) & \boldwidth{3.953} (0.059) & \boldwidth{3.998} (0.208) & 3.954 (0.059) \\
 & 2 & 3.594 (0.149) & \boldwidth{3.410} (0.067) & \boldwidth{3.457} (0.197) & 3.954 (0.059) \\
 & 3 & 3.335 (0.154) & \boldwidth{3.152} (0.079) & \boldwidth{3.204} (0.197) & 3.954 (0.059) \\
 & 5 & 3.089 (0.161) & \boldwidth{2.909} (0.097) & \boldwidth{2.968} (0.204) & 3.954 (0.059) \\
 & 8 & 2.929 (0.168) & \boldwidth{2.757} (0.110) & \boldwidth{2.820} (0.211) & 3.954 (0.059) \\
\addlinespace
\multirow{5}{*}{$\hat e_1$} & 1.001 & 5.767 (0.128) & \boldwidth{3.953} (0.059) & \boldwidth{3.983} (0.096) & 3.954 (0.059) \\
 & 2 & 5.014 (0.140) & \boldwidth{3.410} (0.067) & \boldwidth{3.447} (0.099) & 3.954 (0.059) \\
 & 3 & 4.633 (0.152) & \boldwidth{3.152} (0.079) & \boldwidth{3.190} (0.105) & 3.954 (0.059) \\
 & 5 & 4.245 (0.169) & \boldwidth{2.909} (0.097) & \boldwidth{2.945} (0.115) & 3.954 (0.059) \\
 & 8 & 3.982 (0.183) & \boldwidth{2.757} (0.110) & \boldwidth{2.790} (0.124) & 3.954 (0.059) \\
\addlinespace
\multirow{5}{*}{$\hat\mu$} & 1.001 & 4.138 (0.156) & 4.565 (0.061) & \boldwidth{4.048} (0.095) & 4.567 (0.061) \\
 & 2 & 3.594 (0.149) & 3.734 (0.073) & \boldwidth{3.476} (0.115) & 4.567 (0.061) \\
 & 3 & 3.335 (0.154) & 3.331 (0.086) & \boldwidth{3.212} (0.136) & 4.567 (0.061) \\
 & 5 & 3.089 (0.161) & 2.942 (0.097) & \boldwidth{2.969} (0.162) & 4.567 (0.061) \\
 & 8 & 2.929 (0.168) & 2.725 (0.101) & \boldwidth{2.818} (0.182) & 4.567 (0.061) \\
\addlinespace
\multirow{5}{*}{$\hat p$} & 1.001 & 4.198 (0.090) & 4.031 (0.054) & 4.001 (0.227) & 4.032 (0.054) \\
 & 2 & 3.717 (0.084) & 3.555 (0.052) & 3.447 (0.224) & 4.032 (0.054) \\
 & 3 & 3.480 (0.085) & 3.334 (0.060) & 3.191 (0.228) & 4.032 (0.054) \\
 & 5 & 3.248 (0.090) & 3.127 (0.072) & 2.954 (0.238) & 4.032 (0.054) \\
 & 8 & 3.096 (0.095) & 2.996 (0.081) & 2.805 (0.245) & 4.032 (0.054) \\
\addlinespace
\bottomrule
\end{tabular}}
\par\smallskip\begin{minipage}{\textwidth}\scriptsize {Entries are Monte Carlo means (empirical SEs across 500 repetitions). PIG is independent of $\Lambda$ and $\hat Q_{z\pm}$, and targets the PI value rather than a lower bound.} \end{minipage}
\end{table}

\clearpage
\begin{table}[tbp]
\centering
\caption{Method-specific maximized value estimates across sensitivity parameters, $n=2000$, $\hat Q_{z\pm}$ misspecified, over 500 repetitions. See the caption of Table~\ref{tab:sim_estimated_2000_0} for a description of the shared elements.%
  % Draft return reference; comment out this line for camera-ready.
}
\label{tab:sim_estimated_2000_1}
\setlength{\tabcolsep}{2.5pt}
\scriptsize
\resizebox{\textwidth}{!}{%
\begin{tabular}{llrrrr}
\toprule
\multirow{2}{*}{Misspec.} & \multirow{2}{*}{$\Lambda$} & IPW & OR & DR & PIG \\
\cmidrule(lr){3-3}\cmidrule(lr){4-4}\cmidrule(lr){5-5}\cmidrule(lr){6-6}
 & & Estimated LB & Estimated LB & Estimated LB & Estimated PI value \\
\midrule
\multirow{5}{*}{None} & 1.001 & 4.138 (0.156) & \boldwidth{3.953} (0.059) & \boldwidth{3.998} (0.208) & 3.954 (0.059) \\
 & 2 & 3.511 (0.122) & \boldwidth{3.347} (0.070) & \boldwidth{3.398} (0.182) & 3.954 (0.059) \\
 & 3 & 3.225 (0.122) & \boldwidth{3.072} (0.084) & \boldwidth{3.121} (0.167) & 3.954 (0.059) \\
 & 5 & 2.950 (0.133) & \boldwidth{2.823} (0.103) & \boldwidth{2.853} (0.156) & 3.954 (0.059) \\
 & 8 & 2.769 (0.143) & \boldwidth{2.668} (0.115) & \boldwidth{2.678} (0.152) & 3.954 (0.059) \\
\addlinespace
\multirow{5}{*}{$\hat e_1$} & 1.001 & 5.767 (0.128) & \boldwidth{3.953} (0.059) & \boldwidth{3.983} (0.096) & 3.954 (0.059) \\
 & 2 & 4.951 (0.141) & \boldwidth{3.347} (0.070) & \boldwidth{3.403} (0.096) & 3.954 (0.059) \\
 & 3 & 4.538 (0.155) & \boldwidth{3.072} (0.084) & \boldwidth{3.129} (0.100) & 3.954 (0.059) \\
 & 5 & 4.117 (0.174) & \boldwidth{2.823} (0.103) & \boldwidth{2.869} (0.110) & 3.954 (0.059) \\
 & 8 & 3.831 (0.189) & \boldwidth{2.668} (0.115) & \boldwidth{2.708} (0.120) & 3.954 (0.059) \\
\addlinespace
\multirow{5}{*}{$\hat\mu$} & 1.001 & 4.138 (0.156) & 4.565 (0.061) & \boldwidth{4.048} (0.095) & 4.567 (0.061) \\
 & 2 & 3.511 (0.122) & 3.662 (0.078) & \boldwidth{3.414} (0.107) & 4.567 (0.061) \\
 & 3 & 3.225 (0.122) & 3.227 (0.096) & \boldwidth{3.131} (0.118) & 4.567 (0.061) \\
 & 5 & 2.950 (0.133) & 2.815 (0.105) & \boldwidth{2.860} (0.130) & 4.567 (0.061) \\
 & 8 & 2.769 (0.143) & 2.613 (0.107) & \boldwidth{2.684} (0.138) & 4.567 (0.061) \\
\addlinespace
\multirow{5}{*}{$\hat p$} & 1.001 & 4.198 (0.090) & 4.031 (0.054) & 4.001 (0.227) & 4.032 (0.054) \\
 & 2 & 3.669 (0.072) & 3.499 (0.054) & 3.381 (0.206) & 4.032 (0.054) \\
 & 3 & 3.415 (0.075) & 3.261 (0.064) & 3.103 (0.194) & 4.032 (0.054) \\
 & 5 & 3.165 (0.083) & 3.044 (0.078) & 2.843 (0.183) & 4.032 (0.054) \\
 & 8 & 3.000 (0.090) & 2.908 (0.088) & 2.679 (0.178) & 4.032 (0.054) \\
\addlinespace
\bottomrule
\end{tabular}}
\par\smallskip\begin{minipage}{\textwidth}\scriptsize {Entries are Monte Carlo means (empirical SEs across 500 repetitions). PIG is independent of $\Lambda$ and $\hat Q_{z\pm}$, and targets the PI value rather than a sensitivity lower bound.} \end{minipage}
\end{table}

\clearpage
\begin{table}[tbp]
\centering
\caption{True worst-case policy performance at $\Lambda=5$, $n=5000$, over 500 repetitions. The oracle maximum over linear policies is 2.88356. See the caption of Table \ref{tab:sim_oracle_2000} for a description of the shared elements.%
  % Draft return reference; comment out this line for camera-ready.
}
\label{tab:sim_oracle_5000}
\setlength{\tabcolsep}{2.5pt}
\scriptsize
\resizebox{\textwidth}{!}{%
\begin{tabular}{llrrrrrrrr}
\toprule
\multirow{2}{*}{$\hat Q_{z\pm}$} & \multirow{2}{*}{Misspec.} & \multicolumn{2}{c}{IPW} & \multicolumn{2}{c}{OR} & \multicolumn{2}{c}{DR} & \multicolumn{2}{c}{PIG} \\
\cmidrule(lr){3-4}\cmidrule(lr){5-6}\cmidrule(lr){7-8}\cmidrule(lr){9-10}
 &  & WCV & Gap & WCV & Gap & WCV & Gap & WCV & Gap \\
\midrule
\multirow{4}{*}{Specified} & None & 2.864 (0.038) & 0.020 & \boldwidth{2.877} (0.010) & \boldwidth{0.007} & 2.867 (0.023) & 0.017 & 2.802 (0.020) & 0.082 \\
 & $\hat e_1$ & 2.866 (0.004) & 0.017 & \boldwidth{2.877} (0.010) & \boldwidth{0.007} & 2.872 (0.018) & 0.012 & 2.802 (0.020) & 0.082 \\
 & $\hat\mu$ & 2.864 (0.038) & 0.020 & 2.766 (0.030) & 0.117 & \boldwidth{2.867} (0.022) & \boldwidth{0.016} & 2.747 (0.000) & 0.137 \\
 & $\hat p$ & 2.863 (0.015) & 0.020 & \boldwidth{2.879} (0.005) & \boldwidth{0.005} & 2.861 (0.029) & 0.023 & 2.802 (0.020) & 0.082 \\
\addlinespace
\multirow{4}{*}{Misspecified} & None & \boldwidth{2.874} (0.010) & \boldwidth{0.009} & \boldwidth{2.874} (0.015) & \boldwidth{0.010} & 2.871 (0.011) & 0.013 & 2.802 (0.020) & 0.082 \\
 & $\hat e_1$ & 2.866 (0.004) & 0.017 & \boldwidth{2.874} (0.015) & \boldwidth{0.010} & 2.863 (0.013) & 0.021 & 2.802 (0.020) & 0.082 \\
 & $\hat\mu$ & \boldwidth{2.874} (0.010) & \boldwidth{0.009} & 2.768 (0.046) & 0.116 & 2.870 (0.011) & 0.013 & 2.747 (0.000) & 0.137 \\
 & $\hat p$ & 2.865 (0.010) & 0.018 & \boldwidth{2.881} (0.004) & \boldwidth{0.003} & 2.872 (0.011) & 0.011 & 2.802 (0.020) & 0.082 \\
\addlinespace
\bottomrule
\end{tabular}}
\par\smallskip\begin{minipage}{\textwidth}\scriptsize {WCV entries are Monte Carlo means (empirical SEs across 500 repetitions); Gap entries are mean regrets. Numerical integration uses the true conditional normal laws and $\pi_C(X)$.
%; its agreement with adaptive quadrature is within $10^{-8}$. PIG is trained under principal ignorability, but evaluated under the same $\Lambda=5$ sensitivity model. Boldface is descriptive, not a significance test.
}\end{minipage}
\end{table}

\clearpage
\begin{table}[tbp]
\centering
\caption{Method-specific maximized value estimates across sensitivity parameters, $n=5000$, $\hat Q_{z\pm}$ correctly specified, over 500 repetitions. See the caption of Table \ref{tab:sim_estimated_2000_0} for a description of the shared elements.%
  % Draft return reference; comment out this line for camera-ready.
}
\label{tab:sim_estimated_5000_0}
\setlength{\tabcolsep}{2.5pt}
\scriptsize
\resizebox{\textwidth}{!}{%
\begin{tabular}{llrrrr}
\toprule
\multirow{2}{*}{Misspec.} & \multirow{2}{*}{$\Lambda$} & IPW & OR & DR & PIG \\
\cmidrule(lr){3-3}\cmidrule(lr){4-4}\cmidrule(lr){5-5}\cmidrule(lr){6-6}
 & & Estimated LB & Estimated LB & Estimated LB & Estimated PI value \\
\midrule
\multirow{5}{*}{None} & 1.001 & \boldwidth{4.050} (0.090) & \boldwidth{3.943} (0.036) & \boldwidth{3.977} (0.135) & 3.944 (0.036) \\
 & 2 & \boldwidth{3.501} (0.082) & \boldwidth{3.395} (0.041) & \boldwidth{3.427} (0.128) & 3.944 (0.036) \\
 & 3 & \boldwidth{3.244} (0.084) & \boldwidth{3.134} (0.051) & \boldwidth{3.169} (0.127) & 3.944 (0.036) \\
 & 5 & \boldwidth{3.002} (0.092) & \boldwidth{2.891} (0.065) & \boldwidth{2.932} (0.131) & 3.944 (0.036) \\
 & 8 & \boldwidth{2.847} (0.100) & \boldwidth{2.741} (0.075) & \boldwidth{2.783} (0.136) & 3.944 (0.036) \\
\addlinespace
\multirow{5}{*}{$\hat e_1$} & 1.001 & 5.738 (0.078) & \boldwidth{3.943} (0.036) & \boldwidth{3.963} (0.060) & 3.944 (0.036) \\
 & 2 & 4.984 (0.084) & \boldwidth{3.395} (0.041) & \boldwidth{3.419} (0.060) & 3.944 (0.036) \\
 & 3 & 4.602 (0.091) & \boldwidth{3.134} (0.051) & \boldwidth{3.158} (0.063) & 3.944 (0.036) \\
 & 5 & 4.214 (0.101) & \boldwidth{2.891} (0.065) & \boldwidth{2.913} (0.072) & 3.944 (0.036) \\
 & 8 & 3.951 (0.110) & \boldwidth{2.741} (0.075) & \boldwidth{2.759} (0.080) & 3.944 (0.036) \\
\addlinespace
\multirow{5}{*}{$\hat\mu$} & 1.001 & \boldwidth{4.050} (0.090) & 4.563 (0.037) & \boldwidth{4.000} (0.061) & 4.564 (0.037) \\
 & 2 & \boldwidth{3.501} (0.082) & 3.729 (0.041) & \boldwidth{3.433} (0.076) & 4.564 (0.037) \\
 & 3 & \boldwidth{3.244} (0.084) & 3.325 (0.049) & \boldwidth{3.171} (0.090) & 4.564 (0.037) \\
 & 5 & \boldwidth{3.002} (0.092) & 2.928 (0.057) & \boldwidth{2.931} (0.106) & 4.564 (0.037) \\
 & 8 & \boldwidth{2.847} (0.100) & 2.699 (0.064) & \boldwidth{2.782} (0.118) & 4.564 (0.037) \\
\addlinespace
\multirow{5}{*}{$\hat p$} & 1.001 & 4.141 (0.046) & 4.022 (0.034) & 3.980 (0.143) & 4.022 (0.034) \\
 & 2 & 3.663 (0.043) & 3.543 (0.033) & 3.413 (0.141) & 4.022 (0.034) \\
 & 3 & 3.428 (0.045) & 3.321 (0.038) & 3.150 (0.142) & 4.022 (0.034) \\
 & 5 & 3.197 (0.048) & 3.113 (0.046) & 2.912 (0.146) & 4.022 (0.034) \\
 & 8 & 3.045 (0.052) & 2.981 (0.053) & 2.765 (0.151) & 4.022 (0.034) \\
\addlinespace
\bottomrule
\end{tabular}}
\par\smallskip\begin{minipage}{\textwidth}\scriptsize {Entries are Monte Carlo means (empirical SEs across 500 repetitions). PIG is independent of $\Lambda$ and $\hat Q_{z\pm}$, and targets the PI value rather than a sensitivity lower bound.}\end{minipage}
\end{table}

\clearpage
\begin{table}[tbp]
\centering
\caption{Method-specific maximized value estimates across sensitivity parameters, $n=5000$, $\hat Q_{z\pm}$ misspecified, over 500 repetitions. See the caption of Table \ref{tab:sim_estimated_2000_0} for a description of the shared elements.%
  % Draft return reference; comment out this line for camera-ready.
}
\label{tab:sim_estimated_5000_1}
\scriptsize
\resizebox{\textwidth}{!}{%
\begin{tabular}{llrrrr}
\toprule
\multirow{2}{*}{Misspec.} & \multirow{2}{*}{$\Lambda$} & IPW & OR & DR & PIG \\
\cmidrule(lr){3-3}\cmidrule(lr){4-4}\cmidrule(lr){5-5}\cmidrule(lr){6-6}
 & & Estimated LB & Estimated LB & Estimated LB & Estimated PI value \\
\midrule
\multirow{5}{*}{None} & 1.001 & \boldwidth{4.050} (0.090) & \boldwidth{3.943} (0.036) & \boldwidth{3.977} (0.135) & 3.944 (0.036) \\
 & 2 & \boldwidth{3.430} (0.062) & \boldwidth{3.328} (0.043) & \boldwidth{3.377} (0.118) & 3.944 (0.036) \\
 & 3 & \boldwidth{3.151} (0.068) & \boldwidth{3.052} (0.056) & \boldwidth{3.099} (0.110) & 3.944 (0.036) \\
 & 5 & \boldwidth{2.882} (0.080) & \boldwidth{2.806} (0.069) & \boldwidth{2.826} (0.104) & 3.944 (0.036) \\
 & 8 & \boldwidth{2.705} (0.088) & \boldwidth{2.651} (0.077) & \boldwidth{2.648} (0.101) & 3.944 (0.036) \\
\addlinespace
\multirow{5}{*}{$\hat e_1$} & 1.001 & 5.738 (0.078) & \boldwidth{3.943} (0.036) & \boldwidth{3.963} (0.060) & 3.944 (0.036) \\
 & 2 & 4.922 (0.084) & \boldwidth{3.328} (0.043) & \boldwidth{3.378} (0.059) & 3.944 (0.036) \\
 & 3 & 4.508 (0.092) & \boldwidth{3.052} (0.056) & \boldwidth{3.101} (0.061) & 3.944 (0.036) \\
 & 5 & 4.087 (0.103) & \boldwidth{2.806} (0.069) & \boldwidth{2.836} (0.070) & 3.944 (0.036) \\
 & 8 & 3.801 (0.112) & \boldwidth{2.651} (0.077) & \boldwidth{2.674} (0.079) & 3.944 (0.036) \\
\addlinespace
\multirow{5}{*}{$\hat\mu$} & 1.001 & \boldwidth{4.050} (0.090) & 4.562 (0.037) & \boldwidth{4.000} (0.061) & 4.564 (0.037) \\
 & 2 & \boldwidth{3.430} (0.062) & 3.656 (0.044) & \boldwidth{3.382} (0.072) & 4.564 (0.037) \\
 & 3 & \boldwidth{3.151} (0.068) & 3.219 (0.054) & \boldwidth{3.101} (0.079) & 4.564 (0.037) \\
 & 5 & \boldwidth{2.882} (0.080) & 2.792 (0.063) & \boldwidth{2.828} (0.086) & 4.564 (0.037) \\
 & 8 & \boldwidth{2.705} (0.088) & 2.583 (0.069) & \boldwidth{2.649} (0.091) & 4.564 (0.037) \\
\addlinespace
\multirow{5}{*}{$\hat p$} & 1.001 & 4.141 (0.046) & 4.021 (0.034) & 3.980 (0.143) & 4.022 (0.034) \\
 & 2 & 3.623 (0.044) & 3.485 (0.034) & 3.358 (0.130) & 4.022 (0.034) \\
 & 3 & 3.373 (0.046) & 3.247 (0.041) & 3.077 (0.123) & 4.022 (0.034) \\
 & 5 & 3.124 (0.050) & 3.030 (0.051) & 2.813 (0.118) & 4.022 (0.034) \\
 & 8 & 2.960 (0.054) & 2.893 (0.058) & 2.644 (0.115) & 4.022 (0.034) \\
\addlinespace
\bottomrule
\end{tabular}}
\par\smallskip\begin{minipage}{\textwidth}\scriptsize {Entries are Monte Carlo means (empirical SEs across 500 repetitions). PIG is independent of $\Lambda$ and $\hat Q_{z\pm}$, and targets the PI value rather than a sensitivity lower bound.}\end{minipage}
\end{table}

\clearpage
\begin{figure}[!htbp]
\centering
\captionsetup{skip=4pt}
\includegraphics[width=\textwidth]{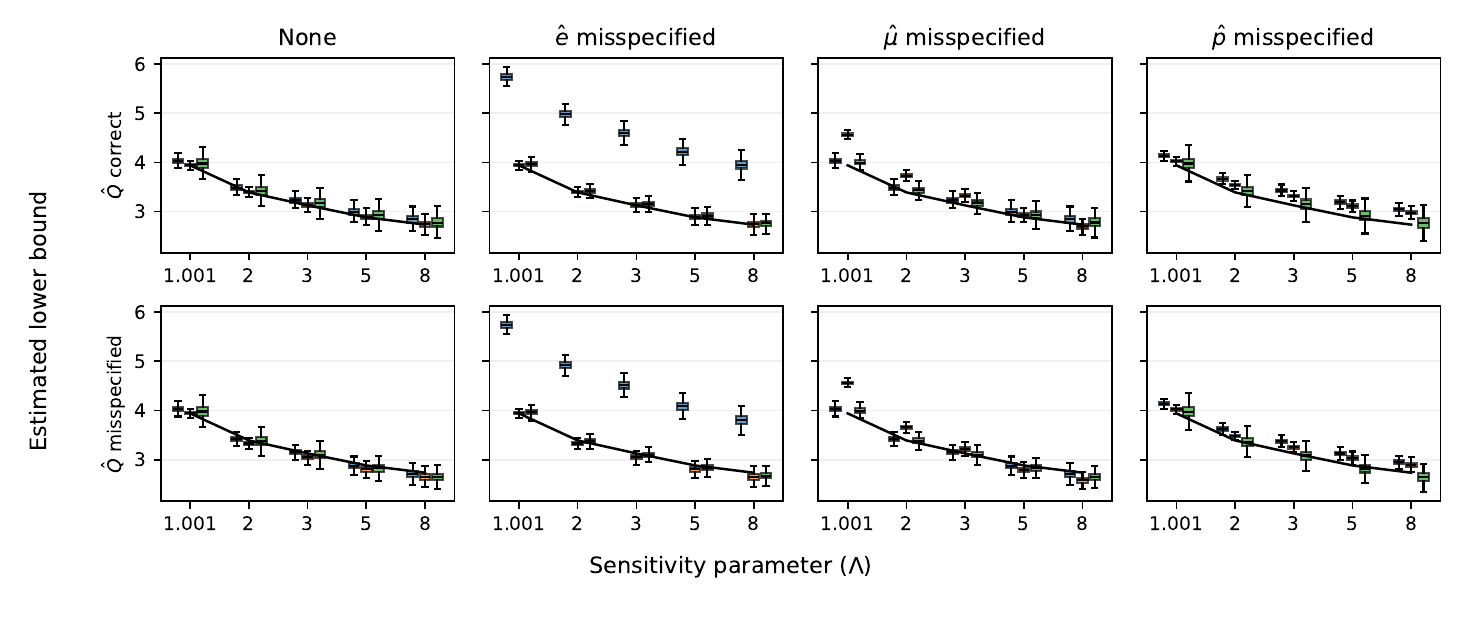}
\includegraphics[width=\textwidth]{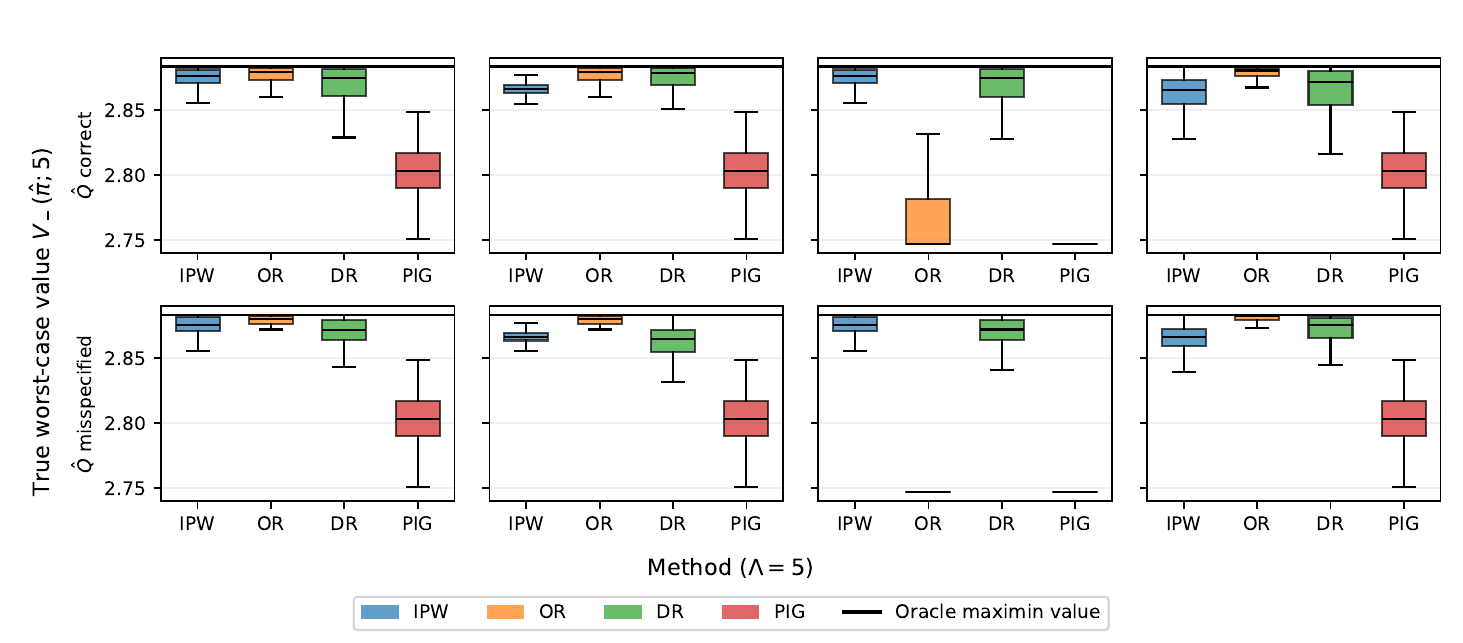}
\caption{Policy learning at $n=5000$. See the caption of Figure~\ref{fig:sim_policy_2000} for a description of the shared elements.%
  % Draft return reference; comment out this line for camera-ready.
}\label{fig:sim_policy_5000}\label{fig:sim_oracle_5000}
\end{figure}
\clearpage

\section{Supplementary OHIE application details}\label{app:oregon_impl}
\subsection{Data and nuisance function estimation}
For the application in Section~\ref{sec:oregon}, we use ten individual baseline variables: age-group indicators \texttt{AG2} and \texttt{AG3}; \texttt{Gender}, \texttt{White}, \texttt{hispanic}, and \texttt{EnglishPreferred}; pre-lottery depression and diabetes indicators \texttt{PREdep} and \texttt{PREdiabetes}; and household lottery-list-size indicators \texttt{houseSize2} and \texttt{houseSize3}. The pre-lottery diagnoses were reported retrospectively. For the variable \texttt{EnglishPreferred}, one denotes a request for materials in a language other than English, and zero denotes a request for English materials. In the analysis sample, survey-weighted enrollment proportions were 0.294 among selected respondents and 0.033 among nonselected respondents; their weighted mean outcomes were 45.102 and 44.387, respectively.

To define household-level policies, let $X^{\mathrm{ind}}_{hi}$ denote the individual baseline vector and let $H_h$ contain its averages across retained survey respondents in household $h$. Each policy is a depth-one tree that splits on one coordinate of $H_h$ and assigns a treatment to each leaf. Nuisance regressions use both levels of information, $X_{hi}=(X^{\mathrm{ind}}_{hi},H_h)$. Although assignment $Z_h$ is shared within the household, potential receipt and outcomes are individual; household members may therefore belong to different principal strata. The value averages over individual compliers using survey weights.

We fit all nuisance regressions using survey weights. The logistic propensity model uses the two lottery-list-size indicators, with ridge penalty $10^{-6}$ and fitted probabilities clipped to $[0.05,0.95]$. For the principal scores, a weighted logistic model includes all covariates and assignment, with penalty 0.1. We project its predictions to $0.01\le\hat p_{0,1}(X)\le0.98$ and $\hat p_{0,1}(X)+0.01\le\hat p_1(X)\le0.99$. Across fits, the maximum proportions of predictions affected by principal-score projection and propensity clipping were 11.79\% and 0.37\%, respectively. 
Within each assignment--receipt group, random forests for outcome means and hinge regressions use 200 trees, maximum depth five, and minimum leaf size 15. Gradient boosting for quantiles uses 100 trees, learning rate 0.05, maximum depth two, and minimum leaf size 15. Quantile regressions are fitted at 33 levels: 31 equally spaced levels from 0.01 to 0.99 and the additional levels 0.001 and 0.999. We rearrange the predicted quantile curves into nondecreasing order and interpolate at the required levels.

\subsection{Sensitivity bound estimation}
To estimate bounds on $\psi_C$, we use the entire sample with household-level three-fold cross-fitting, repeated over 100 randomized fold assignments. Let $\hat N_{iz\pm}^{\mathrm{DR}}(\Lambda)$ and $\hat D_i^{\mathrm{DR}}$ be the DR numerator and denominator estimating functions. The weighted estimate is
\[
 \widehat\psi_{C\pm}^{\mathrm{DR}}(\Lambda)=
 \frac{\sum_i w_i\{\hat N_{i1\pm}^{\mathrm{DR}}(\Lambda)-\hat N_{i0\mp}^{\mathrm{DR}}(\Lambda)\}}
 {\sum_i w_i\hat D_i^{\mathrm{DR}}}.
\]
Here, $\hat D^{\mathrm{DR}}$ is the fitted version of \eqref{eq:D}. The finer sensitivity grid includes 1, 1.001, and increments of 0.01 from 1.01 to 1.06. On this grid, the estimated DR interval contained zero in 22\% of cross-fits at $\Lambda=1.04$ and 91\% at $\Lambda=1.05$.

\begin{table}[!htbp]
\centering
\caption{DR bounds on the complier treatment effect in SF-8 mental-component-score units. Entries give median endpoints across 100 household cross-fits, with median within-cross-fit bootstrap SEs in parentheses. Each cross-fit uses 500 household draws with all nuisance functions fixed. Width is the median difference between upper and lower endpoints.}
\label{tab:oregon_effect}
\begin{tabular}{rccc}
\toprule
$\Lambda$ & Lower & Upper & Width \\
\midrule
1 & 0.294 (0.317) & 0.294 (0.317) & 0.000 \\
1.1 & -0.363 (0.316) & 0.955 (0.319) & 1.317 \\
1.3 & -1.494 (0.315) & 2.119 (0.324) & 3.612 \\
1.5 & -2.434 (0.316) & 3.113 (0.330) & 5.545 \\
2 & -4.213 (0.321) & 5.058 (0.344) & 9.270 \\
3 & -6.407 (0.333) & 7.586 (0.369) & 13.992 \\
\bottomrule
\end{tabular}
\end{table}
\clearpage

\subsection{Maximin policy learning}
For policy learning, we instead evaluate learned policies on held-out households. Each of 100 outer splits assigns 80\% of households to training and 20\% to testing, with household-level three-fold cross-fitting within training. The survey-weighted training criterion is
\[
 \widehat V_{-,\mathrm{train}}^{\mathrm{DR}}(\pi;\Lambda)=
 \frac{\sum_{i\in\mathcal I_{\mathrm{train}}}w_i\{(1-\pi_i)\hat N_{i0-}^{\mathrm{DR}}(\Lambda)+\pi_i\hat N_{i1-}^{\mathrm{DR}}(\Lambda)\}}
 {\sum_{i\in\mathcal I_{\mathrm{train}}}w_i\hat D_i^{\mathrm{DR}}},\qquad \pi_i=\pi(H_{h(i)}).
\]
We maximize the DR training criterion at each sensitivity level by exhaustive enumeration of split coordinates, cutoffs and leaf assignments, including constant policies. Each learned DR policy is evaluated on the held-out sample at its training sensitivity level. Constant no-access and universal-access policies are evaluated on the same testing samples.

We estimate sampling uncertainty for the DR effect endpoints and held-out DR policy values using 500 household bootstrap draws for each cross-fit or outer split. For policy evaluation, we resample only the held-out households, keeping the trained policy fixed. Every draw retains all members of each sampled household and their survey weights, and recomputes the ratio of the summed numerator contributions to the summed denominator contributions. All fitted nuisance functions remain fixed. We justify this choice by using the asymptotic linearity and error rate conditions in Theorem~\ref{thm:TR}(iii)--(iv), applied to the DR effect and fixed-policy estimating functions.
For each endpoint or policy value, the bootstrap SE is the sample standard deviation of its 500 bootstrap estimates. The tables report the median of these SEs across the 100 cross-fits or outer splits, alongside the median point estimate.

\begin{table}[!htbp]
\centering
\caption{Held-out DR lower-bound values of the learned DR depth-one policies, evaluated at their training sensitivity level. Entries give medians across 100 outer splits, with median within-split household-bootstrap SEs in parentheses. Each split uses 500 draws with the policy and all nuisance functions fixed.}
\label{tab:oregon_policy_se}
\begin{tabular}{rc}
\toprule
$\Lambda$ & DR value \\
\midrule
1 & 44.133 (0.618) \\
1.1 & 43.920 (0.650) \\
1.3 & 43.799 (0.663) \\
1.5 & 43.689 (0.670) \\
2 & 43.486 (0.678) \\
3 & 43.187 (0.694) \\
\bottomrule
\end{tabular}
\end{table}

\clearpage

\end{document}